\documentclass[11pt,a4paper]{article}
\usepackage{jcappub}
\usepackage[utf8]{inputenc}
\usepackage[table]{xcolor}
\usepackage{verbatim}
\usepackage{amsmath,amsfonts,amssymb}
\usepackage{subfloat}
\usepackage{float}
\usepackage{xcolor}
\usepackage{braket}
\usepackage{adjustbox}
\usepackage{graphicx}
\usepackage{hyperref}
\usepackage{array, booktabs, tabularx, makecell, multirow}
\usepackage{multicol}
\usepackage{tikz}
\usetikzlibrary{shapes.geometric, arrows}
\usepackage{tabularx}
\usepackage[normalem]{ulem}
\usepackage{multirow}
\usepackage{xfrac}
\usepackage{makecell}
\usepackage{lmodern}

\newcolumntype{s}{>{\hsize=.5\hsize}X}

\usepackage{pifont}

\newcommand{\gayy}{g_{{\rm a}\gamma\gamma}}
 
\def\mnras{Monthly Notices of the Royal Astronomical Society}
\def\aap{Astronomy and Astrophysics}

\def\apj{Astrophysical Journal}

\def\prd{Physical Review D}
\def\prl{Physical Review Letters}
\def\nat{Nature}
\def\apjl{Astrophysical Journal, Letters}
\def\pasa{Publications of the Australian Society of Astronomy}

\title{Axion signals from neutron star populations}

\author[b]{U.~Bhura}
\author[b]{R.~A.~Battye}
\author[b]{J.~I.~McDonald}
\author[c]{S.~Srinivasan}

\affiliation[b]{Department of Physics and Astronomy, University of Manchester, Oxford Road, Manchester, M13 9PL, United Kingdom}
\affiliation[c]{Universit\"{a}ts-Sternwarte, Fakult\"{a}t f\"{u}r Physik, Ludwig-Maximilians Universit\"{a}t, Scheinerstraße 1, D-81679 M\"{u}nchen, Germany}

\emailAdd{utkarsh.bhura@postgrad.manchester.ac.uk}
\emailAdd{richard.battye@manchester.ac.uk}
\emailAdd{jamie.mcdonald@manchester.ac.uk}
\emailAdd{ssriniva@usm.lmu.de}

\date{Today}

\abstract{
Neutron stars provide a powerful probe of axion dark matter, especially in higher frequency ranges where there remain fewer laboratory constraints. Populations of neutron stars near the Galactic Centre have been proposed as a means to place strong constraints on axion dark matter. One downside of this approach is that  there are very few direct observations of neutron stars in this region, introducing uncertainties in the total number of neutron stars in this ``invisible" population at the Galactic Centre, whose size must be inferred through birth rate modelling. We suggest  this number could also be reduced due to stellar dynamics carrying stars away from the Galactic Centre via large kick velocities at birth.  We attempt to circumvent the uncertainty on the Galactic Centre population size by modelling the axion signal from better understood populations outside the Galactic Centre using {\tt PsrPopPy} which is normalised against pulsar observations. We consider lower-frequency, wider-angle searches for this signal via a range of instruments including MeerKAT and SKA-low  but find that the sensitivity is not competitive with existing constraints. Finally, returning to the Galactic Centre, we compare populations to single objects as targets for axion detection. Using the latest modelling of axion-photon conversion in the Galactic Centre magnetar, we conclude that within astrophysical uncertainties, the Galactic Centre population and the magnetar could give comparable sensitivities to axion dark matter, suggesting one should continue to search for both signals in future surveys.
}

\begin{document}
 
\maketitle

\section{Introduction}
One of the most popular ideas in cosmology at present is that the dark matter may be comprised of axions~\cite{ref:PQ, ref:K, ref:SVZ, ref:DFSZ, ref:Zhit, ref:misalign1, ref:misalign2, ref:misalign3, Arvanitaki:2009fg, Svrcek:2006yi}, whose  detection is the subject of a number of searches~\cite{RBF, Wuensch:1989sa, UF, ADMX:2009iij, ref:ADMX2018, ADMX:2018ogs, ADMX:2019uok, ADMX:2021nhd, ADMX:2021mio, Crisosto:2019fcj, CAPP:Lee_2020cfj, CAPP:Jeong_2020cwz, CAPP:Lee_2022mnc, CAPP:Kim2022, HAYSTAC:Brubaker2017, HAYSTAC:2018rwy, HAYSTAC:2020kwv, Alesini:2019ajt, Alesini:2020vny, ORGAN1, ORGAN2, CAST:2020rlf, TASEH:2022vvu, Grenet:2021vbb}. Complementary to these laboratory experiments are indirect searches for radio signatures produced by axion dark matter converting into photons in the magnetospheres of neutron stars (NSs)~\cite{Pshirkov:2007st, Huang:2018lxq, hook2018, Battye_2020, walters2024axions}.

Predictions for the luminosity from numerical simulations \cite{leroy2020, Battye:2021xvt, Witte:2021arp, McDonald:2023shx} has been made within the framework of the Goldreich-Julian (GJ) model~\cite{goldreich1969} for the star's magnetosphere. These depend on the magnetic field at the surface of the star, $B_0$, the period of rotation, $P$ and the angle between the spin and magnetic axes, $\alpha$, the viewing angle $\theta$ and the dark matter density $\rho_{\rm DM}$ at the location of the star\footnote{In addition there is a very weak dependence on the mass and radius of the NS, $M_{\rm NS}$ and $R_{\rm NS}$, respectively. Throughout the analysis presented here we will take $M_{\rm NS}=1\,M_{\odot}$ and $R_{\rm NS}=10\,{\rm km}$. Even significant variations from these values will only have a minor impact on the predictions.}.

The predicted luminosities for specific individual stars, notably the Galactic Centre Magnetar (GCM), have been used to obtain constraints on the axion-photon coupling, $g_{a\gamma\gamma}$ as a function of axion mass, $m_{\rm a}$. Some searches have used the total flux expected from isolated stars~\cite{foster2020, Darling:2020plz, darling2020apj, Battye2022}, while others have examined the time-dependence with pulse phase \cite{Battye:2023oac}.

An alternative approach of using a population of NSs was suggested in refs.~\cite{Safdi2019, Foster:2022fxn}. The radio signatures for an individual star depend on a number of parameters, namely $P$, $B_0$, $\alpha$ and $\theta$. Some or all of these parameters might not be known exactly. One can effectively marginalise over these parameters by integrating over all stars in a population with appropriate distributions for $(B_0, \, P, \,\theta,\, \alpha)$. In addition, for standard NSs in the population, the magnetosphere is believed to resemble more closely that of the GJ model, while for magnetars, less is known about the precise details of the structure of the magnetosphere. The population approach has been used in the context of Breakthrough-Listen (BL) observations of the Galactic Centre (GC) to deduce a strong upper limit on the axion-photon copuling $g_{a\gamma\gamma}\lesssim 10^{-11}\,{\rm GeV}^{-1}$~\cite{Foster:2022fxn}. This constraint applies to a significant portion of the range $16 \,\mu{\rm eV}\lesssim m_{\rm a}\lesssim 33 \,\mu{\rm eV}$, which corresponds to the observing frequency range of 4-8~GHz, sometimes known as the C-band~\footnote{In what follows we will use L-band to refer to the frequency range 1-2~GHz and C-band to 4-8~GHz.}.

However, this approach brings in additional uncertainties associated with the modelling of the NSs population itself, particularly a putative ``invisible population'' very close to the centre of the Galaxy. The objective of this paper is to investigate these uncertainties more closely and then examine whether they can be circumvented by probing different sub-populations of NSs within our Galaxy. For convenience we summarise our principle conclusions below.

\begin{itemize}
    \item \textbf{Uncertainties on the properties of the  ``invisible GC NS population''}\\
    We point out there have been almost no (or, at best, only a few) normal pulsars observed in the central $1{\rm pc}^3$ of the Galaxy. As a result, there is no direct data either on the total number of stars at the GC, or on their population statistics, i.e., their magnetic fields and periods. We are, therefore, essentially blind to this population of GC NSs and have no direct measure of the total number of NSs, $N_{\rm GC}$, in that region. Instead, this population has been inferred indirectly by modelling its birth rate and subsequent evolution of the periods and magnetic fields of these stars. We also point out that owing to high kick velocities at birth, these stars can propagate away from the GC, with the fraction of stars remaining at the GC being highly sensitive to the depth of the gravitational potential and the distribution of their initial velocities (see appendix \ref{appendix:diffusion}). For the fiducial models considered in this work, we find the fraction of stars remaining in the GC is $\mathcal{O}(\text{few} \%)$ of the total number birthed, though the fraction can be higher in different models \cite{Jurado:2023lkg}. We suggest, therefore, that this necessitates a closer study of stellar dynamics and birth rates near the GC in order to better infer the number of stars at the GC. 
    \item \textbf{Probing axions with normal pulsar populations and {\tt PsrPopPy}}\\
    In light of the uncertainty in the size of the population at the GC, we instead investigate whether regions close to, but outside the very central region could be used to probe axions. To achieve this we employ the open-source package {\tt PsrPopPy} used to model NS populations across the Galaxy. Crucially, the population statistics, distribution and overall number of stars are calibrated using real observations from, e.g., the Parkes Multi Beam Survey (PMBS)~\cite{Manchester2001_PMBS}. As a result, we obtain a robust population model, circumventing the uncertainties inherent in the obscured GC region.
  \item \textbf{Bespoke and future instruments for detection of axion signals from populations}\\
  With the simplest existing telescopes, constraints from the wider galactic NS population would already be excluded by existing lab searches. We therefore consider both bespoke instruments and next-generation telescopes. 
  
   Even with a best-case scenario based on future facilities like SKA-low, constraints from the wider galactic NS population would be complementary to constraints from pulsar caps \cite{Noordhuis:2022ljw} derived from existing telescopes or with lab searches. We therefore conclude that if one is to leverage the power of NS populations to get competitive constraints, one is likely forced to use the hidden NS population at the GC and confront the current uncertainties in its size and properties.

    \item \textbf{Populations vs.~Individual Stars \& the Galactic Centre Magnetar}\\
    In light of the uncertainties concerning the GC populations, and the uncertainties on the precise magnetosphere structure and observing direction of the Galactic Centre magnetar (GCM), we find that, within these uncertainties, the signal strengths of the GC population and the GCM could be comparable in size. Note this includes the latest modelling of axion signatures from the Magnetar considered in \cite{Tjemsland:2023vvc}. Furthermore, in light of the diffusion of the GC population into the wider Galactic potential, the number of stars in the GC is likely to be weaker than previously thought. By precisely how much will require further studies of stellar dynamics near the GC. As a result, when carrying out observations (including future programmes with the SKA) one should perform searches for both signals, keeping an open mind as to which is strongest.
\end{itemize}

The structure of this paper is organised as follows. In section~\ref{sec:luminosity} we describe the fundamentals of radio signals from NSs and introduce our approach for calculating the signal from a pulsar which is contributing to a population. Rather than using the ray-tracing approach that is computationally expensive, this uses the flux from the star averaged over all orientations. In addition, we outline our treatment of cyclotron resonances and re-conversion of photons into axions, both of which occur for more strongly magnetised stars leading to a suppression of the signal. In section~\ref{sec:populations} we discuss the various components of our model which includes (i) the background axion signal from the most conservative population of NSs in the Galaxy using the open-source package {\tt PsrPopPy}; (ii) the signal from a putative `invisible" NS population located in the GC and (iii) the observed GCM.  In section~\ref{sec:Constraints} we combine results from both these proceeding sections to compute the signal strengths from various population models and compare this to the signal from the GCM. We deduce constraints on $\gayy$ for a range scenarios and also discuss an idea for a bespoke axion radio telescope designed to be sensitive to the easier to predict signal expected due to ``normal pulsar" modelled using {\tt PsrPopPy} \footnote{Note that we use Natural Units where $\hslash = c = 1$. Therefore, care must be taken when applying our equations to compute observational quantities like brightness temperature and flux density to put back in the appropriate powers of $c$.}. Finally, in section~\ref{sec:conclusions} we summarise our conclusions.  

\section{Modelling radio signatures from axions in neutron stars}\label{sec:luminosity}

Computing radio signatures produced by axion dark matter in NS magnetospheres has been the subject of a number of studies in recent years. This has involved numerical modelling of photon transport \cite{leroy2020,Witte:2021arp,Battye:2021xvt,McDonald:2023shx,Tjemsland:2023vvc} using high-performance computing, and analytic \cite{Battye_2020,Carenza:2023nck,McDonald:2023ohd,McDonald:2024uuh} and numerical \cite{Gines:2024ekm} calculations of the production process itself\footnote{It has been suggested~\cite{Noordhuis:2022ljw} that there can be signals from regions such as the polar caps of pulsars. These are not the main focus of this work, but would only serve to tighten the constraints on the coupling constant and are discussed, briefly, in section~\ref{sec:future}.}.  These approaches involve ray-tracing individual photons from the magnetosphere in order to reconstruct the signal flux in different observing direction. However, below, since we aim to marginalise over population parameters, we can essentially circumvent ray-tracing by averaging over observing directions. This allows one (up to absorbative effects discussed below) to essentially circumvent ray-tracing, compute the surface flux from the star, and divide by $4\pi$. Ray-tracing then only contributes to the statistical error incurred from the sample of size $N$, which depends on the variance of the signal with observing direction and the number of stars in the sample. This will be discussed more below, but for now we proceed to describe our procedure for computing the marginalised flux from a stellar population. 

\subsection{Calculation of the axion signal averaging over orientations}

As pointed out in ref.~\cite{foster2020}, for normal active NSs \cite{Philippov:2014mqa,Hu:2021nxu}, we expect the Goldreich-Julian  (GJ) \cite{goldreich1969} model to be a reasonable description for the magnetosphere. The expectation would be that  magnetars (see the discussion in ref.~\cite{McDonald:2023ohd}) may differ more from the GJ model. As a result, for a population study containing large numbers of normal pulsars the GJ model should provide a good approximation. More concretely, the GJ model gives a distribution of charge carriers of the form $n_{\mathrm{GJ}} (r, \theta, \phi) = 2\mathbf{\Omega}\cdot\mathbf{B}/e$
where, 
\begin{align} \label{B-eq1}
    B_r (r, \theta, \phi) & = B_0\left(\frac{R}{r}\right)^3(\cos \alpha \cos \theta+\sin \alpha \sin \theta \cos \psi)\,, \\ \label{B-eq2}
    B_\theta (r, \theta, \phi) & =\frac{B_0}{2}\left(\frac{R}{r}\right)^3(\cos \alpha \sin \theta-\sin \alpha \cos \theta \cos \psi)\,, \\ \label{B-eq3}
    B_\phi (r, \theta, \phi) & =\frac{B_0}{2}\left(\frac{R}{r}\right)^3 \sin \alpha \sin \psi\,,
\end{align}
is the magnetic field of an inclined rotating magnetic dipole, $B_0$ is the strength at the poles, and $\alpha$ is the inclination angle of the magnetic dipole relative to the spin axis. Here, $(r, \theta, \phi)$ are coordinates in the spherical system, $t$ is the time, and $\psi = \phi - \Omega t$. Coordinates are chosen so that $\boldsymbol{\Omega} = \Omega \hat{\textbf{z}}$ and  relativistic effects have been excluded. The plasma frequency is then given by
\begin{align}
    \omega_{\rm P} = & \sqrt{\frac{4 \pi \alpha_{\rm EM} |n_{\rm GJ}|}{m_{\rm e}}}\,,
\end{align}
where $\alpha_{\rm EM}=1/137$ is the Fine-Structure constant and $m_{\rm e}$ is the mass of the electron. Axions and photons can mix resonantly in regions where $\omega_{\rm P} \simeq m_{\rm a}$, which describes a 2D surface (or, more strictly speaking, a foliation of 2D surfaces in phase space \cite{McDonald:2023ohd,McDonald:2023shx}, labelled by $\Sigma_\textbf{k}$, where $\textbf{k}$ is the 3-momentum of the axion) within the plasma. By summing the contribution from all photons produced at this surface, and transporting them through the plasma, one can obtain predictions for the emitted power of NSs. 

When performing single object observations, one has to compute the power flowing into a particular solid angle $d \mathcal{P}(\theta,\phi)/d\Omega$ pointing away from the star in the direction of the observer located along the line of sight in the direction $(\theta, \phi)$. Computing the angular power $d\mathcal{P}/d\Omega$ in a given $(\theta,\phi)$ direction requires detailed plasma ray-tracing to track photons from their emission to detection \cite{Battye_2020,Battye:2021xvt,McDonald:2023ohd}. However, since we are dealing here with populations, integrating over different orientations of stars within the population is equivalent to averaging over all $(\theta,\phi)$ the total flux, $F$, from the population, measured on Earth, is given by
\begin{align}\label{eq:PopulationSum}
  F & =  \sum_{\textbf{x}_i,B_i,P_i} \ \sum_{\theta_i,\phi_i} \frac{1}{\left| \textbf{x}_i\right|^2}\frac{d\mathcal{P}(\theta_i,\phi_i,\textbf{x}_i,B_i,P_i)}{d\Omega} n_{\rm ns}(\textbf{x}_i,B_i,P_i)  \nonumber \\[5pt]
  &\simeq \sum_{\textbf{x}_i,B_i,P_i}  \frac{L(\textbf{x}_i,B_i,P_i)}{\left| \textbf{x}_i\right|^2} n_{\rm ns}(\textbf{x}_i,B_i,P_i)\,,
\end{align}
where $L(\textbf{x}_i,B_i,P_i)$ is the integrated luminosity of a star with magnetic field and period given by $B_i$ and $P_i$, respectively, and $n_{\rm ns}$ is the corresponding number density of NSs at the point $\textbf{x}_i$ with those values. Of course, this assumes that the flux is collected from an infinitesimal volume across which the dark matter density can be treated as constant, and where all stars can be treated as being the same distance from Earth. This allows the averaging over $(\theta_i,\phi_i)$ to be performed in a separable way from other population parameters. 

The luminosity of a single star from resonant conversion of axions has been computed in ref.~\cite{Witte:2021arp,McDonald:2023ohd} and is given by
\begin{eqnarray}\label{eq:IntegratedPower}
  L =  \int d^3 \textbf{k} \int d  \boldsymbol{\Sigma}_\textbf{k}\cdot \textbf{v}_{\rm a} P_{{\rm a} \gamma} \, \omega f_{\rm a}\,.
\end{eqnarray}
where 
\begin{eqnarray}\label{eq:Pag}
    P_{{\rm a} \gamma}  = \frac{\pi}{2} \frac{\gayy^2\left| \textbf{B} \right|^2E_\gamma ^4 \sin ^2 \theta_B }{\cos ^2 \theta_B
   \, \omega _{\rm P}^2 \left(\omega _{\rm P}^2-2 E_\gamma ^2\right)+E_\gamma ^4} 	\frac{1}{\left| \textbf{v}_a \cdot \nabla_\textbf{x} E_\gamma  \right|}\,,
\end{eqnarray}
is a conversion probability which describes the ratio of axion and photon phase-space densities at the point of conversion. $\theta_B$ is the angle between $\textbf{k}$ and $\textbf{B}$, and $E_\gamma = \omega$ is the photon energy, which for the modes relevant for transport (the Langmuir Ordinary (LO) mode) is given by
\begin{eqnarray}
    E_\gamma^2 = \frac{1}{2} \Bigg[ |\textbf{k}|^2 + \omega_{\rm P}^2 + \sqrt{|\textbf{k}|^4  + \omega_{\rm P}^4 + 2\omega_{\rm P}^2 |\textbf{k}|^2(1 - 2 \cos^2 \theta_B)} \Bigg]\,.
\end{eqnarray}
The gradient in the denominator of \eqref{eq:Pag} is understood to act on $\textbf{B}(x)$ contained implicitly in the angle $\theta_B$, as well as the plasma frequency $\omega_{\rm P}^2(x)$. The axion phase-space density $f_a$,\footnote{Not to be confused with the axion decay constant which is often denoted in exactly the same way.} is taken from refs.~\cite{hook2018,Battye:2021xvt,Witte:2021arp,leroy2020,McDonald:2023shx} and we use the fact that the axion density is enhanced due to the gravitational field of the star, so that 
\begin{eqnarray}
  	f_{\rm a}(\textbf{x},\textbf{k}) =
  v_{\rm a} n_{{\rm a,c}} \frac{\delta\left( \omega - \omega_c
  \right)}{4 \pi k^2 }  \, , \qquad   n_{\rm a} \simeq n_{\rm a}^\infty \sqrt{ \frac{2GM_{\rm NS}}{\left|\textbf{x}\right|} } \frac{1}{v_0}\,,
\end{eqnarray} 
where $M_{\rm NS}$ is the mass of the star, $v_{\rm a}$ is the phase velocity at the point $\textbf{x}$ around the star and  $n_{\rm a}^\infty$ is the number density of axions in the halo where the star is located. We show the signal variation as a function of pulsar parameters and frequency in fig.~\ref{fig:lvbpam}. Note that in previous work \cite{Witte:2021arp}, there was some speculation as to whether refraction of photons relative to incoming axions could lead to suppression of the conversion probability. However, it has now been shown that no such effects occur, and that refraction of the photon is already accounted for by the energy gradients appearing in Eq.~\eqref{eq:Pag}. The validity of this formula has been further confirmed through numerical simulations~\cite{Gines:2024ekm} and  independent analytic calculations~\cite{McDonald:2024uuh}, giving three cross-checks of this result.

It is clear from \eqref{eq:PopulationSum}  that instead of needing to carry out ray tracing for each value $(\theta_i,\phi_i)$ and then re-summing the results,  one should formally marginalise over unknown parameters ($\theta_i,\phi_i,\textbf{B}_i, P_i$), compute the average signal and then quote the statistical uncertainty on this quantity, which scales as $\sigma/\sqrt{N}$, where $\sigma$ is the variance due to the distribution of these parameters, and $N$ is the number of stars in the sample. This provides a rigorous basis on which to perform a stochastic analysis for populations of stars. The contribution to the statistical error from $(\theta,\phi)$ (which necessitates a ray-tracing study) is discussed more in appendix~\hyperref[sec:rayTracing]{B.2}. 

\subsection{Absorbative effects}
\label{sec:absorb}

There are two further more subtle caveats to the discussion above which relate to the most strongly magnetised NSs. Namely that photons in strongly magnetised NSs experience reabsorption both from standard model processes (in the form of the Cyclotron resonance) and from re-conversion into axions when the conversion probability becomes so large the conversion is strongly adiabatic~\cite{Tjemsland:2023vvc}. We shall now discuss each of these in turn, with a view to how these phenomena play out at a population level, since this is all that is relevant in the present paper. 

\subsubsection{Cyclotron resonance}\label{sec:Cyclotron}

Firstly, for more strongly magnetised stars, radio signals will experience attenuation due to the Cyclotron resonance~\cite{Witte:2021arp,McDonald:2023ohd}. In which case the total luminosity of the star is not equal to the surface luminosity in \eqref{eq:IntegratedPower} due to reabsorption by plasma before detection. Formally, when such effects are important one would have yet again to trace photons through the plasma and integrate along their worldlines to determine the degree of Cyclotron damping. We will instead approximate this effect following arguments in refs.~\cite{Witte:2021arp,McDonald:2023shx} by applying an overall damping factor $e^{-\tau}$ to the flux, where $\tau \simeq \frac{\pi}{3} \left(\frac{\omega_{\rm P}^2} {\omega}\right) r$
is the optical depth, and all quantities are understood to be evaluated at the Cyclotron resonance $\omega = \omega_{\rm c}$, where the Cyclotron frequency is given by $\omega_{\rm c} = e B/m_{\rm e}$. Estimating $B \simeq (R/r)^3 B_0$ and using a GJ-like plasma distribution $\omega^2_{\rm P} = 4\pi \alpha_{\rm EM} n_{\rm e}/m_e$ with $n_{\rm e} \simeq 2\Omega B/e$, we can estimate 
\begin{eqnarray}
    \tau \simeq 1.4 \left( \frac{R}{10 {\rm km}}\right)
    \left(\frac{1 {\rm s}}{P}
    \right)\left( \frac{B_0}{10^{14} \; {\rm G
    }}\right)^{1/3} \left( \frac{ \mu {\rm eV}}{\omega}\right)^{1/3}\,.
    \label{eq:Cylotron}
\end{eqnarray}

\begin{figure}
    \centering
    \includegraphics[width = \textwidth]{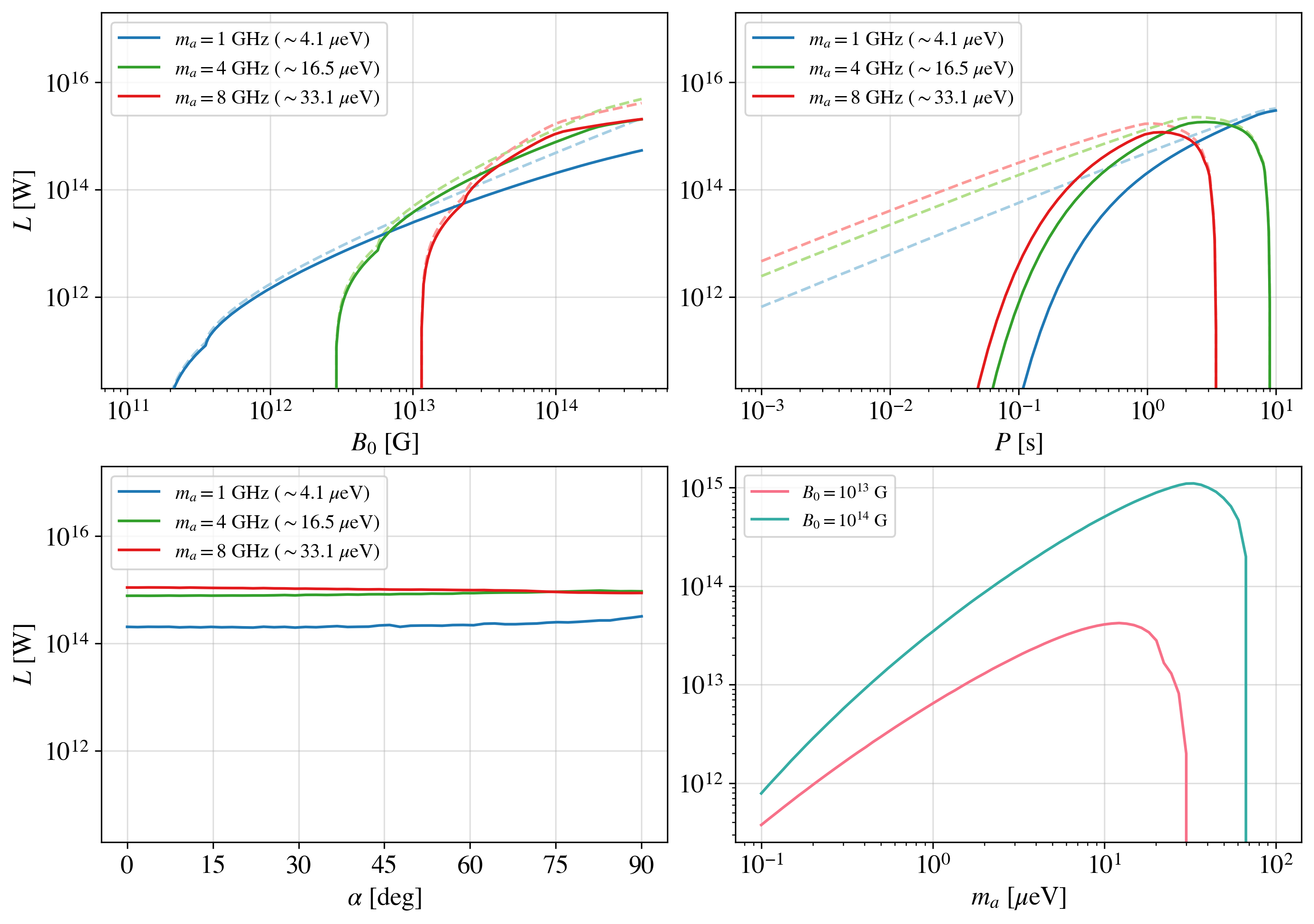}
    \caption{The integrated luminosity, $L$ - see \eqref{eq:IntegratedPower} - of the axion signal for a single star as a function of parameters $B_0$, $P$, $\alpha$ and $m_{\rm a}$. We fix the fiducial values of these parameters to $B_0=10^{14}\,{\rm G}$, $P=1\,{\rm s}$, $\alpha=0^\circ$. The coupling constant is held fixed at $\gayy=10^{-11}\,{\rm GeV}^{-1}$ and the dark matter density is fixed to a local value of $\rho_{\rm DM}|_{\rm local}=0.45\,{\rm GeV}\,{\rm cm}^{-3}$. The solid/dashed lines show $L$ with/without the suppression pre-factor $e^{-\tau}$ due to the Cyclotron resonance discussed in section~\ref{sec:Cyclotron}. The panel in the bottom-right shows the variation of the averaged luminosity \eqref{eq:IntegratedPower} as a function of axion mass $m_{\rm a}$ for two values of the magnetic field. The relatively sharp cut-off in the luminosity is due to the fact that the resonant conversion surface disappears inside the NS surface.}
    \label{fig:lvbpam}
\end{figure}
Clearly the damping is more sensitive to the period than the magnetic field and so slightly less magnetised stars with shorter periods could have significant $\tau$. That being said, it is unlikely to have a star with simultaneously large magnetic fields and short periods, owing to spin-down. To model this effect, we pre-multiply the integrated luminosity \eqref{eq:IntegratedPower} by $e^{- \tau}$. This effect is insignificant for most stars within the Galactic population and even when it is sizeable, it is typically only an ${\cal O}(1)$ factor and has only a minor effect on the total luminosity due to axions, as shown later in fig.~\ref{fig:bfield}. 

\subsubsection{Reconversion of photons into axions}\label{subsubsec:Reconversion}
The other sense in which strongly magnetised stars can modify the arguments at the beginning of this section, is that axion-photon conversion can become hyper-efficient in strongly magnetised environments when the combination $\left|\textbf{B} \right|\gayy$ becomes sufficiently large. In that case, one must use an expression for the conversion probability that is re-summed to all orders in $\gayy$. At present, the working hypothesis in the literature~\cite{Foster:2022fxn,Tjemsland:2023vvc} is that this should be a Landau-Zener-like formula~\cite{Battye_2020} which replaces
\begin{equation}
    P_{\rm a \gamma} \rightarrow 1 - \exp(- P_{\rm a \gamma})\,. 
\end{equation}
In this case, the conversion becomes strongly adiabatic leading to close to order one conversion probabilities between axions and photons. When axion photon mixing becomes strongly adiabatic, photons can be converted back into axions again, weakening the signal. 

In this regime, the emitted power is not given simply by a surface integral, but requires detailed ray-tracing to track not only the production, but survival and re-conversion of photons in a complicated branching tree which has recently been studied in ref.~\cite{Tjemsland:2023vvc}. To circumvent this, ref.~\cite{Foster:2022fxn} applied a weighting to each axion of the form $e^{- P_{\rm a \gamma}} (1 - e^{-P_{\rm a \gamma }})$, that is, they assigned each axion the probability of producing a photon, times the survival probability of that photon as it re-crosses the resonant surface. This is a very conservative approximation, assuming that every photon re-encounters the conversion surface again, which need not be the case. More recently, it has been possible to carry out detailed branching tree simulations~\cite{Tjemsland:2023vvc}, which reveal that this assumption is indeed overly conservative, dramatically underestimating the signal in the adiabatic regime. Furthermore, this approach was able to compute the luminosity explicitly for an aligned $(\alpha = 0)$ rotator for a strongly magnetised star with $B_0 = 10^{14}\,{\rm G}$. It revealed that the key to avoiding large suppression in the non-perturbative regime, is to ensure that an initial axion undergoes an \textit{odd} number of conversions before leaving the magnetosphere, such that its final state is a photon. This becomes easier when the critical surface obscures less of the star, which occurs at higher masses. Here, adiabatic mixing suppresses the signal less than at lower masses, and the signal is suppressed by a factor of only a view relative to the naive assumption of a single level crossing with $1 - e^{-P_{\rm a\gamma}}$.

At present the algorithm in ref.~\cite{Tjemsland:2023vvc} is computationally intensive, so that while it provides a means to tackle the adiabatic regime for individual stars, applying it to populations is at present computationally prohibitive. To circumvent this, ref.~\cite{Tjemsland:2023vvc} offered an approximate scaling formula based on the size of the critical surface, however, caution is needed when applying this to larger populations, as it was inferred only for a single value of $B_0=10^{14}\,{\rm G}$ and $\alpha=0$. 

\begin{figure}
    \centering
    \includegraphics[width = \textwidth]{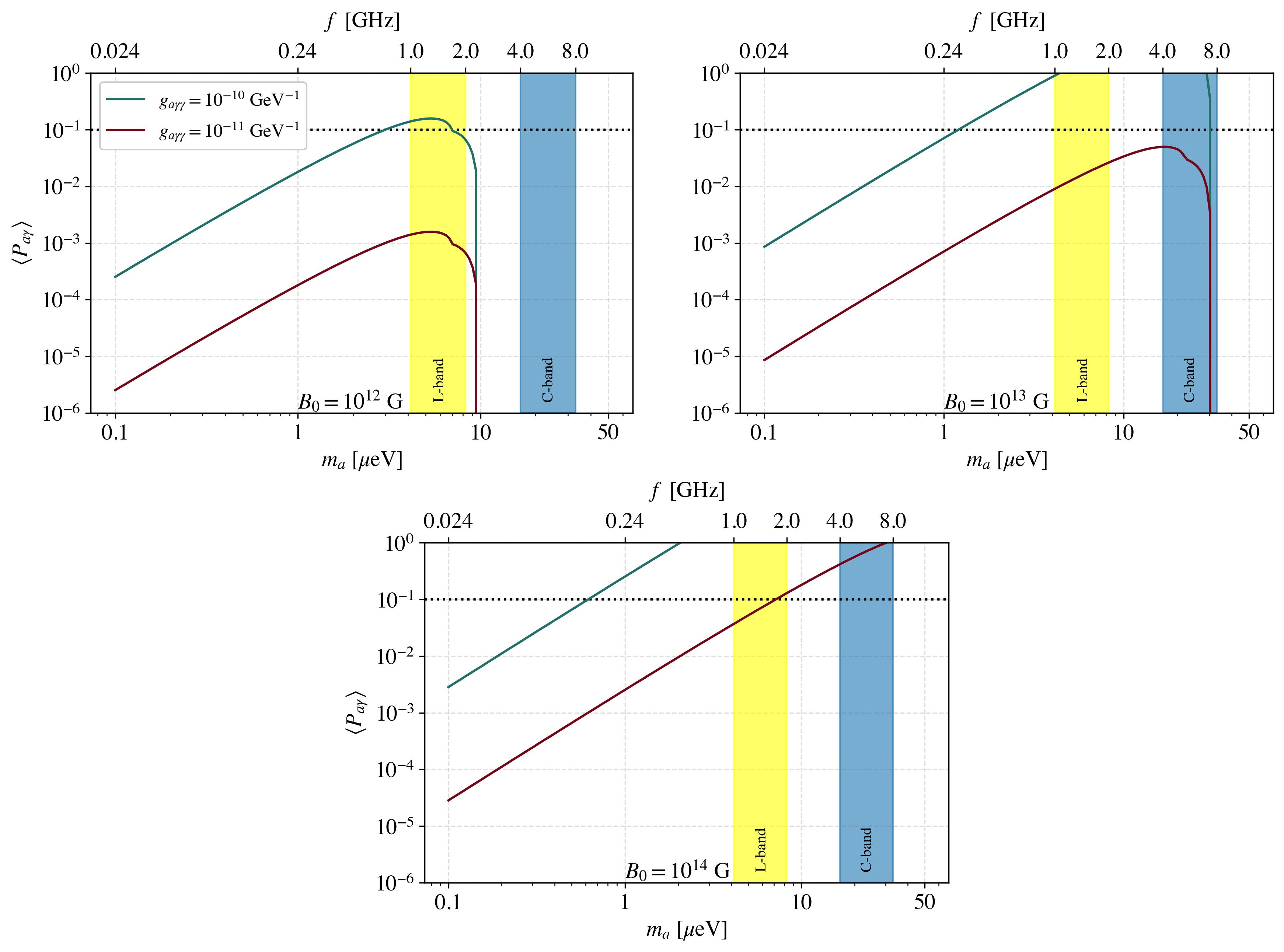}
    \caption{Size of the average axion-photon conversion probability \eqref{eq:PAvg} as a function of axion mass $m_{\rm a}$ for different magnetic fields ($B_ 0= 10^{12}, 10^{13}, 10^{14}$ G). Each panel displays two values of $g_{a \gamma \gamma}$. We also display the L- and C-bands. The horizontal dotted line indicates the adiabatic regime where the re-conversion of photons back into axions becomes important.}
    \label{fig:AdiabaticPagam}
\end{figure}

In summary, then, the adiabatic regime remains somewhat of an open question. The most conservative approach would, therefore, be to excise those stars from our analysis for which conversion is strongly adiabatic. In this spirit, we define phase space averaged conversion probability
\begin{eqnarray}\label{eq:PAvg}
    \braket{P_{a \gamma }} \equiv \frac{\int d \Omega_\textbf{k} \int d\Sigma_\textbf{k} \cdot \hat{\textbf{v}}_a P_{a \gamma \gamma}}{\int d\Omega_\textbf{k} \int d\Sigma_\textbf{k}}\,.
\end{eqnarray}
This provides a proxy for when a star enters the adiabatic regime. The results are plotted in fig.~\ref{fig:AdiabaticPagam}. Here, we see that the importance of adiabatic conversion depends crucially on the observing frequency, the strength of the magnetic field, and the value of $g_{a \gamma \gamma}$.  

For smaller couplings $\gayy\approx 10^{-11}\,{\rm GeV}^{-1}$, only the most strongly magnetized stars are affected, and even then this occurs at higher frequencies in, for example, the C-band. At these couplings, we therefore find that at a population level (see section~\ref{sec:Constraints}), conservatively removing these adiabatic stars from our analysis has little impact on the size of the population signal. However, at larger couplings $10^{-11}\,{\rm GeV}^{-1}\lesssim\gayy \lesssim 10^{-10}\,{\rm GeV}^{-1}$, things are more subtle. Towards the upper end of the C-band $(\omega \approx 2\pi\times{\rm 8 GHz} \approx 33\,\mu{\rm eV})$, removing all stars with $\braket{P_{\rm a \gamma}}>0.1$ from the population would suppress the signal in an overly conservative way since from the analysis of ref.~\cite{Tjemsland:2023vvc} adiabatic effects lead to milder signal suppression above $m_{\rm a} \gtrsim 26\,\mu{\rm eV}$. Tentatively, above $m_{\rm a} \gtrsim 26\,\mu \mathrm{eV}$, one might hope to retain adiabatic stars in the population and simply ignore multiple level-crossings using the naive result $1 - e^{- P_{\rm a \gamma}}$. This is a good approximation to the numerically generated branching tree to within factors of a few. However, the most problematic regime is for combinations of couplings $\gayy\lesssim 10^{-10}\,{\rm GeV}^{-1}$ and frequencies in the lower end of the C-band. Here, multiple level-crossings cannot be ignored for a large proportion of stars, but excising such a large number of stars with $\braket{P_{\rm a\gamma}} > 0.1$ would dramatically underestimate the signal, making this excision prescription too conservative. 

In the absence of an ability to model the adiabatic contribution in a population at the simultaneous parameter combinations $\gayy \lesssim 10^{-10}\,{\rm GeV}$ and masses $10\,\mu{\rm eV} \lesssim m_{\rm a} \lesssim 24\,\mu{\rm eV}$, we must decide how to proceed. We shall opt to continue our conservative excision prescription for $\braket{P_{\rm a\gamma}} > 0.1$, which means that we may be unable to exclude larger values of $\gayy$ in higher mass ranges. We hope this can be rectified with more efficient code in the future. However, since the purpose of the present paper is ultimately to study the modelling of the NS population itself, we content ourselves with being able to place limits at values closer to $\gayy\approx 10^{- 11}\,{\rm GeV}^{-1}$, which we can do across the full spectrum of axion dark matter frequencies by removing adiabatic stars form our analysis. This does not incur a larger penalty on signal strength. We will illustrate more explicitly how our excision of adiabatic stars and the Cyclotron resonance impacts signals from NS populations in section~\ref{sec:Constraints}. 

Putting the effects discussed in this section together, we can comment on the behaviour of the plots in fig.~\ref{fig:lvbpam}. In the top left panel, we see that there is a sharp cutoff in the luminosity at some critical value of $B_0$, below which the conversion surface no longer protrudes outside the stellar surface. Similar features are also seen for $P$ (top right panel), with both these effects arising from the scaling  $n_{\rm GJ} \propto \Omega B_0$. The same behaviour is also seen as $m_{\rm a}$ increases (bottom right panel).  We also see that there is little variability in $\alpha$ (bottom left panel), owing to the fact that since we integrate over the whole surface, although the location of strong emission regions may vary with $\alpha$, the characteristic size of the total \textit{integrated} luminosity, therefore, does not change much with $\alpha$. 

\section{Relevant constituents of the neutron star population}\label{sec:populations}

The number, distribution and characteristics of NSs in the Galaxy is very uncertain. There are various populations that have been identified observationally. Most notable of these are what we will term ``active" NSs or pulsars that are believed to be highly magnetized spinning NSs and these are prima-facie the most obvious population that will contribute to the signal from axions. As pulsars evolve in time it is thought that their magnetic fields will reduce in amplitude via  spin-down  and their periods will increase to the point where they become ``dead". There is very little direct observational evidence for this dead population, rather its presence is inferred from models for the evolution of pulsars and some estimate for the age of Galaxy. There could be a very large population of ``dead'' NSs, but it is also likely that it will not contribute significantly to the axion signal because the magnetic fields are expected to be relatively low. For this reason we will concentrate in this paper on the active pulsar population.

In what follows, the approach we will take is to attempt to model the minimum axion signal from the NS population that one might expect based on the best information that is available. The minimum signal will give the most reliable upper bound on $g_{a\gamma\gamma}$ that one can achieve from a given set of observations. We will consider three contributions which from the point of view of our modelling takes as being distinct: (i) ``Normal pulsars" whose total population we infer by calibrating against pulsar surveys which detect those normal pulsars visible from Earth. Crucially, this population extends beyond the GC; (ii) a more speculative contribution from the GC - which we define to be the central $1\,{\rm pc}^3$ - for which there is little direct evidence; (iii) the already detected GCM (PSR J1745–2900) which has already been extensively discussed in the context of axion signals~\cite{foster2020,Darling:2020plz,darling2020apj,Battye:2021xvt,Battye:2023oac}. These will be discussed in the subsequent subsections. Note that the contribution from (ii) (unlike (i)) is not directly observable, but must be inferred indirectly through modelling the birth rates of NSs and their subsequent evolution. 

\subsection{``Normal" pulsar population}
\label{sec:normal_pulsar}

The standard tool for predicting the number of pulsars that would be found in a survey is {\tt PsrPopPy}~\cite{Bates:2013uma} which evolved from \cite{lorimer_pmps} and which is also compatible with the modelling in ref.~\cite{Faucher-Giguere:2005dxp}. We have used the default implementation of this tool as described in appendix \hyperref[ref:Appendix C]{C}. Briefly, it predicts a synthetic pulsar population that is normalised using the observed number of pulsars in a user-defined survey. We will use the Parkes Multi Beam Survey (PMBS)~\cite{Manchester2001_PMBS} which found 1206 pulsars in survey of $\sim 1500\,{\rm deg}^2$ (more details in ref.~\cite{Manchester2001_PMBS}) to a flux limit of $\approx 0.15\,{\rm mJy}$ for ``long-period" pulsars~\cite{Manchester2001_PMBS}. By doing this, within the context of the modelling, the synthetic population can be thought of as normalised to observations. This relies heavily on the luminosity function for the pulsars, modelling of the gravitational potential of the Galaxy and the evolution of $B_0$, $P$ and $\alpha$ from initial distributions. However, these are based on the state-of-the-art modelling of the pulsar population. 

\begin{figure}
    \centering
    \includegraphics[width = 0.7\textwidth]{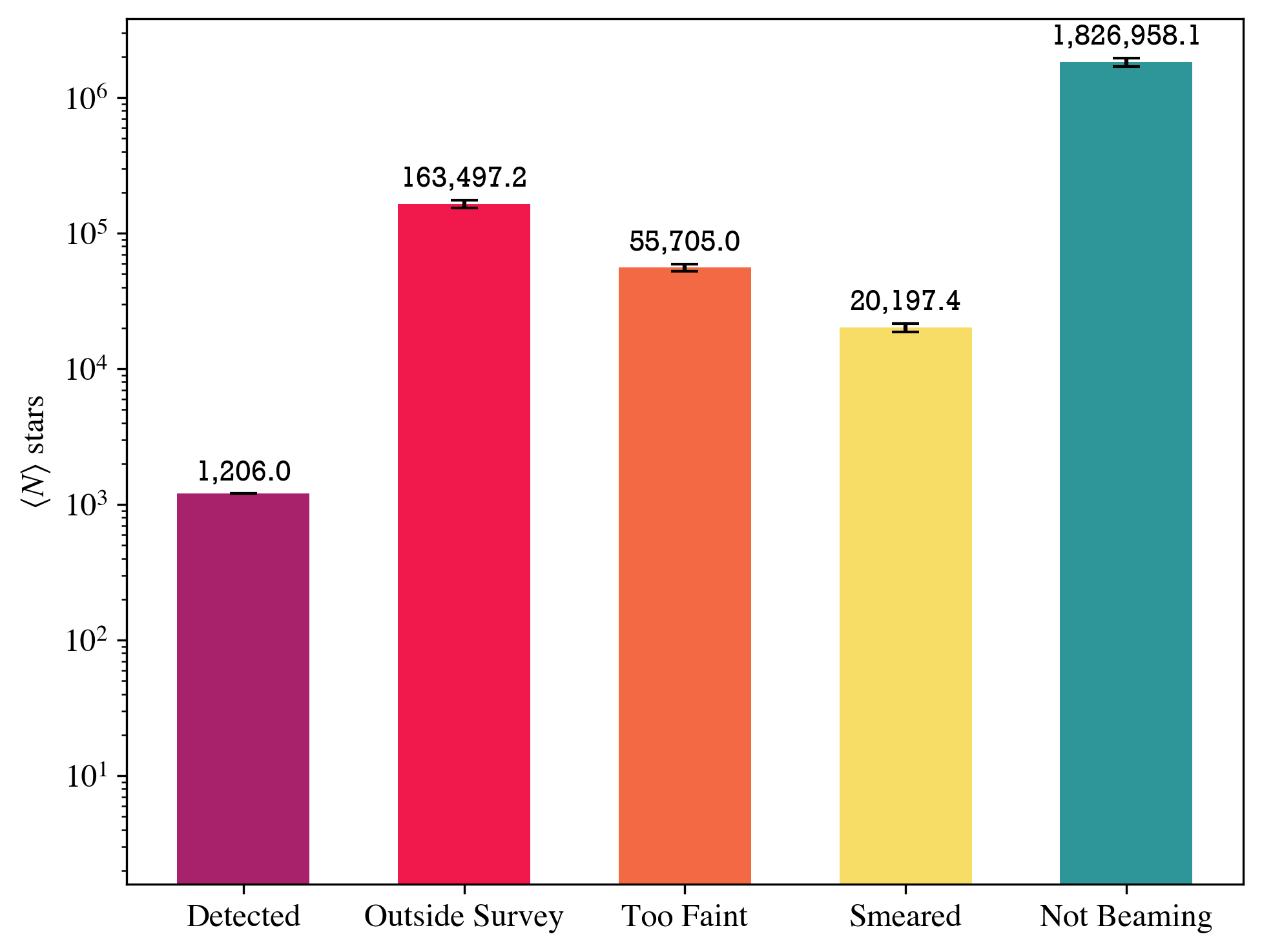}
    \caption{The number of pulsars generated into five categories by {\tt PsrPopPy} from 100 realizations. The normalising survey is the PMBS which found 1206 pulsars and, therefore, the detected number is this each time. The uncertainties on the numbers quantified by the error bars, which show the maximum and minimum values in each category, are very small for such large numbers of objects. Around 90\% of the pulsars are not beaming toward us. Pulsars whose signal-to-noise is intrinsically too low to detect are too faint, while those whose pulses are smeared by scattering due to the interstellar medium are separately classified here. }
    \label{fig:num_pulse}
\end{figure}

In fig.~\ref{fig:num_pulse} we present the distributions of NSs predicted by {\tt PsrPopPy} across 5 different categories. The last category (``not beaming") corresponds to all pulsars whose beam is directed away from Earth - we do not sub-divide this category any further. Usually this category is ignored in pulsar studies, but here, we are still interested in this sub-population since it can still contribute to the axion signal. The remaining categories in fig.~\ref{fig:num_pulse} correspond to pulsars which are beaming towards Earth, and these are further sub-divided into four types. The first is a population that would be found in a region covered by a simulated survey akin to PMBS (``detected") - this sets the overall normalisation of the population across all categories. Then there are those pulsars which would give a high enough signal-to-noise ratio (SNR) to be detected on Earth, but lie outside the survey region (``outside survey"). In addition there are those pulsars (both inside and outside the survey region) whose signal to noise is too low to be detected either because it is ``too faint" or because their pulses have  been ``smeared" by scattering in the Interstellar Medium (ISM).

These are the results of 100 realizations of the population simulated by \texttt{PsrPopPy} and include some estimate of the spread, quantified by the maximum and minimum values obtained in each category, which can be seen to be small. The standard deviation of the samples obtained were also calculated and found to be relatively small. Overall there are $\approx 2.1\times 10^{6}$ objects in each synthetic catalogue. The number detected each time is that found by the actual survey. Within the survey there are $\approx 5.5\times 10^{4}$ which are too faint to be detected by PMBS and a further $\approx 2.0\times 10^4$ whose pulse would be smeared by the ISM. There are $\approx 1.6\times 10^{5}$ outside the survey area, but the largest fraction $\sim 90\%$ corresponding to $\approx 1.8\times 10^6$ are those which do not beam towards Earth. These non-beaming pulsars are initially ignored by \texttt{PsrPopPy}, but we modified the code to include them. This is quantified using the model of ref.~\cite{1998MNRAS.298..625T}. We argue that the minimum axion signal from the NS population in the Galaxy can be modelled using this synthetic catalogue.

\begin{figure}
    \centering
    \includegraphics[width = 0.8\textwidth]{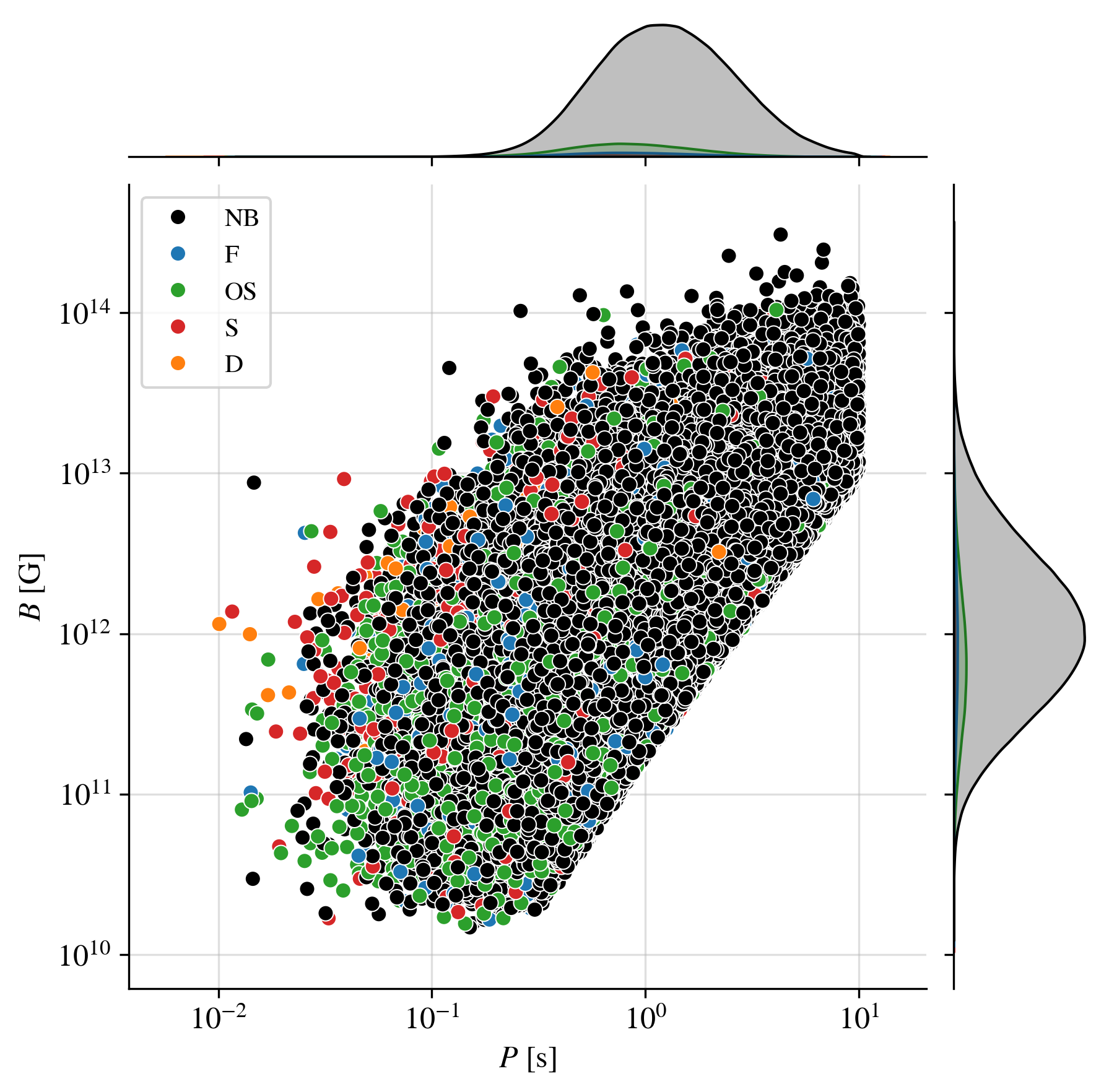}
    \caption{The distribution of the pulsars in the $(B_0,P)$ produced from one realization from {\tt PsrPopPy}. The points are colour coded as detectable ('D'), too faint ('F'), outside the survey ('OS'), smeared ('S') and not beaming towards us ('NB'). The distributions of $B$ and $P$ have been explained in section \ref{BPalpha_distribution}.}
    \label{fig:cpdiag}
\end{figure}

The modelling of evolution of $B_0$, $P$ and $\alpha$ is discussed in appendix \hyperref[ref:Appendix C]{C}.  In fig.~\ref{fig:cpdiag}, we show the $B_0$ and $P$ values for a typical realization.  We present the $B_0-P$ plane for one realization which shows the death line $B_0/P^2\approx 1.7\times 10^{11} \;{\rm G}\,{\rm s}^{-2}$. As described in appendix \hyperref[ref:Appendix C]{C}, there is good agreement between the observed pulsars and those generated by {\tt PsrPopPy}, which is to be expected as the initial distributions were chosen so that, given the modelling, there is some agreement with the observations. However, this is the best way one has of modelling the very many objects which have not yet been detected, or indeed which can never be detected since they do not pulse toward us.

\begin{figure}
    \centering
    \includegraphics[width = 0.8\textwidth]{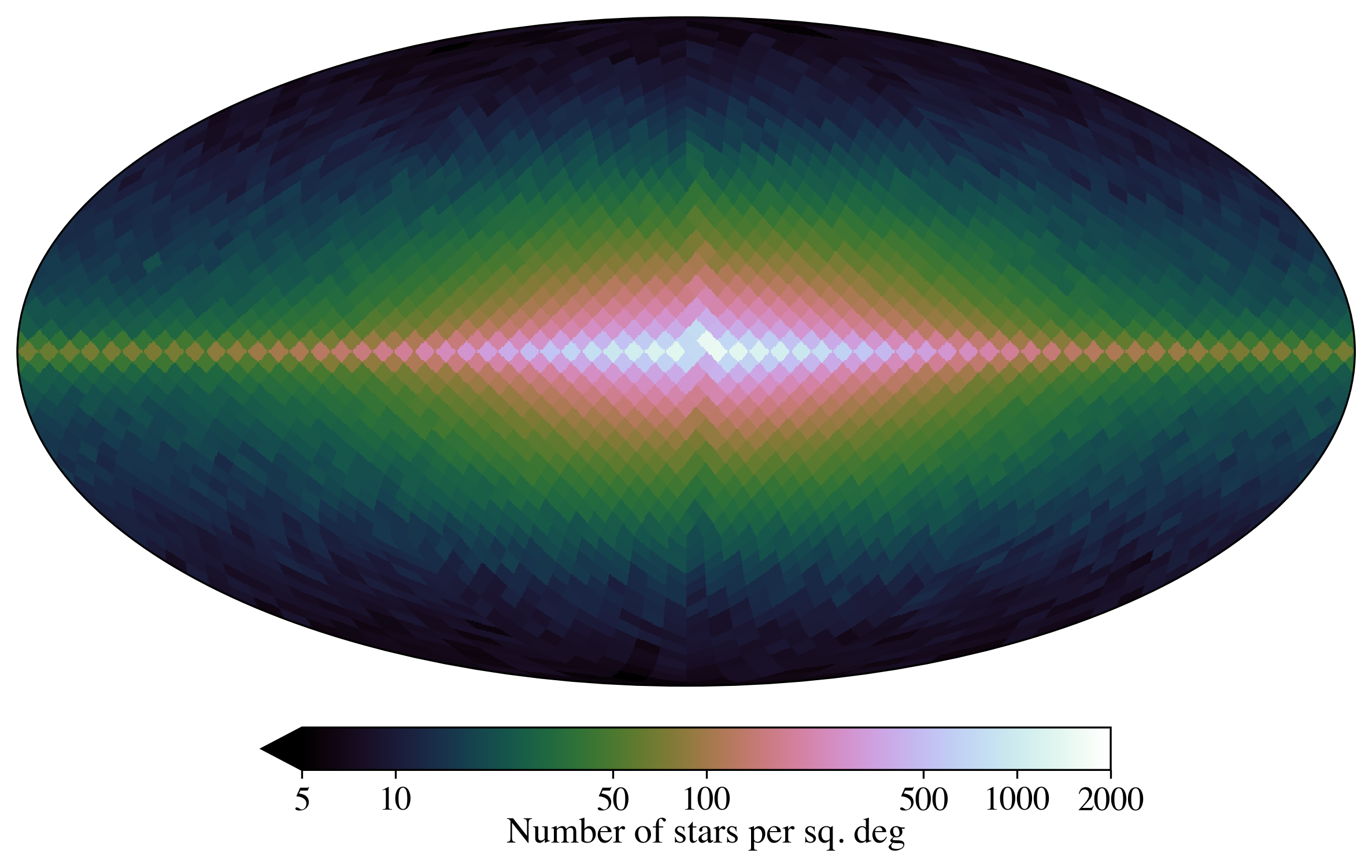}
    \caption{The predicted number of NSs in Galactic coordinates from a {\tt PsrPopPy} realization using a {\tt Healpix} map with ${\tt nside}=64$ which corresponds to $\theta_{\rm pix}\approx 3.7\,{\rm deg}$. The pulsars are clearly clustered toward the GC (relative to the resolution in this map, see sec.~\ref{sec:comment} for a discussion on the lack of detection of pulsars in the inner regions of the GC) and in the plane of the Galactic disk as one would expect. The majority of pixels in any direction of interest for our calculation have $>100$ objects in them which makes them amenable to the approach for calculating the axion signal outlined in section~\ref{sec:luminosity}.}
    \label{fig:dist_pulse}
\end{figure}

We can take the positions of the pulsars and bin them into sky pixels generated using the {\tt Healpix} package~\cite{Gorski:2004by}. In fig.~\ref{fig:dist_pulse} we display the results of doing this using a pixel size $\theta_{\rm  pix}\approx 3.7\,{\rm deg}$ in Galactic coordinates $(l,b)$. It is clear that the modelling preferentially places the pulsars in the Galactic Disk and indeed there are many more in the direction of the GC as one would expect. As we will show in section~\ref{sec:normal_pulsar} realizations of the sky produced using {\tt PsrPopPy} provide all the information needed to make reliable predictions of the signal expected from the population of ``normal" pulsars.

\subsection{Comment on the non-detection of pulsars in the Galactic Centre}
\label{sec:comment}

 Prior to the detection of the GCM in 2013 no pulsars had been detected in the central $\sim 1{\rm pc}^3$.  Circa 2012 there were a total of 5 pulsars detected within $15\,{\rm arcmin}$, of Sgr A*  with the closest being $\sim 11\,{\rm arcmin}$~\cite{2006MNRAS.373L...6J,Deneva:2009mx,Bates:2010wb} corresponding to distances $\sim 10$ pc. Since then, there have been a few more ``normal" pulsars detected near the GC, but the increase is not sufficient to have a significant impact on the argument presented here. Any normal pulsars close to (but not in) the GC would be accounted for by the modelling of {\tt PsrPopPy} whose population size is normalised by calibrating against PMBS. Of course, the non-detection of pulsars does not mean that there are no NSs, simply that they cannot be observed directly, and if they do exist their population must be inferred through birth rate modelling. This is the approach taken in ref.~\cite{Foster:2022fxn}, but it should be clear that it is an inferred, as opposed to detected population of stars.

Stars of masses $\sim 8-20\,M_{\odot}$, that is, those assumed in ref.~\cite{Foster:2022fxn} to become NSs, are expected to have nuclear lifetimes $\sim 10\,{\rm Myr}$ and the density of such stars is observed to be high in the nuclear star cluster at the GC where monitoring has been used to establish the existence of a black hole with $M\approx 4\times 10^{6}M_\odot$ \cite{GRAVITY_coll_2022, EHT_SagA}. It would be reasonable to expect that these stars will produce NSs according to the arguments presented in ref.~\cite{Foster:2022fxn}  and reproduced in detail in appendix \hyperref[ref:Appendix A]{A}. Several other authors have made estimates for the number of pulsars that might be detectable in the GC~\cite{Pfahl:2003tf,Dexter:2013xga,Zhang:2014kva} and these are broadly compatible with ref.~\cite{Foster:2022fxn}.

The alternative is a more observation-based approach~\cite{Wharton:2011dv,Chennamangalam:2013zja} that refers only to the NSs that are pulsars. In particular, ref.~\cite{Chennamangalam:2013zja} states that the number of \textit{observable} pulsars in the GC is constrained to be $\lesssim 200$ by observations, which imposes an upper limit on the \textit{total} number of NSs at the GC satisfying $N_{\rm GC}\lesssim 2000$ assuming that $\sim 10\%$ of pulsars beam toward us. Again this is compatible with ref.~\cite{Foster:2022fxn} without necessarily supporting it quantitatively.

The detection of a population of pulsars in the GC would be very exciting since they would allow more precise tests of the gravity around the central black hole~\cite{Pfahl:2003tf}. However, prior to the detection of the GCM~\cite{2013ApJ...770L..23M,2013ApJ...770L..24K} it was believed that the significant background of electrons and turbulence leads to a smearing of the pulsar signal making them impossible to detect. This temporal smearing is characterised by a timescale at frequency $f$ of~\cite{Cordes:1996bt}
\begin{equation}
\tau_{\rm ISM}\sim 10^3\,{\rm s}\left({\frac{f}{1\,{\rm GHz}}}\right)^{-4}\,,
\end{equation}
for a screen around $\sim 100\,{\rm pc}$ from the GC. The pulsed emission from the spinning neutron star would be impossible to detect for $P\lesssim\tau_{\rm ISM}$ using standard techniques, and this was believed to be the reason why no pulsars had been detected, when large numbers of NSs were predicted.

The detection of the GCM~\cite{2013ApJ...770L..23M,2013ApJ...770L..24K} put this picture into question: the temporal smearing was measured to be $\tau_{\rm ISM}\approx (1.3\pm 0.2)\,{\rm s}$ at $f\approx 1\,{\rm GHz}$~\cite{Spitler:2013uva}, three orders of magnitude lower than the predicted value. This means that at least along that line-of-sight the smearing of the pulsars, while still relatively large compared to other directions, is much lower than predicted. If one extrapolates this to be relevant to the whole GC then it could have some bearing on the number of NSs at the GC since virtually none have been found there, even at relatively high frequencies.

A deep survey of the GC at $f=15\,{\rm GHz}$ was performed using the GBT and no convincing pulsar candidates were found~\cite{Macquart:2010vf}. Similar results have been found in other surveys~\cite{2013IAUS..291...57S,2013Natur.501..391E,EHT:2023hcj}. At this frequency $\tau_{\rm ISM}\sim 10^{-2}\,{\rm s}$ and using a model for the pulsar luminosity, they were able to rule out a region which contains at least half, possibly more, of the ``normal'' pulsar population in the $L_{1.4\,{\rm GHz}}-P$ diagram. From this it seems reasonable to conclude that the NS population in the GC could be different from the ``normal" pulsar population (modelled by {\tt PsrPopPy}). Similar, but less constraining observations had been previously been performed at $f=5\,{\rm GHz}$~\cite{2006MNRAS.373L...6J}. It seems reasonable to believe that, if there are pulsars in the GC, they could be very different from the ``normal" pulsar population modelled by {\tt PsrPopPy}, and also as implicitly assumed in ref.~\cite{Foster:2022fxn}.

At least two possible ideas have been suggested to explain this conundrum: (i) magnetars are more prevalent than ``normal" pulsars in the GC~\cite{Dexter:2013xga} or (ii) the GC NS population are millisecond pulsars which are ``recycled normal pulsars"~\cite{Macquart:2015jfa}. If correct, or even partially correct, both would imply that the assumptions made in ref.~\cite{Foster:2022fxn} may need re-examining and the NSs in the GC could be both lower in number than assumed and have different distributions in $B_0$ and $P$.

\subsection{Additional neutron stars in the Galactic Centre}
\label{sec:gcdist}

When normalised to the PMBS, {\tt PsrPopPy}~\cite{Bates:2013uma} predicts $<1$ pulsar with a radius of $1\,{\rm pc}$ of the GC and $5-10$ within the Green Bank beam which has a FWHM of $\approx 2\,{\rm arcmin}$ (projected diameter $\sim 5\,{\rm pc}$) at $6\,{\rm GHz}$. This is compatible with the non-observation of pulsars in the GC as described above and another recent assessment~\cite{Xie:2024rru}, but is contrary to what is deduced in ref.~\cite{Foster:2022fxn}.

As explained in appendix \hyperref[ref:Appendix A]{A}, ref.~\cite{Foster:2022fxn} estimated the number of NSs in the GC to be  $N_{\rm GC}=\psi T$ where $\psi$ is an inferred birth rate and $T\approx 30\,{\rm Myr}$ is the star lifetime. Their value of $\psi\approx 5.4\times 10^{-3}\,{\rm century}^{-1}$ is calculated using a measured star-formation rate for the GC region and a power law Initial Mass Function (IMF) and has no connection to any observations of pulsars. Using estimates for $\psi$ they conclude that $N_{\rm GC}\approx 1600$ that we argue in appendix \hyperref[ref:Appendix A.1]{A.1} is uncertain at the level of at least a factor of ten\footnote{In addition to leading to a weaker constraint just on the basis of the reduced number of stars, if $N_{\rm GC}\lesssim 100$ then the approach of integrating over the observer position when calculating the axion induced flux is probably not justified.}. There are a number of factors which are probably only known at the factor of two level and the power-law of the IMF can have a significant impact. Nonetheless, on the basis of these it is reasonable to say that there could be a population of NSs in the GC that is not modelled by {\tt PsrPopPy} since only very extreme values of the parameters reduce $N_{\rm GC}$ to that predicted for ``normal" pulsars. However, it is clear that the actual number of objects is difficult to estimate at anything more than a factor of ten level. In what follows we will investigate the possible amplitude of a signal coming from the central ${\rm pc}^3$ of the Galaxy based on such a population.

In addition to the size of the birth rate itself,  kinematic arguments concerning the velocity of the NSs born in the central region of the GC also suggest that the number of stars could  be lower than one might expect unless the stars are strongly bound in some way to the GC. In appendix \hyperref[appendix:diffusion]{B.2}, we argue that the degree to which stars are confined to the GC depends crucially on the velocity distribution of stars and the depth of the GC potential. As a result, it is difficult to infer the number of stars at the GC based on birth rates alone since precise details of the velocity distribution of NSs and the size of the nature of the gravitational potential near the GC are not known. In particular, for stars with velocities of ${\cal O}(100\,{\rm km}\,{\rm sec}^{-1})$ and percent level variations in the GC potential can vary the confinement radius by ${\cal O}(100\,{\rm pc})$.

We also note that ref.~\cite{Foster:2022fxn} argued that the existence of $\mathcal{O}(1)$ strongly magnetised stars predicted by birth rates could be viewed as a sanity check that the modelling was correct. This presupposes that magnetars are simply a sub-poulation of normal neutron stars modelled by the birth rate calculations described above. However, if the process of magnetar formation is markedly different, some caution may be needed, since it is not obvious that the high-B tail of the normal pulsar distribution is one and the same thing as the galactic centre magnetar population.

Given the uncertainties outlined in the previous two paragraphs, we will parameterise the strength of the signal from the GC population in terms of the number $N_{\rm GC}$ of neutron stars inside the telescope beam. By doing this, we avoid the necessity to model the very uncertain radial dependence of the number density and assume that every object in our population has a radial distance of $0.5\,{\rm pc}$  from the GC. Here, and throughout the remainder of the paper we use the Navarro-Frenk-White (NFW) profile~\cite{NFW, 1997ApJ...490..493N} to describe the dark matter profile of the Galaxy. In particular, 
\begin{align}\label{eq:NFW}
    \rho_{\mathrm{NFW}}(r)=\frac{\rho_0r_0^3}{r\left(r_0+r\right)^2},
\end{align}
where $r$ represents the distance from the GC, $\rho_0$ is a constant and $r_0$ is the scale radius. We will use $r_0 =(8.1 \pm 0.7)\,{\rm  kpc}$ and $\rho_0 = (1.99 \pm 0.172)\,{\rm GeV}\,{\rm cm}^{-3}$~\cite{lin2019dark}. The local value of the density is assumed to be $\rho_{\rm DM}|_{\rm local}\approx 0.45\,{\rm GeV}\,{\rm cm}^{-3}$ and the modelling gives us $\rho_{\rm DM}|_{\rm GCM}/\rho_{\rm DM}|_{\rm local}\approx 3.6\times 10^5$ and $\rho_{\rm DM}|_{\rm GC}/\rho_{\rm DM}|_{\rm local}\approx 7.2 \times 10^4$ where we have defined the radius of the GC to be $0.5\,{\rm pc}$.

We will also assume that the observing bandwidth $\Delta f_{\rm obs}> v_{\rm gal}f_{\rm obs}/c$ where $v_{\rm gal}\approx 200\,{\rm km}\,{\rm s}^{-1}$ is the Galactic velocity dispersion which we assume governs the spread of the signal.  This avoids modelling the, again very uncertain, velocity distribution of the population in the GC. 

We stress that neither of these assumptions - about the radial and velocity distribution - needs to be absolutely true if $N_{\rm GC}$ is considered to be the effective number of NSs that a given observation is sensitive to. Our position is that it is best to just specify $N_{\rm GC}$ and with future more detailed understanding of the GC it might be possible to make a more precise prediction of this value. These assumptions will allow us to assess the accuracy of any estimate of the signal from a population of stars in the GC as a function of pulsar parameters, independent of the Galactic dark matter density profile. 

Rather than trying to model the distribution of $B_0$, $P$ and $\alpha$ values we will use specific distributions and vary the parameters governing these distributions. For $B_0$ and $P$ we will use log-normal distributions which are similar to both the observed distributions and those in the synthetic catalogues produced by {\tt PsrPopPy}. The mean and variance of $b_0=\log_{10}(B_0/{\rm G})$ are $\mu_B$ and $\sigma_B$, respectively, whereas the equivalents for $p_0=\log_{10}(P/{\rm s})$ are $\mu_P$ and $\sigma_P$. In what follows we will vary these parameters, but will also use fiducial values of $(\mu_B,\sigma_B)=(12.1,0.54)$ and $(\mu_P,\sigma_P)=(-0.23,0.36)$ that are obtained by fits to the ATNF catalogue of pulsars~\cite{Manchester2001_PMBS}.  The distribution of $\alpha$ is assumed to be uniform in $\cos\alpha$. Such estimates have the feature that one can, in a relatively controlled way, understand the uncertainties. The values of $B_0$ are inferred from the spin-down rate of pulsars and hence we would expect the values of $B$ and $P$ to be correlated. Hence formally, the distributions of $B$ and $P$ should not be independent. Using the data from the ATNF catalogue we have found that, for a covariance matrix for $B_0$ and $P$ defined by
\begin{equation}
\begin{pmatrix}\sigma_{B}^2 & r_{\mathrm{corr}} \sigma_B\sigma_P\cr r_{\mathrm{corr}} \sigma_B\sigma_P& \sigma_P^2\end{pmatrix}\,,
\end{equation}
the data is compatible with $r_{\mathrm{corr}} \approx 0.44$. We discuss how this may affect the PDF for $P$, $B_0$ and the size and nature of axions signals from the population in more detail in appendix \hyperref[appendix:BPCorellaiton]{B}. The conclusion is that this could at the very most an uncertainty of a factor of three each way, but probably much less (see eq.~\eqref{lum_gc}).

\subsection{Galactic Centre Magnetar}
\label{sec:gcm}

The GCM has been observed and its properties measured~\cite{2013ApJ...770L..24K} and so it is natural to compare its contribution to the galactic centre signal against the more speculative contribution from the population of ``hidden" neutron stars described in preceding sections. In this subsection, we therefore compare the signal from the GC NS population discussed in the previous section to that of the GCM, whose existence, at least, is not in doubt.

\begin{figure}
    \centering
    \includegraphics[width = 0.7\textwidth]{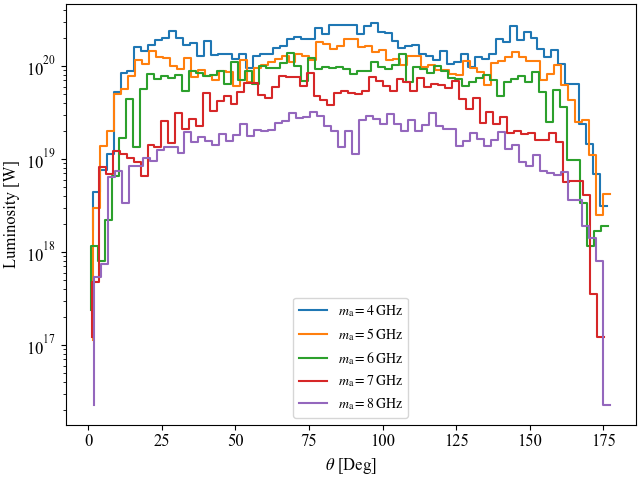}
    \caption{Estimates of the radiated power from the GCM as a function of the viewing angle $\theta$ for an aligned rotator, that is, $\alpha=0$ for $\gayy = 10^{-11}\,{\rm GeV}^{-1}$ from the multi-branch simulations presented in ref.~\cite{Tjemsland:2023vvc}. The dark matter density relative to the local value at the position of the GCM is enhanced by a factor $\rho_{\rm DM}|_{\rm GCM}/\rho_{\rm DM}|_{\rm local} \sim 3.6\times 10^5$ assuming the NFW~\cite{NFW} profile for the Galactic dark matter halo. This is to be compared to the estimate in \eqref{lum_gc}, but bearing in mind that this signal is narrower in frequency width. From this, we can conclude that the signal from the GCM is of comparable brightness to the signal from the GC populations for $N_{\rm GC}=1000$ and is likely brighter when $N_{\rm GC}$ is lower.}
    \label{fig:GCM_Power}
\end{figure}

 We have calculated the predicted luminosity assuming GCM values $B_0=2\times 10^{14}\,{\rm G}$ and $P=3.76\,{\rm s}$. The values of $\alpha$ and $\theta$ are unknown and hence any conclusion could be affected by a specific choice. As previously mentioned, for large values of $B_0$, it is important to take into account the suppression of the flux due to re-conversion of the photons into axions when they encounter a resonant point before they escape the magnetosphere. We use the full branching-tree simulations developed in ref.~\cite{Tjemsland:2023vvc} that account for the adiabatic suppression for the case of $\alpha=0$ which should yield a lower bound on the expected signal. We computed the time-averaged flux as a function of $\theta$ at frequencies across the C-band, i.e., 4, 5, 6, 7 and 8 GHz, which correspond to $m_{\rm a} = 16.5, 20.6, 24.2, 28.8$ and $33.0\,\mu{\rm eV}$, respectively. The results are presented in fig.~\ref{fig:GCM_Power} and they show that the luminosity from the GCM is similar to the estimate from the population in the GC except for the extreme values near $\theta\approx 0^{\circ}$ and $\approx 180^{\circ}$ (see eq.~\eqref{lum_gc} in appendix \hyperref[appendix:BPCorellaiton]{B}). Note that we have assumed an NFW profile for the dark matter halo in order to compute the dark matter density at the position of the GCM, resulting in an enhancement relative to the local value of $\sim 3.6\times 10^5$. Later, in fig.~\ref{fig:constraint} we compare limits from the GCM and neutron star populations, from which we see that comparable limits can be obtained from the GCM with a GC population of $\sim 1000$ stars. 

\section{Axion signals from Populations}\label{sec:Constraints}

In this section we integrate the signals predicted for NSs in section~\ref{sec:luminosity} over the stellar populations models discussed in section~\ref{sec:populations}. We shall discuss the baseline signal from the population of normal pulsar populations throughout the Galaxy and the point-source like signal coming from the more speculative populations of invisible pulsars which reside within 1~pc of the GC. 

\subsection{Predicted Galactic signal due to ``normal pulsars"}

In section~\ref{sec:normal_pulsar} we discussed the modelling of the ``normal pulsar" population using {\tt PsrPopPy}. This is the tool used in pulsar astronomy to predict the number of objects detected by a given survey and, hence, when normalised to a survey such as the PMBS and tuned to given the correct properties, it can be used to calculate the minimum axion signal from different locations in the Galaxy. In this section we will quantify the nature of this signal more precisely.

\subsubsection{All-sky signal}
\label{sec:allsky}

\begin{figure}
    \centering
   \includegraphics[width = 0.5\textwidth,height=0.26\textheight]{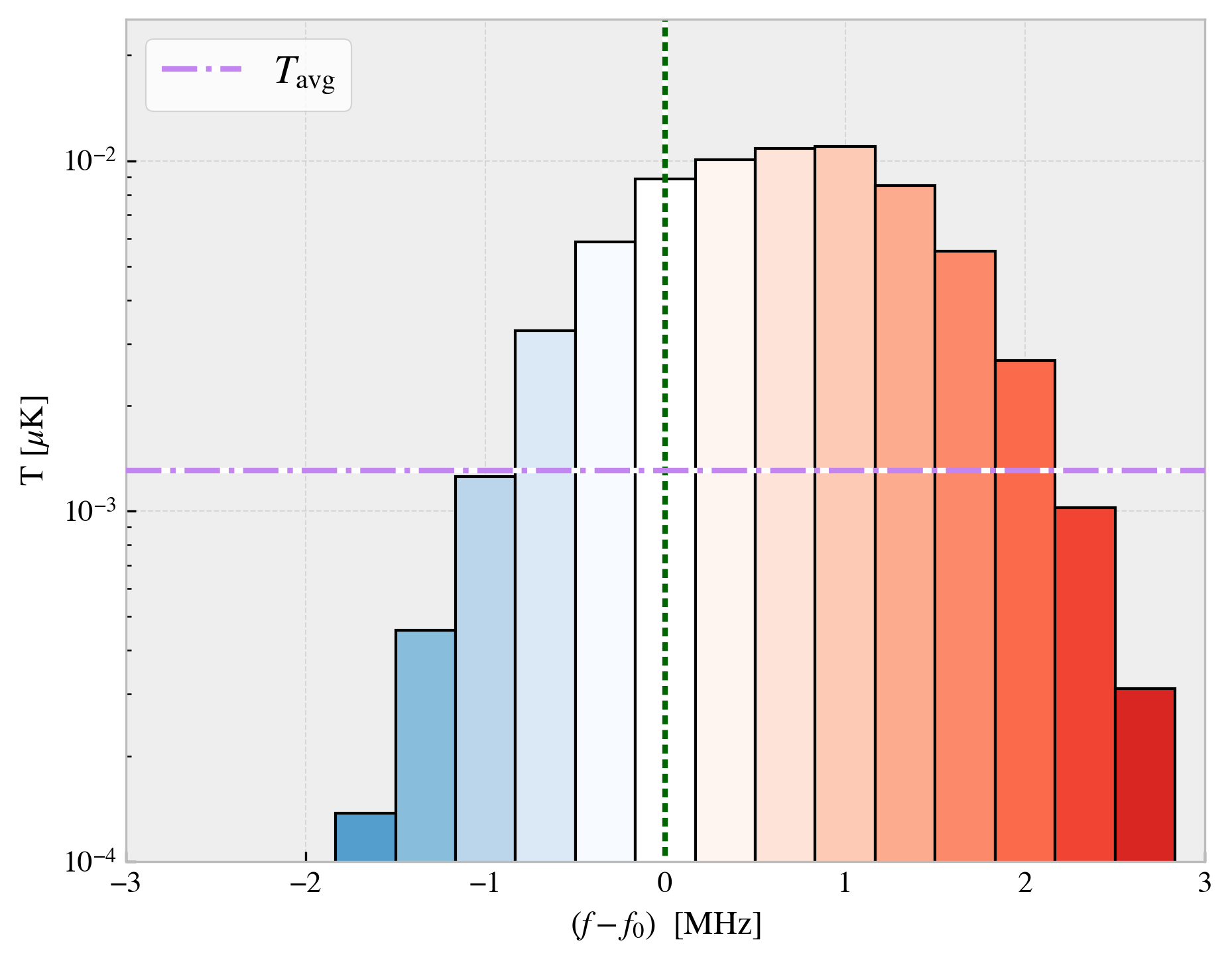}\includegraphics[width = 0.5\textwidth]{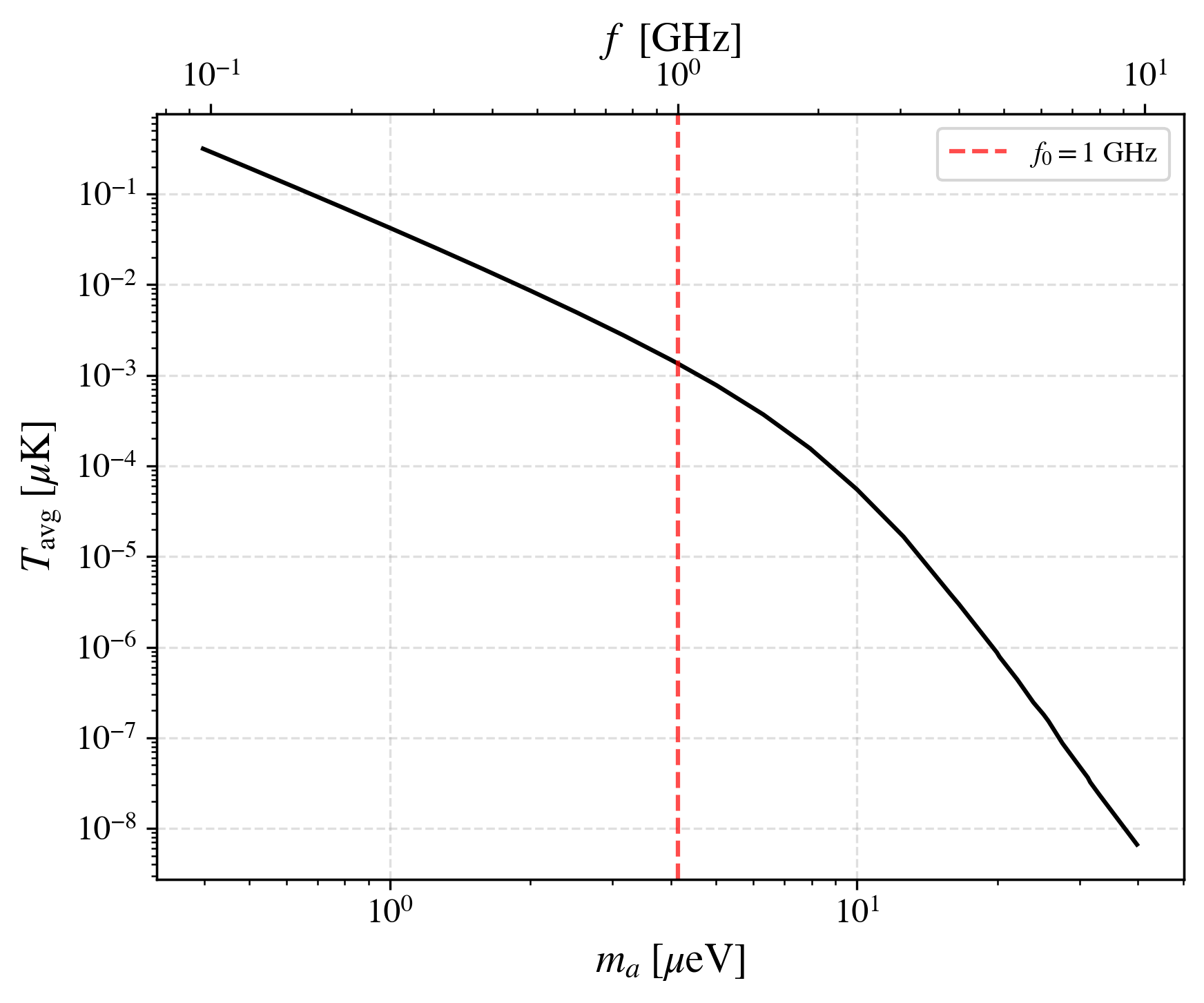}
    \caption{The left-hand figure is the brightness temperature, $T$, summed over all NSs predicted by {\tt PsrPopPy} as a function of frequency with a horizontal line representing the average temperature, $T_{\rm avg}$, over $\Delta f_{\rm obs}/f_{\rm obs}=10^{-3}$ about $f_{\rm obs}=1\,{\rm GHz}$ which corresponds to  $m_{\rm a}=4.1\,\mu{\rm eV}$ and we have assumed $\gayy=10^{-11}\,{\rm GeV}^{-1}$.  The dispersion arises due to the motion of the stars relative to the Solar System. For any specific realization the  peak signal is not centred on $f_{\rm obs}= 1\,{\rm GHz}$ to the systemic velocity of the Solar System. The right-hand  figure shows the relationship between average  brightness temperature and axion mass, $m_{\rm a}$. Notably, there is a significant drop around $m_{\rm a} \approx 10\,\mu {\rm eV}$ due to the critical radius for resonant axion-photon conversion being inside stars with sufficiently small magnetic fields.}
    \label{fig:allspec}
\end{figure}

To start with we will calculate the ``all-sky" signal obtained by binning in frequency the Rayleigh-Jeans brightness temperature, $T$, defined in terms of the intensity $I=2k_{\rm B}f_{\rm obs}^2T$ at observing frequency $f_{\rm obs}$ for all pulsars predicted by {\tt PsrPopPy}. This is presented for $m_{\rm a}=4.1\,\mu{\rm eV}$ corresponding a central frequency of $f_{\rm obs}=1\,{\rm GHz}$ in left-hand panel in fig.~\ref{fig:allspec} for $g_{a\gamma\gamma}=10^{-11}\,{\rm GeV}^{-1}$. We note that, due to the motion of the pulsars relative to the position of the Solar System, the spectrum is centred at $f=m_{\rm a}c^2(1+v_{\rm centre}/c)/h$ where $v_{\rm centre}$ is the systemic velocity of GC. This will typically be $\sim 200\,{\rm km}\,{\rm s}^{-1}$  which will be realization dependent and the average over multiple realizations will be zero~\cite{Bates:2013uma}. The frequency width of the spectrum is given by $\Delta f_{\rm obs}/f_{\rm obs} \approx \langle v_{\rm gal}^2\rangle^{1/2}/c \approx 10^{-3}$ where $\langle v_{\rm gal}^2\rangle^{1/2}$  is the r.m.s. velocity dispersion of the Galaxy. The overall `sky-averaged' signal, calculated approximately to be $10^{-2}\,\mu{\rm K}$ at $f_{\rm obs}=1\,{\rm GHz}$, is weak and is expected to scale like $\gayy^2$ for $\gayy<10^{-11}\,{\rm GeV}$ while there will be corrections due to photons turning back into axions for larger values of $g_{a \gamma \gamma}$, as described in section~\ref{sec:luminosity}. It is important to note that the contribution to the spectral width from this dispersion, resulting from stellar motion, dominates over the contribution caused by plasma-effects~\cite{Witte:2021arp,Battye:2021xvt} which shift photon frequencies as they propagate through the time-dependent magnetosphere.

\begin{figure}
    \centering
    \includegraphics[width = \textwidth]{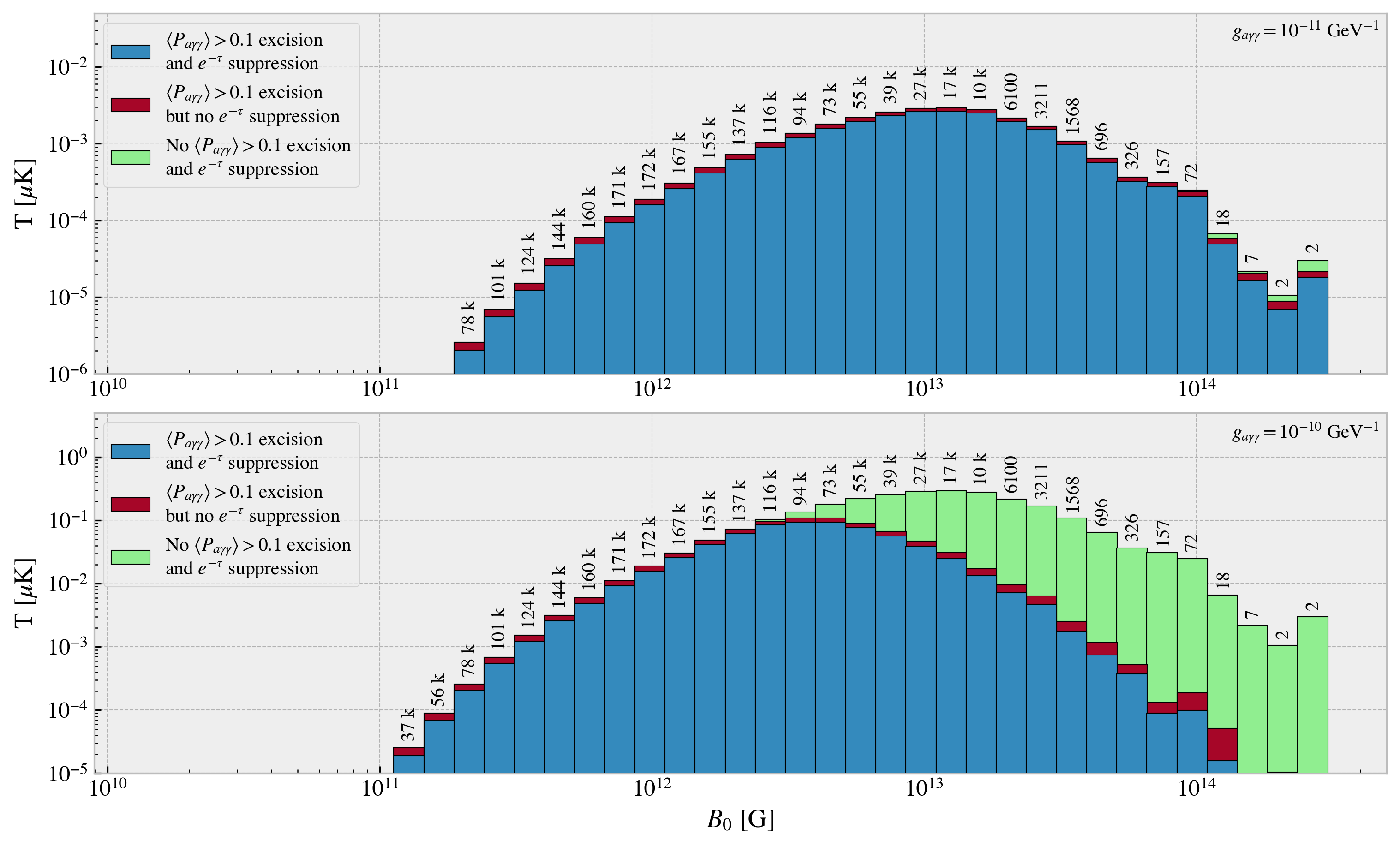}
    \caption{Contributions to the brightness temperature, $T$ for $f_{\rm obs}=1\,{\rm GHz}$ ($m_{\rm a}=4.1\,\mu{\rm eV}$) as a function of the NS's magnetic field using $\gayy=10^{-11}\,{\rm GeV}^{-1}$ (top) and $\gayy=10^{-10}\,{\rm GeV}^{-1}$ (bottom). In addition there is a number associated with each bar which is the number of stars contributing to that bin. We consider three scenarios for photon attenuation (green) (i) no attenuation, (ii) excision of stars with $\braket{P_{a\gamma}}>0.1$ and (red) (iii)  an additional suppression due to the cyclotron resonance (see \eqref{eq:Cylotron}) (blue). We see that the attenuation has little effect for $\gayy=10^{-11}\,{\rm GeV}^{-1}$ and the signal is dominated by NSs with $B_0\sim 10^{13}\,{\rm G}$. The excision of stars with $\braket{P_{a\gamma}}>0.1$ is significant for $\gayy=10^{-10}\,{\rm GeV}^{-1}$ and the signal becomes dominated by stars with $B_0\sim 10^{12.5}\,{\rm G}$.}
    \label{fig:bfield}
\end{figure}

It is instructive to examine the contributions to this all-sky signal as a function of the magnetic field of the stars parameterized by $B_0$, as illustrated in fig.~\ref{fig:bfield} where we present the relative contributions for both $g_{a\gamma\gamma}=10^{-10}\,{\rm GeV}^{-1}$ and $\gayy=10^{-11}\,{\rm GeV}^{-1}$ again for $f_{\rm obs}= 1\,{\rm  GHz}$. Included also are the number of stars contributing to  each magnetic field  bin. We see that the largest contribution to this comes from stars with $B_0\approx 10^{13}\,{\rm G}$ when the signal is not affected by excision of stars with $\braket{P_{a\gamma}}>0.1$ and cyclotron absorption. These affects are not important when $\gayy=10^{-11}\,{\rm GeV}^{-1}$, but we see that the non-perturbative effects modelled by the excision of stars with $\langle P_{a\gamma}\rangle>0.1$ are important for $\gayy=10^{-11}\,{\rm GeV}^{-1}$  and this significantly reduces the number of stars with large values of $B_0$ that contribute to the all-sky signal. This has the effect of moving the dominant value to $B_0\approx 3\times 10^{12}\,{\rm G}$. The impact of cyclotron absorption is much weaker. The discussion here is compatible with that in section~\ref{sec:absorb}.

In the right-hand panel of fig.~\ref{fig:allspec} we present the dependence of the average temperature $T_{\rm avg}$,calculated by averaging the spectrum in the left-hand panel over $\Delta f_{\rm obs} =  10^{-3}f_{\rm obs}$, as a function of $m_{\rm a}$. We see that the signal decays as $m_{\rm a}$ increases and this becomes much stronger for $m_{\rm a}>10\,\mu{\rm eV}$. There are two effects at play here: the first is that the luminosity emitted by a single star is increases with $m_{\rm a}$ when it is sufficiently large that the critical radius for resonance is outside the star as illustrated in fig.~\ref{fig:lvbpam}. For sufficient low values of $m_{\rm a}$ and $B_0\sim 10^{13}\,{\rm G}$, where the largest contribution to the signal comes from, this can be approximated as being close to $L\propto m_{\rm a}$. Hence, $T\propto L/ (f_{\rm obs}^2\Delta f_{\rm obs})\propto L/f_{\rm obs}^3\propto m_{\rm a}^{-2}$ which is close to what is observed. The second important effect is that as $m_{\rm a}$ increases the critical radius comes closer to the surface and ultimately disappears. This means that stars with low magnetic fields no longer contribute and we see an effect similar to that observed at larger values of $B_0$ due the excision of $\braket{P_{a\gamma}}>0.1$ in fig.~\ref{fig:bfield}, but this time for lower values of $B_0$. 

\subsubsection{Spatial variation of the signal in the sky }
\label{sec:spatial}

\begin{figure}
    \centering
    \includegraphics[width = 0.5\textwidth]{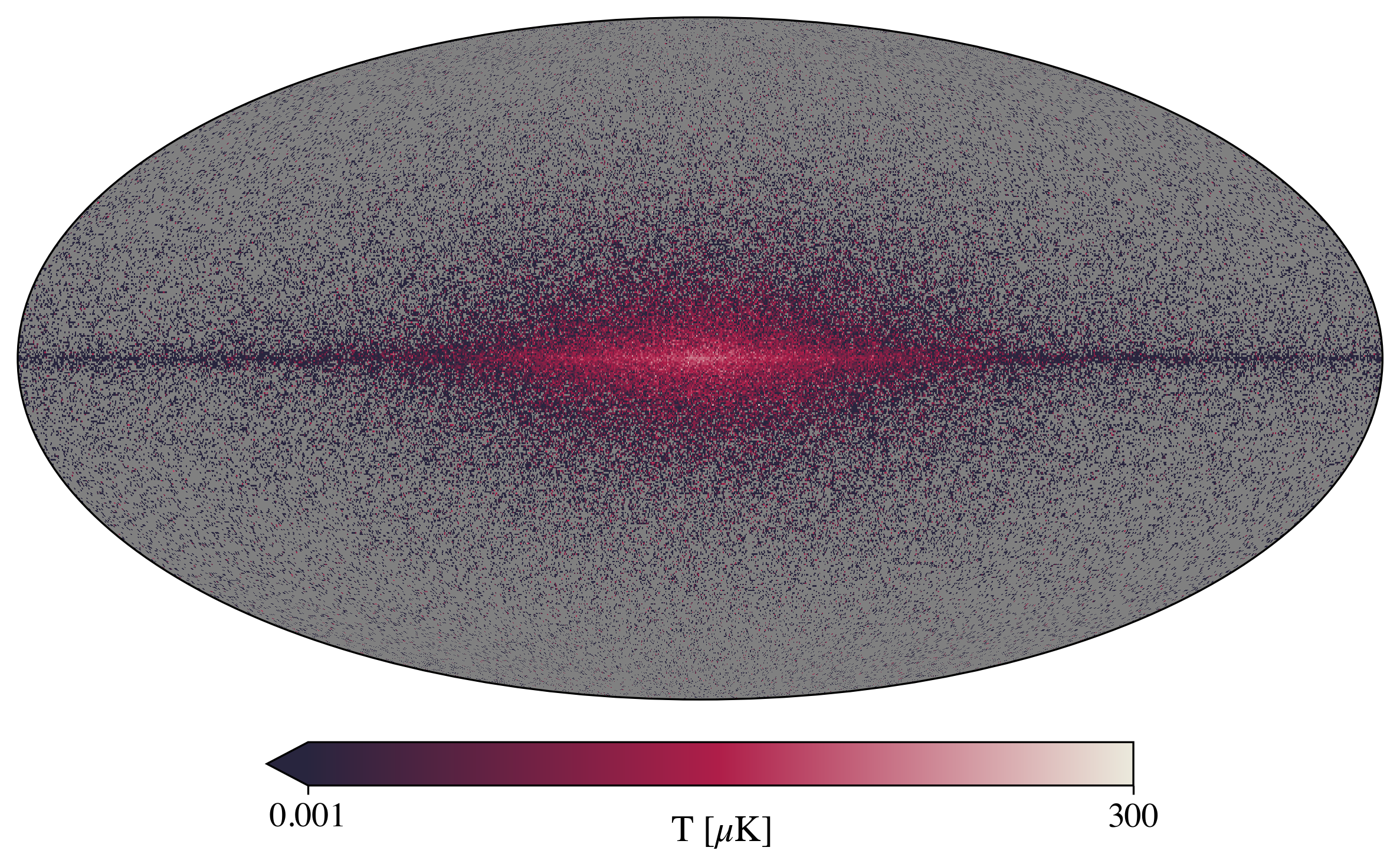}\includegraphics[width = 0.5\textwidth]{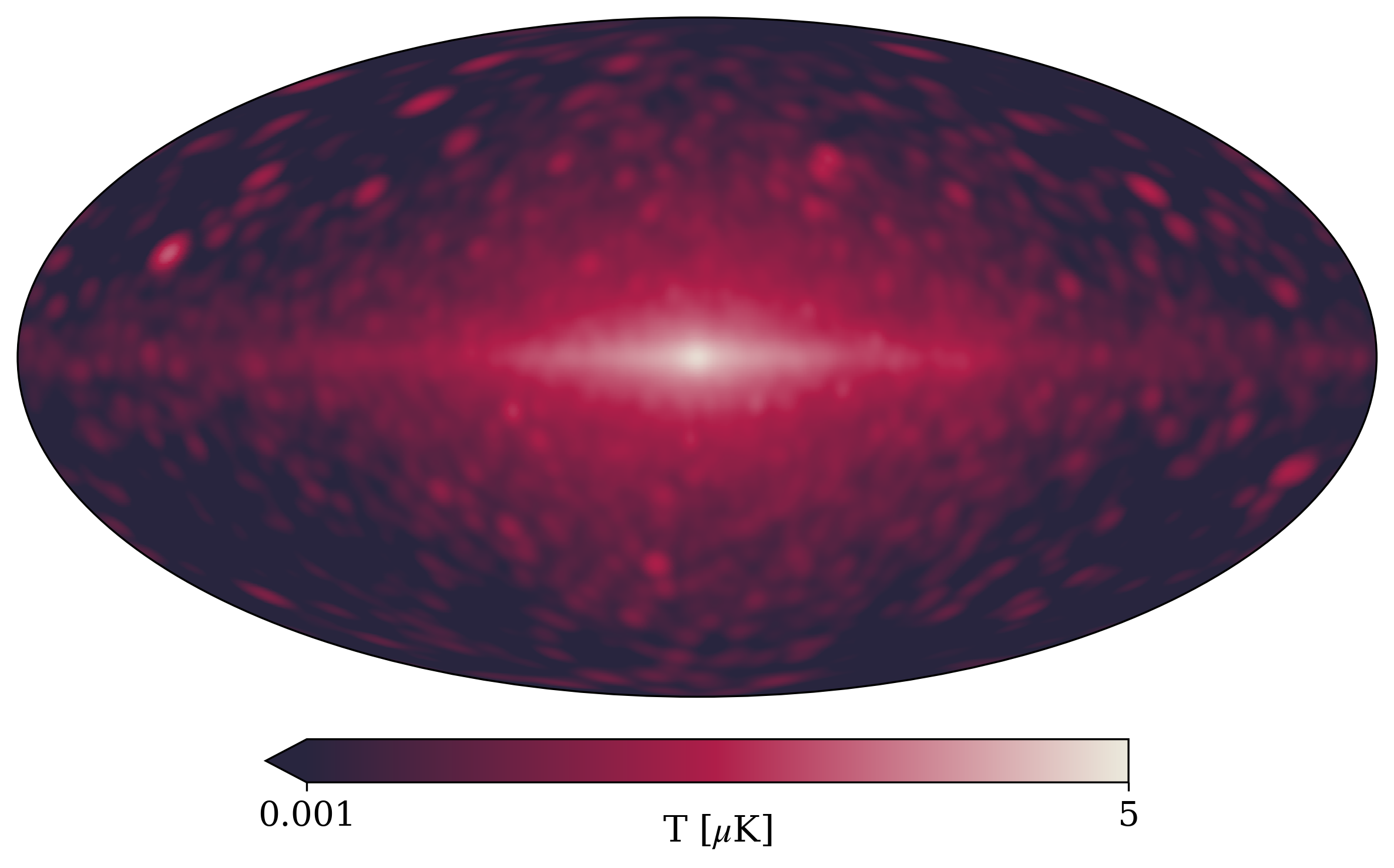}
    \caption{Sky maps of the brightness temperature due to emission from axions using ${\tt Healpix}$ for the NS distribution predicted by {\tt PsrPopPy} where $m_{\rm a}=4.1\,\mu{\rm eV}$ and $\gayy=10^{-11}\,{\rm GeV}^{-1}$. On the left $\theta_{\rm pix}\approx 0.11\,{\rm deg}$ corresponding to ${\tt nside=512}$ which is the base level resolution. The right figure is the same map, but smoothed using a Gaussian filter with $\theta_{\rm FWHM}=3.7\,{\rm deg}$. The axion signal from the population of NSs is clearly peaked in the GC and along the Galactic Plane.}
    \label{fig:skymaps}
\end{figure}

We now consider the spatial variation of the signal within the galactic field of view integrating over the width of the line $\Delta f_{\rm obs}/f_{\rm obs}\approx 10^{-3}$. In order to do this we will bin the pulsar population predicted by ${\tt PsrPopPy}$ in {\tt Healpix} maps as described in section~\ref{sec:normal_pulsar}. We will use a base resolution of $\theta_{\rm pix}=0.46\,{\rm deg}$ corresponding to ${\tt nside=128}$. Let us consider an infinitesimal volume in which the distance to stars and the dark matter density can be treated as effectively constant. The validity of our approach relies on the number of stars in this volume being sufficiently large that one can effectively marginalise over the viewing angles of these stars without incurring a large statistical error. This effectively allows one to replace the line-of-sight power, with an average luminosity for each star. That is opposed to ray-tracing (see discussion in sec.~\ref{sec:luminosity}) which requires a detailed numerical calculation of the line-of-sight power for each star. Examination of fig.~\ref{fig:dist_pulse} suggests that the number of stars in a map with $\theta_{\rm pix}=3.7\,{\rm deg}$ and ${\tt nside=64}$ is $>100$ along the Galactic Plane (GP) and within around $30^{\circ}$ of the GC. We have chosen a finer resolution, but then we smooth to different resolutions with a Gaussian filter with FWHM $\theta_{\rm FWHM}$. We expect this process to work well for resolutions $\theta_{\rm FWHM}>0.5\,{\rm deg}$. Example maps of the sky temperature predicted using the NFW profile for $\rho_{\rm DM}$ are given in fig.~\ref{fig:skymaps}.

\begin{figure}
    \centering
    \includegraphics[width = 0.8 \textwidth]{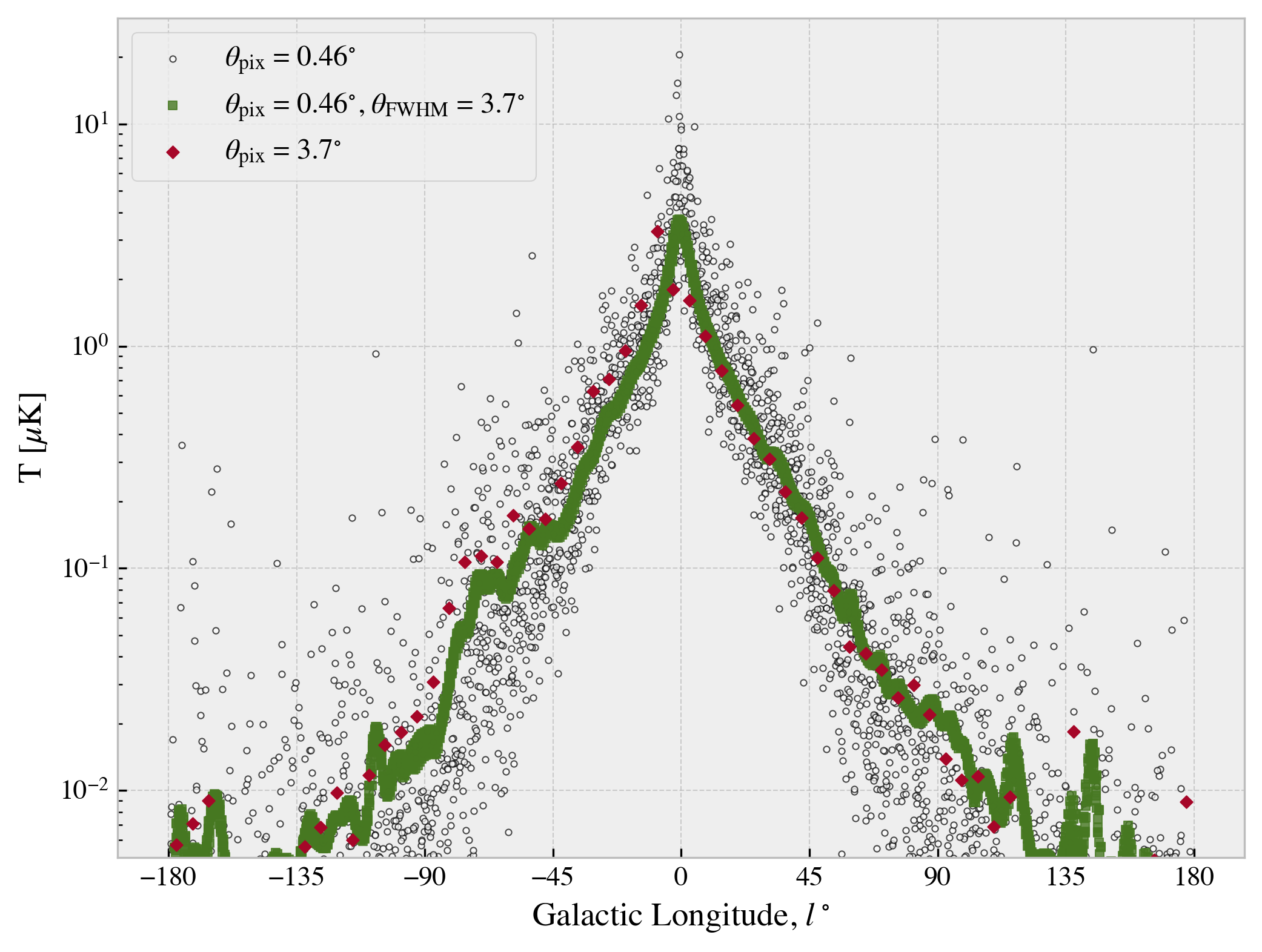}
    \caption{The brightness temperature as a function of Galactic Longitude, $l$, for a patch of Galactic Latitude, $|b| \leq 1^\circ$. The unfilled circles are for the original {\texttt{Healpix}} map with $\theta_{\rm pix}=0.46\,{\rm deg}$. There is significant scatter in this quantity. When we smooth the map using a Gaussian filter with $\theta_{\rm FWHM}=3.7\,{\rm deg}$ we obtain the green curve. This can be compared to the using a {\tt Healpix} map with $\theta_{\rm pix}=3.7\,{\rm deg}$  which is very similar, but shows more discrete features, as one would expect.}
    \label{fig:gpscan}
\end{figure}

\begin{figure}
    \centering
    \includegraphics[width = 0.7\textwidth]{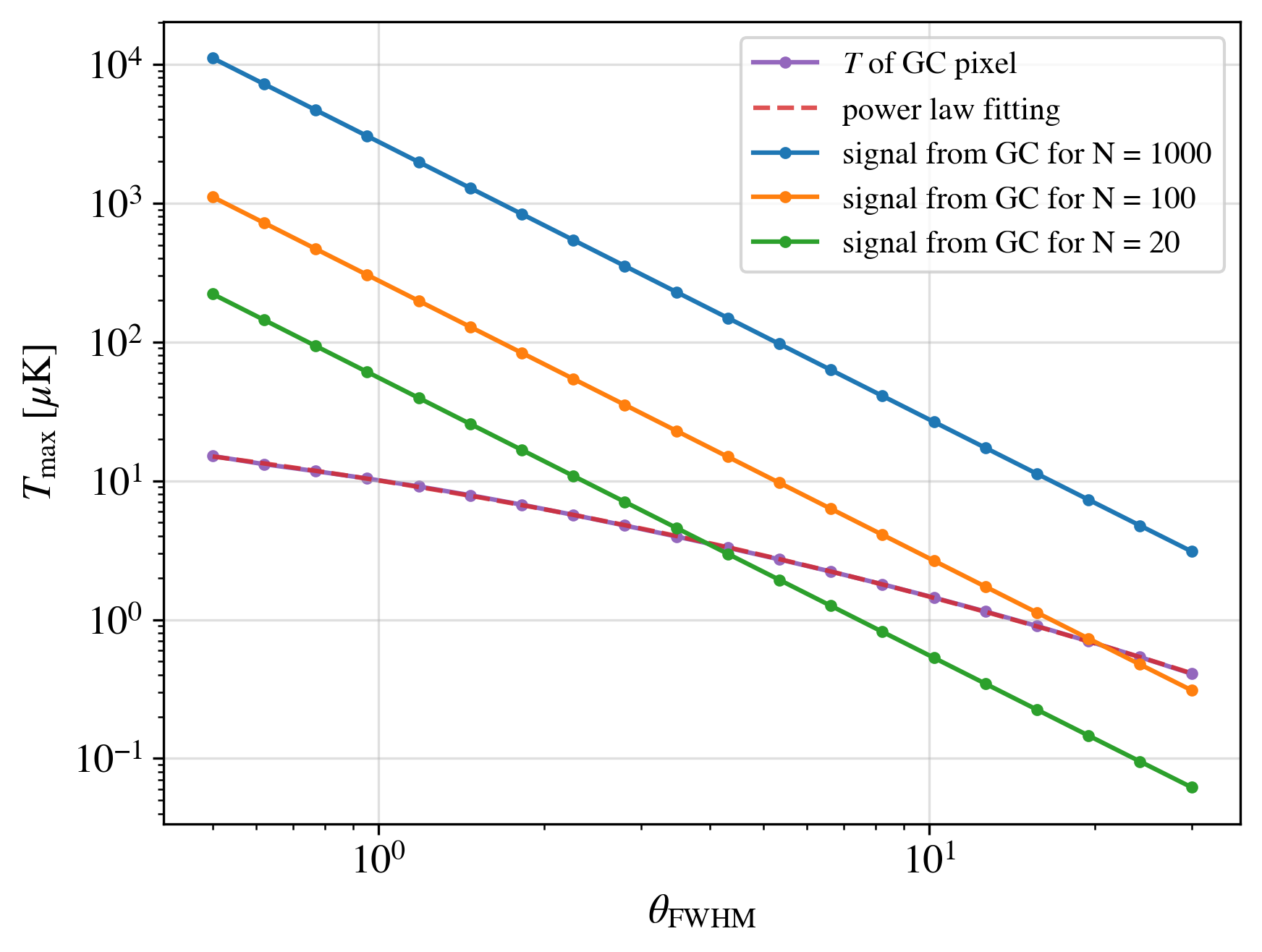}
    \caption{The maximum brightness temperature, $T_{\rm max}$, found at the centre of the sky map when it is smoothed with a Gaussian filter with $\theta_{\rm FWHM}$ between $30\,{\rm arcmin}$ and $10\,{\rm deg}$. The analysis is carried out at $f_{\rm obs}=1\,{\rm GHz}$ which corresponds to $m_{\rm a}\sim 4.1\,\mu{\rm eV})$. Included also is the best fit power law given by (\ref{tmax}). We see that for values of $N_{\rm GC}\lesssim 100$ there is a value of $\theta_{\rm FWHM}$ where the signal predicted by {\tt PsrPopPy} is larger than that from the GC. Note that we have not included the GCM which definitely exists and could lead to a strong axion signal, but it is uncertain because we do not know $\alpha$ and $\theta$.}
    \label{fig:tvtheta}
\end{figure}

In fig.~\ref{fig:gpscan} we present the temperature for scans along the Galactic Plane as a function of Galactic Longitude, $l$. Using the high resolution map, $\theta_{\rm pix}=0.46\,{\rm deg}$, smoothed using a Gaussian filter with $\theta_{\rm FWHM}=3.7\,{\rm deg}$ - equivalent to the {\tt Healpix} map with {\tt nside}=64 - revealing a smooth profile increasing to $\approx 10\,\mu{\rm K}$. Included also are the original pixels which illustrate the discreteness of the signal along lines-of-sight with strong sources. It is clear that the maximum temperature found in the direction of the GC will increase as one reduces $\theta_{\rm FWHM}$ and the maximum brightness temperature, $T_{\rm max}$, as a function of $\theta_{\rm FWHM}$ is presented in fig.~\ref{fig:tvtheta}. We see that $T_{\rm max}$ decreases and in fact if $\theta_{\rm FWHM}\rightarrow\infty$ it becomes that described in the previous section and presented in the right-hand panel of fig.~\ref{fig:allspec}. The best fit to the curve for $T_{\rm max}(\theta_{\rm FHWM})$ is a broken power law in the range $30\,{\rm arcmin}<\theta_{\rm FWHM}<10\,{\rm deg}$ given by 
\begin{equation}
T_{\mathrm{max}} \approx 1.0 \, \mu \mathrm{K} \left(\frac{\theta_{\mathrm{FWHM}}}{ 14.24 \text{ deg}}\right)^{-\alpha_1}\left\{\frac{1}{2}\left[1+\left(\frac{\theta_{\mathrm{FWHM}}}{ 14.24 \text{ deg}}\right)^{1 / \Delta}\right]\right\}^{\left(\alpha_1-\alpha_2\right) \Delta}
\label{tmax}
\end{equation}
where $\alpha_1 = -0.23, \ \alpha_2 = 2.47, \ \Delta = 3.5$.

\subsection{Signal from the Galactic Centre and present constraints}
\label{sec:gc}

 \begin{figure}
     \centering
     \includegraphics[width = \textwidth]{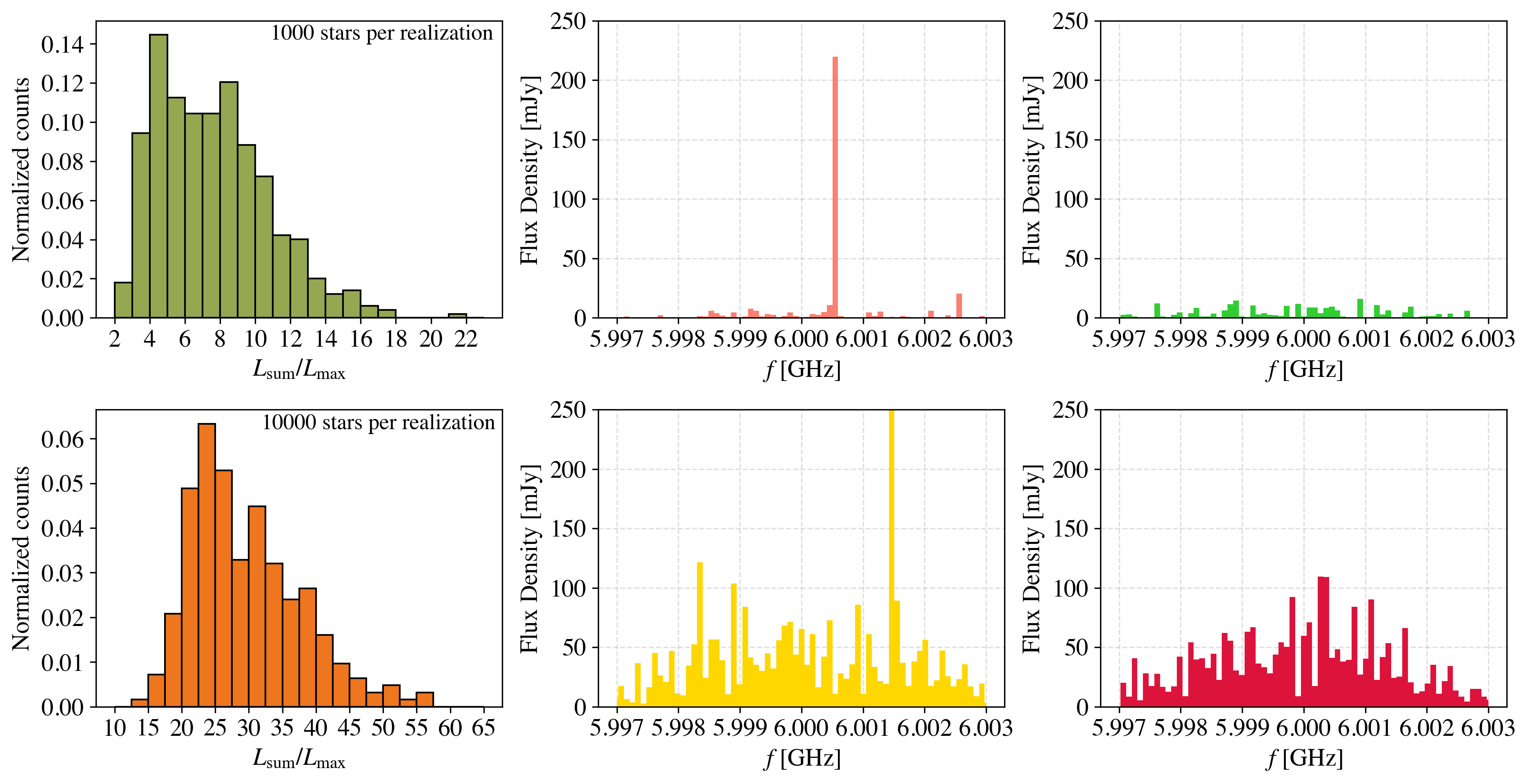}
     \caption{The spectral properties of the signal derived from the GC using $N_{\mathrm{GC}}$ of 1000 (depicted in the top row) and 10000 (depicted in the bottom row) at $f_{\rm obs}=6\,{\rm GHz}$ and $\gayy = 10^{-11} {\rm GeV}^{-1}$ using the fiducial model for the values of $B_0$ and $P$. In both rows, the left-most figure is a histogram of the ratio total luminosity of all stars, $L_{\rm sum}$ and the luminosity of the brightest star, $L_{\rm max}$ for a given realization. The other two figures in each panel are the spectra of the flux density with channel width $\Delta f_{\rm obs}=91.6\,{\rm kHz}$ for the minimum and maximum values of $L_{\mathrm{sum}} / L_{\mathrm{max}}$ - the middle two and right-most panels, respectively. We see that when the ratio is low, one (or a small number) of stars dominate the spectrum, whereas for larger value the spectrum is much closer to the ``top-hat" function with $\Delta f_{\rm obs}/f_{\rm obs}\approx 10^{-3}$ assumed when calculating our constraints. The impact of individual stars with larger fluxes is more prevalent for lower values of $N_{\rm GC}$.}
     \label{fig:spectra}
 \end{figure}

In section~\ref{sec:gcdist} we have argued that there could be a population of the NSs in the GC that can, subject to some caveats, be modelled using its size $N_{\rm GC}$ and log-normal distributions with parameters $\mu_B$, $\sigma_B$, $\mu_P$, $\sigma_P$ for the magnetic fields and periods. In this subsection we examine the size of the signal from this population. 

The region of the GC we are considering has a diameter $\approx 1\,{\rm pc}$ and hence an angular size of $\approx 0.5\,{\rm arcmin}$. We are now considering a point source for most telescopes, as opposed to a background field, and therefore we have calculated the flux density 
\begin{equation}
    S=\frac{L}{4\pi D^2\Delta f_{\rm obs}}\,,
\end{equation}
using the distance to the GC $D=D_{\rm GC}\approx 8.3\,{\rm kpc}$. The Green Bank Telescope (GBT) observations of the Breakthrough Listen (BL) project used in ref.~\cite{Foster:2022fxn} derived limits on the flux density for a bandwidth of $\Delta f_{\rm obs}=91\,{\rm kHz}$ imposing a limit of $S_{\rm lim} \lesssim 4\,{\rm mJy}$ across a range $4\,{\rm GHz} \lesssim f_{\rm obs} \lesssim 8\,{\rm GHz}$. In this paper we have argued that the appropriate signal bandwidth is $\Delta f_{\rm obs}/f_{\rm obs} \approx 10^{-3}$ which corresponds to $\Delta f_{\rm obs} \approx 6\,{\rm MHz}$ in the centre of the band. In order get an appropriate limit we will assume that the signal is a ``top-hat" function with this width and artificially reduce the limit by a factor $(6\,{\rm MHz}/91.6\,{\rm kHz})^{1/2}$ which means that $S_{\rm lim}\approx 0.5\,{\rm mJy}$.

Since we do not have the raw data we are not able to search for signals with this increased bandwidth and, therefore, one should see our limits as indicative rather than actual limits. The reason why this was not done in ref.~\cite{Foster:2022fxn} is that they suggested that $\Delta f_{\rm obs}=91.6\,{\rm kHz}$ is more appropriate arguing that the spectrum is dominated by a single star for their modelling. We have investigated this issue in fig.~\ref{fig:spectra}. Initially, we plot a histogram of $L_{\rm sum}/L_{\rm max}$ which is the ratio of the total luminosity from all the stars to that of the largest luminosity. Low values of this indicate that one star dominates the spectrum from the GC, while for larger values spectra look much more like a ``top-hat" with a width of $\Delta f_{\rm obs}/f_{\rm obs}\approx 10^{-3}$ as we have assumed. Also we see that the larger the value of $N_{\rm GC}$ then the more likely the spectrum is best modelled as a ``top-hat". It is clearly possible for the spectrum to be dominated by a single star, but this appears to less prevalent given our modelling. We are unsure of the reason for this discrepancy, but we speculate that it could be related to the modeling of the stars' positions. We placed all the stars at $r=0.5\,{\rm pc}$, while \cite{Foster:2022fxn} used a probability distribution that depends on $r$. Given that we have argued that the overall number of stars is not well known, their distribution is even less well known. Moreover, the SNR is $\propto \Delta f_{\rm obs}^{-1/2}$ and so choosing the lower value could possible artificially inflate the constraints. Our conclusion is that one should use $\Delta f_{\rm obs}/f_{\rm obs}=10^{-3}$ corresponding to the marginalised signal width in the large $N_{\rm GC}$ limit. 

\begin{figure}
     \centering
     \includegraphics[width = \textwidth]{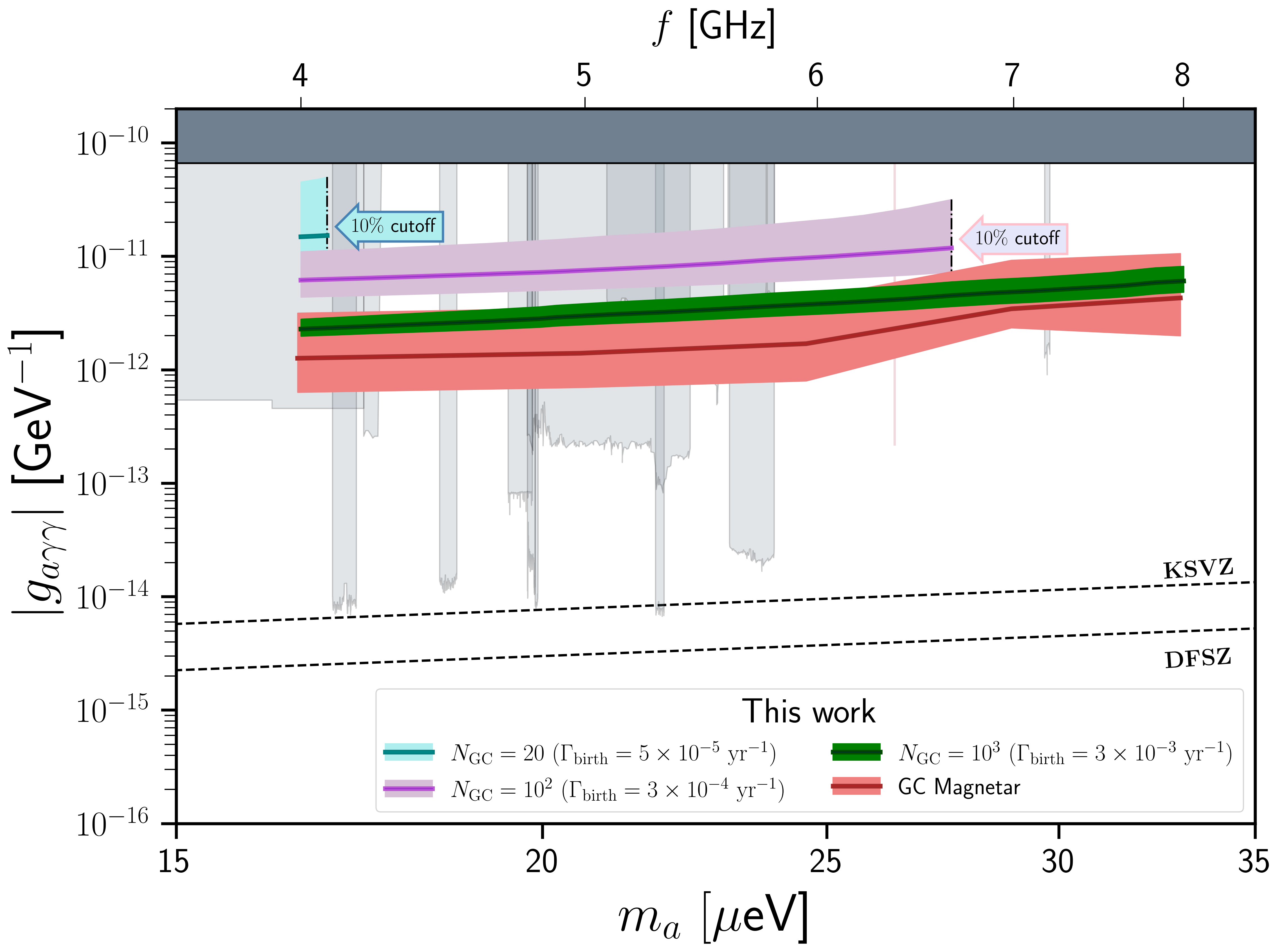}
     \caption{We present constraints on $\gayy$ based on $S_{\rm lim}<0.5\,{\rm mJy}$ over a bandwidth of $\Delta f_{\rm obs}/f_{\rm obs}=10^{-3}$ which is an extrapolation of the limit from the BL data~\cite{Foster:2022fxn} as explained in the text for $N_{\rm GC}= 20, 100$ and $1000$. Note the relation between birth rates and actual numbers at the GC includes loss due to diffusion of stars away from the GC (see appendix \hyperref[appendix:diffusion]{A.2}). We observe that the limits follow the proportionality $\propto N_{\rm GC}^{-1/2}$, as expected, with the bands for populations corresponding to the relative error $\propto \sigma/\sqrt{N}_{\rm GC}$ where $\sigma$ is the standard deviation of the luminosity. We impose cuts when the probability that the population does not emit rises above $10\%$, beyond which we do not impose a limit. This phenomenon is further illustrated in fig.~\ref{fig:rc_cut}. Grey bands show haloscope experiments \cite{ADMX:2009iij,ref:ADMX2018,ADMX:2009iij,ref:ADMX2018,ADMX:2019uok,ADMX:2021nhd,ADMX:2018ogs,ADMX:2021mio,Crisosto:2019fcj, HAYSTAC:Brubaker2017, HAYSTAC:2018rwy, HAYSTAC:2020kwv,CAPP:Jeong_2020cwz, CAPP:Lee_2020cfj,CAPP:Lee_2022mnc, CAPP:Kim2022, Adair_2022, RBF, UF,ORGAN1, ORGAN2} and we also display two QCD axion models \cite{ref:K, ref:SVZ, ref:DFSZ} as dashed lines.}
     \label{fig:constraint}
 \end{figure}

In fig.~\ref{fig:constraint} we compute limits using the fiducial values of $\mu_B$, $\mu_P$, $\sigma_B$ and $\sigma_P$ where we have computed the average flux from a population of $N_{\rm GC}$ stars in the Galactic Centre. We also display the relative statistical error bars defined by $\sim \pm \sigma/\sqrt{N}$ where $\sigma$ is the variance of the one-star luminosity at a given mass. 

\begin{figure}
    \centering
    \includegraphics[width = 0.7\textwidth]{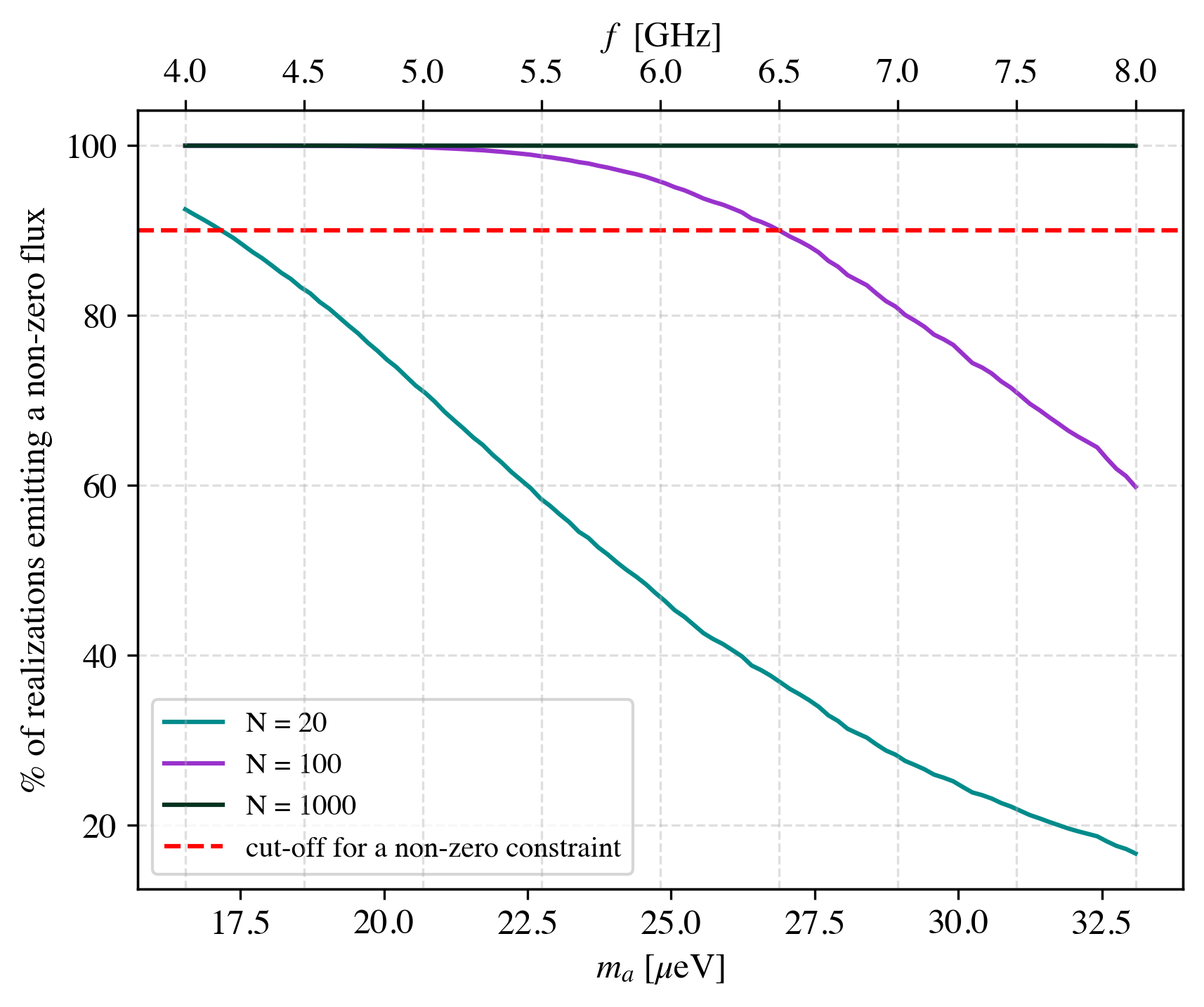}
    \caption{The figure shows the percentage of realizations emitting a non-zero flux as a function of $m_{\rm a}$. We note that as $N_{\rm GC}$ increases, our results are insensitive to the cut-off choice. On the other hand, the constraint does depend on this choice for low values of $N_{\rm GC}$, which simply demonstrates that the statistical method of computing the constraint breaks down as the properties of individual stars in the sample are much more important. These results are further complemented by the constraint plot in fig.~\ref{fig:constraint}.}
    \label{fig:rc_cut}
\end{figure}

Note that we must also confront the fact that for a shrinking population size, it becomes increasingly likely that the population does not emit at all. If the probability that a \textit{star} does not emit is $p_{\rm null} = p_{\rm null} \; (m_{\rm a}) $  then the probability the \textit{population} does not emit is given simply by the binomial probability $p_N = p_{\rm null}^N$. Now, $p_{\rm null}$ grows with $m_{\rm a}$ as the critical radius is more likely to be found inside the star and  $p_{\rm null}^N$ grows for decreasing $N$. Hence, for low values of $N$ and high values of $m_{\rm a}$ it becomes increasingly likely there is no signal at all from the population, making any constraint increasingly unreliable. To reflect this, we impose a cut on the mass when $p_N$ for the population reaches the $10\%$ level, beyond which the likelihood of there being no signal forbids a constraint being imposed.  This is shown in figs.~\ref{fig:constraint} and \ref{fig:rc_cut}.

\subsection{Obtaining a constraint on \texorpdfstring{$\gayy$}{gayy} from the background pulsar population}
\label{sec:future}

In the previous section we have calculated constraints on $\gayy$ imposed by a population in the GC using our modelling and the data from ref.~\cite{Foster:2022fxn}. We have highlighted that constraint is $\propto N_{\rm GC}^{-1/2}$, but that $N_{\rm GC}$ could also be very low, possibly to the point where the constraint is either non-existent, or is, at the very least, no stronger that the CAST, bound given the uncertainties. In particular, for the fiducial pulsar velocities and galactic potential models used in appendix \hyperref[appendix:diffusion]{B.2} we find $N_{\rm GC}\approx 20$ (which is less than $N_{\rm GC}\approx 1000$ used in ref. \cite{Foster:2022fxn}) though we caution that more work is needed to properly model stellar dynamics and diffusion away from the GC. 

The uncertainty in the population in the central $1\,{\rm pc}^3$ is a key problem in deriving a reliable constraint using this approach. We have also explained that the population predicted by {\tt PsrPopPy} is likely much more reliable. In what follows we will discuss the possibility of detecting this diffuse, more easily predicted signal and how one might use this to derive constraints on $\gayy$. It is also more in keeping with the original philosophy~\cite{Safdi2019} of using the population to reduce the uncertainties in the magnetosphere modelling which one might worry about in studies of individual stars~\cite{foster2020,Darling:2020plz,darling2020apj,Battye2022,Battye:2023oac}.

To obtain the constraint plot in fig.~\ref{fig:constraint}, we compute the total luminosity defined by
\begin{equation}
   L_{\rm GC}=N_{\rm GC} \langle L\rangle  = \int L \; n(B,P,\alpha)d(\log_{10} B)d(\log_{10} P)d(\cos\alpha)\,,
\end{equation}
so that the population is drawn from a bivariate log-normal distribution with parameters $(\mu_B,\sigma_B)=(12.1,0.54)$, $(\mu_P,\sigma_P)=(-0.23,0.36)$, and correlation parameter (see appendix \hyperref[appendix:BPCorellaiton]{B}) $r_{\mathrm{corr}} = 0.44$. We compute this integral via Monte-Carlo sampling of the population and compared the resultant flux from BL, which we have rescaled to $\simeq 0.5 {\rm mJy}$ by using a larger bin width $\Delta f / f = 10^{-3}$, as per the discussion above. Fig.~\ref{fig:rc_cut} shows the probability that the population has non-zero flux (i.e. the percentage of population realizations for which at least one star emits) for different values of values of $N_{\mathrm{GC}}$. As stated above, we demand that the probability $p_N$ that the population does not emit satisfies\footnote{One could of course use a different threshold value, which is a matter of convention, but the general principle remains the same, namely it is not reasonable to impose a constraint if it becomes unacceptably likely that there is no signal.} $p_N < 0.1$, or else we do not impose a limit, since it becomes unacceptably likely that there is no signal form the population at all, despite a non-zero average. 

We now turn to the detectability of this signal. We first compare the signal predicted by {\tt PsrPopPy} with that from the GC using the simple estimate for the luminosity (\ref{lum_gc}). If we were to assume that all the flux from GC population contributes only to the central pixel of a \texttt{Healpix} map, then we can calculate the brightness temperature as $T= S/(2k_{\rm B}f_{\rm obs}^2\theta_{\rm pix}^2)$. If we added this flux to the centre of a {\tt Healpix} map and employ a smoothing as in the section on the signal predicted by {\tt PsrPopPy}, for $f_{\rm obs}=1\,{\rm GHz}$ and $\gayy=10^{-11}\,{\rm GeV}^{-1}$ we can estimate the contribution from the GC as  
\begin{equation} 
T_{\rm GC}\approx 2.8\,{\rm mK}\left(\frac{\theta_{\rm FWHM}}{1\,{\rm deg}}\right)^{-2}\left(\frac{N_{\rm GC}}{1000}\right)\,,
\end{equation}
for $\Delta f_{\rm obs}/f_{\rm obs}=10^{-3}$. We have 
included this estimate in fig.~\ref{fig:tvtheta} for $N_{\rm GC}=20$, 100 and 1000. For $N_{\rm GC}=1000$ the signal is always dominated by GC population, whereas there is a crossover for lower values and in particular this takes place at $\theta_{\rm FWHM}\approx 20\,{\rm deg}$ for $N_{\rm GC}=100$ and $\theta_{\rm FWHM}\approx 4\,{\rm deg}$ for $N_{\rm GC}=20$. So if we were to design instruments with apertures corresponding to these values, we argue that we would be able to make accurate  predictions of the minimum signal. 

Note the average constraint from the GCM gives the strongest constraint, since the associated flux is larger and is also associated to a narrower bandwidth \cite{Battye:2021xvt}, but is more uncertain due to the variation of the signal as a function of $(\alpha, \theta)$. In addition, one can increase the SNR associated to the GCM by adding time-domain information \cite{Battye:2023oac} (this statement holds even without knowing the specific pulse templates, which are challenging to model in general, and especially for the magnetar whose magnetosphere is poorly understood). We leave a more sophisticated observational analysis of the GCM to future work.

\begin{figure}
    \centering
    \includegraphics[width = \textwidth]{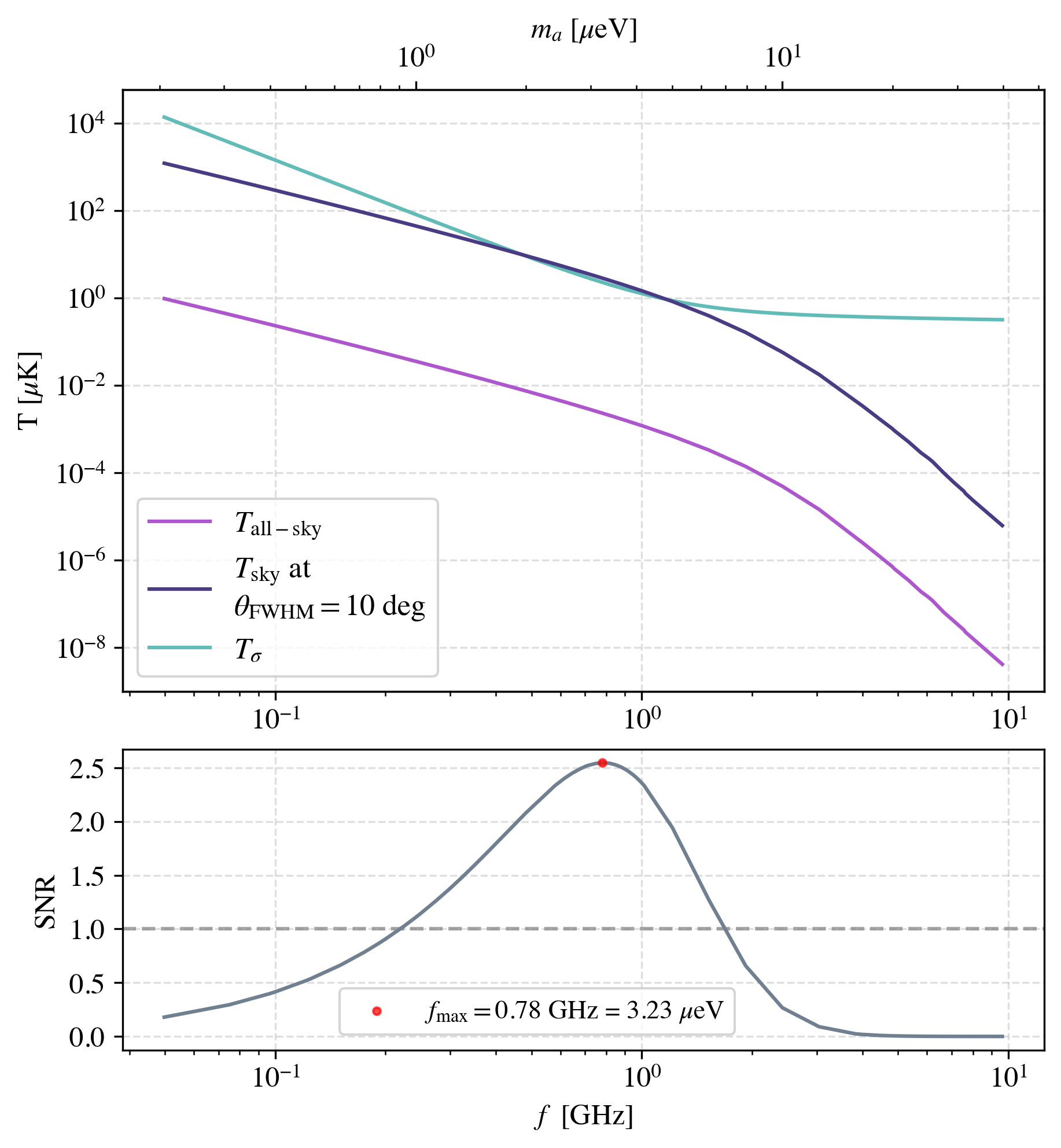}
    \caption{In the top panel we present, $T_{\rm source}$ at $\theta_{\rm FHWM}=10\,{\rm deg}$ for $\gayy=10^{-11}\,{\rm GeV}^{-1}$ which is an estimate of the axion signal with $\Delta f_{\rm obs}/f_{\rm obs}=10^{-3}$ that might be detected by the ``strawman" horn array discussed in the text where we have used {\tt PsrPopPy} to model the population of pulsars. We have also included $T_{\rm all-sky}$ which is the all-sky signal presented in fig.~\ref{fig:allspec}. Notice that the shape as a function of frequency (or $m_{\rm a}$) is almost universal. We have also plotted the noise level we would expect using the instrument model described in the text using $t_{\rm int}=1\,{\rm year}$ and $N_{\rm H}=100$. In the bottom panel we have calculated ${\rm SNR}=T_{\rm source}/T_\sigma$ to give an idea of the optimum observing frequency which appears to be at $f_{\rm max} \approx 750 \,{\rm MHz}$.}
    \label{fig:future_probe}
\end{figure}

Let us now consider a simple ``strawman'' design for an instrument that is capable of detecting the wider signal predicted by {\tt PsrPopPy}  before considering specific instruments already available or proposed that might be capable of doing this. What we are looking for is an instrument with a relatively large value of $\theta_{\rm FWHM}$, and hence a relatively small aperture, along with plenty of sensitivity, which will require many detectors. It also has to be relatively cost effective and should have other applications, for example, detection of the redshifted 21cm emission from the Epoch of Reionization (EoR). Imagine that we have $N_{\rm H}$ horns with aperture $5\lambda_{\rm obs}$ which is typical, corresponding to $\theta_{\rm FWHM}\sim 10\,{\rm deg}$ - which is well into the regime where we would expect the diffuse signal to dominate. A horn array of this kind would be relatively cheap to build since it would not require much infrastructure. However, it will need to be capable of tracking the GC for a significant fraction of the time in order to maximise the signal. What we are envisaging is not dissimilar to the kind of arrays that are placed at the focus of telescopes to increase the survey speed. If the system temperature is $T_{\rm sys}$ and all the horns were able to point at the GC all of the time with a single polarization and absolute calibration then one could achieve a noise level 
\begin{equation}
T_{\sigma}\approx\frac{T_{\rm sys}}{\sqrt{\Delta f_{\rm obs} N_{\rm H}t_{\rm int}}}\,,
\end{equation}
where $t_{\rm int}$ is the on-sky integration time. The system temperature is the sum of a number of components 
\begin{equation}
T_{\rm sys}=T_{\rm CMB}+T_{\rm bg}+T_{\rm rx}.
\end{equation}
The background $T_{\rm CMB}=2.726\,{\rm K}$ is always present. The Galactic contribution\footnote{The amplitude of this contribution is taken to be that for a pointing on the GC. This is somewhat larger than the quietest parts of the sky, but the effects of the central part of the GC, where is well known that $T_{\rm sys}$ is somewhat larger, are diluted by the aperture sizes, $\theta_{\rm FWHM}\gtrsim 1\,{\rm deg}$, relevant to this discussion.} increases at low frequencies and can be modelled using the ``Haslam map'', which gives the background contribution of Galactic and extragalactic sources at 408~MHz~\cite{Remazeilles:2014mba}, smoothed to 10~deg giving 
\begin{equation}
T_{\rm bg}\approx 1\,{\rm K}\left(\frac{f}{3.1\,{\rm GHz}}\right)^{-2.75}\,,
\label{t_bg}
\end{equation}
whereas the receiver temperature, $T_{\rm rx}$ would be expected to increase with $f$. For the purposes of this discussion we will take it to be given by 
\begin{equation} 
T_{\rm rx}\approx 10{\rm K}\left(\frac{f}{1\,{\rm GHz}}\right)^{0.4}\,,
\end{equation}
which gives $T_{\rm rx}\approx 40{\rm K}$ at $f_{\rm obs}=30\,{\rm GHz}$~\cite{Cleary:2021dsp} which is approximately the right scaling from $T_{\rm rx}=10{\rm K}$ at $f_{\rm obs}=1\,{\rm GHz}$~\cite{SKA:2018ckk}. The exact values below $f_{\rm obs}\sim 1\,{\rm GHz}$ are relatively unimportant since they are typically dominated by the Galactic contribution. If we consider an array of horns with $N_{\rm H}=100$ and a year of on-source integration time, $t_{\rm int}=1\,{\rm year}$, with a bandwidth $\Delta f_{\rm obs}/f_{\rm obs}=10^{-3}$ we might be able obtain the noise level presented in fig.~\ref{fig:future_probe}. Such an instrument would not be difficult to imagine using present technology and indeed it could be built for very other uses. The key thing to take away from this discussion is that $\gayy=10^{-11}\,{\rm GeV}^{-1}$ could easily be detected/ruled out in the frequency range $f=100\,{\rm MHz}-1\,{\rm GHz}$ using an instrument with $Nt_{\rm int}=100\,{\rm years}$ and limits on $\gayy$ will reduce $\propto (N_{\rm H}t_{\rm int})^{-1/4}$.

Intensity mapping experiments and global 21cm instruments, along with Square Kilometre Array (SKA) and its precursors~\cite{SKA:2018ckk} present opportunities to make observations that might be used to constrain this signal in the near future. Often this will involve using instruments designed as interferometers in ``single dish mode" (as described, for example in ref.~\cite{Bull:2014rha}). Given that the instruments are not designed for these purposes there could be many issues that prevent them from achieving their full potential in terms of noise sensitivity. For the moment, we will ignore these issues for the purposes of this discussion which is aimed at establishing whether there is any possibility of detecting these signals. Below we will consider various examples in order to get some idea of what might be possible with the aim of developing more detailed analyses for options that seem viable.

\begin{itemize} 
\item The Hydrogen Epoch of Reionization Array\footnote{\url{https://www.sarao.ac.za/science/hera/}} (HERA) operates at $f=150-200\,{\rm MHz}$~\cite{HERA:2021bsv}. Using (\ref{t_bg}) we estimate $T_{\rm sys}\approx 2700\,{\rm K}$ for the centre of the band, $f_{\rm obs}=175\,{\rm MHz}$, when pointing at the GC which corresponds to a noise level of $110\,{\rm mK}\,(t_{\rm int}/{\rm hour})^{-1/2}$ for each of the individual dishes.  There are $350\times 14\,{\rm m}$ dishes observing in the Karoo site. The resolution of the individual dishes is $\theta_{\rm FWHM}\approx 7\,{\rm deg}$ at $f_{\rm obs}=175\,{\rm MHz}$, and the signal for $f_{\rm obs}=175\,{\rm MHz}$ and $\theta_{\rm FWHM}\approx 7 \,{\rm deg}$ is estimated to be $T_{\rm signal}\approx 130 \,\mu{\rm K}$ for $\gayy=10^{-11}\,{\rm GeV}^{-1}$. This could be excluded at 95\% confidence detected in $\approx 2\times 10^5\,{\rm hours}$ of on source integration time with one dish. The array is static and fields are only observed at transit meaning a given field-of-view, for example the GC, is only visible for $\approx 40\,{\rm mins}$ per day. If all 350 dishes observed the GC for 40~mins a day that would correspond to $\approx 200\,{\rm hours}$ of integration in a day and therefore the signal for $\gayy=10^{-11}\,{\rm GeV}^{-1}$ could be excluded in a few years. Such a limit would over the mass range $0.6\,\mu{\rm eV}<m_{\rm a}<0.8\,\mu{\rm eV}$ and would be slightly stronger at the high-end and vice-versa at the low-end. 

\item The MeerKAT\footnote{\url{https://www.sarao.ac.za/science/meerkat/about-meerkat}} is an SKA precursor and comprises $64\times 13.5\,{\rm m}$ dishes in South Africa which could track the GC in single dish mode. It will be subsumed into SKA-mid\footnote{\url{https://www.skao.int/en/explore/telescopes/ska-mid}} once it is operational which will involve the inclusion of an extra $133 \times \ 15 \, {\rm m}$ dishes. Let us consider the UHF band of MeerKAT covering $580-1015\,{\rm MHz}$ corresponding to $\theta_{\rm FWHM}\approx 1.7\,{\rm deg}(f_{\rm obs}/750\,{\rm MHz})^{-1}$. The measured values for MeerKAT estimate that is $T_{\rm sys}$ is in the range $20-40\,{\rm K}$ for fields at high Galactic Latitude\footnote{\url{https://skaafrica.atlassian.net/servicedesk/customer/portal/1/article/277315585}}. However, this will be substantially increased in the direction of the GC: (\ref{t_bg}) gives $T_{\rm bg}\approx 50\,{\rm K}$ at $f_{\rm obs}=750\,{\rm MHz}$ and hence $T_{\rm sys}\approx 65\,{\rm K}$ seems a more reasonable value to use. We estimate a noise level of $\approx 150\,\mu{\rm K}(t_{\rm int}/{\rm hour})^{-1/2}$ using all 64 telescopes. The estimated signal is $T_{\rm signal}\approx 16 \,\mu{\rm K}$ for $\gayy=10^{-11}\,{\rm GeV}^{-1}$ and $f_{\rm obs}=750\,{\rm MHz}$. Achieving target noise level of $8\,\mu{\rm K}$ to rule out $\gayy= 10^{-11}\,{\rm GeV}^{-1}$ at 95\% confidence level would take $t_{\rm int}\approx 350\,{\rm hours}$. If one were to use SKA-mid this would take $\approx 100\,{\rm hours}$. 

\item SKA-low\footnote{\url{https://www.skao.int/en/explore/telescopes/ska-low}} will comprise of 512 ``tiles' each formed from 256 antennas that can be thought to  behave like a single aperture with a diameter of $\approx 40\,{\rm m}$ with $\theta_{\rm FHWM}\approx 2\,{\rm deg}\,(f_{\rm obs}/200\,{\rm MHz})^{-1}$ where $f_{\rm obs}=50-350\,{\rm MHz}$. The overall collecting area is expected to be $\approx  4.2\times 10^5\,{\rm m^2}$.  Even though it is a static array, it will the ability to track the GC using beam forming techniques. At $200\,{\rm MHz}$ we have that  $T_{\rm sys}\approx 1900\,{\rm K}$ and hence the noise level will be $T_{\rm noise}\approx 3\,{\rm mK}(t_{\rm int}/{\rm hour})^{-1/2}$ by using all the tiles. The signal is expected to be $\approx 300 \,\mu{\rm K}$ for $\gayy=10^{-11}\,{\rm GeV}^{-1}$ which could be achieved in $\sim 1\,{\rm day}$ of integration.  
\end{itemize}

\begin{figure}
    \centering
    \includegraphics[width = \textwidth]{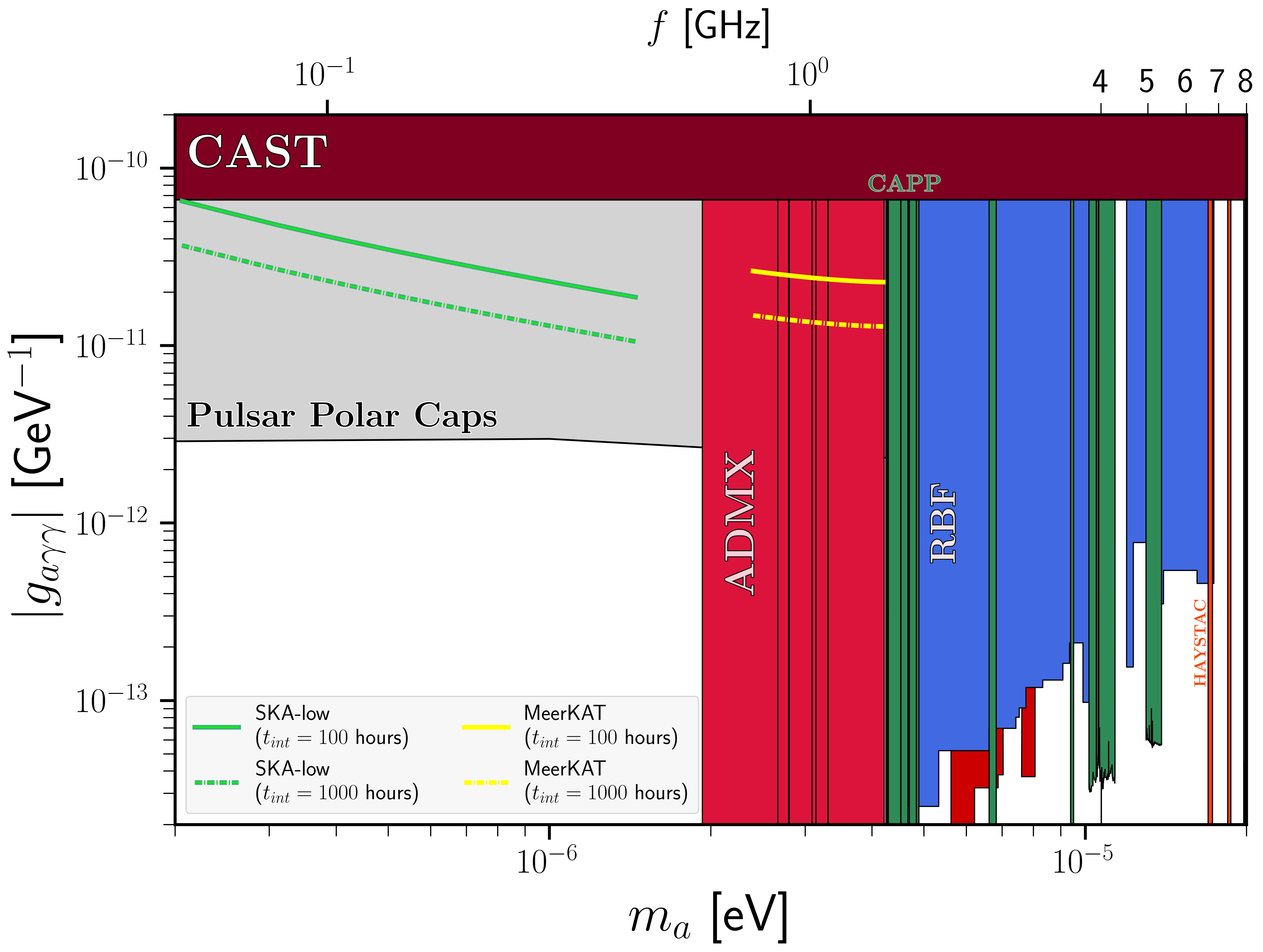}
    \caption{The constraint on the coupling constant $\gayy$ predicted from future SKA-low and MeerKAT surveys. The SKA-low covers a broad range of frequency, starting from 50 MHz to 350 MHz while the MeerKAT survey covers a range of 580 MHz to 1015 MHz. Coloured bands show haloscopes \cite{ADMX:2009iij,ref:ADMX2018,ADMX:2009iij,ref:ADMX2018,ADMX:2019uok,ADMX:2021nhd,ADMX:2018ogs,ADMX:2021mio,Crisosto:2019fcj, HAYSTAC:Brubaker2017, HAYSTAC:2018rwy, HAYSTAC:2020kwv,CAPP:Jeong_2020cwz, CAPP:Lee_2020cfj, CAPP:Lee_2022mnc, CAPP:Kim2022, Adair_2022, RBF, UF}, and the gray region indicates constraints from  pulsar polar caps \cite{Noordhuis:2022ljw}.}
    \label{fig:lowfreqconstraints}
\end{figure}

The conclusion appears to be that HERA, MeerKAT/SKA-mid and SKA-low possibly have the ability to beat the CAST bound. We illustrate some projected sensitivities in fig.~\ref{fig:lowfreqconstraints}. It is conceivable that these observations could be made commensually with other observations of the GC. Qualitatively each of these cases can be thought of being similar to our ``strawman" instrument albeit with different values of $N_{\rm H}t_{\rm int}$.

\section{Discussion and conclusions}\label{sec:conclusions}

In this paper we have critically examined NS populations as probes of axion dark matter proposed in refs.~\cite{Safdi2019,Foster:2022fxn} as an alternative to using single pulsars~\cite{foster2020,Darling:2020plz,darling2020apj,Battye:2021xvt,Battye:2023oac}. Using {\tt PsrPopPy}, we have found that the population modelled through calibrating against detected pulsars beaming towards Earth predicts a much lower number of stars near the GC than that inferred from birth rates alone. Indeed, very few pulsars have been detected in the GC making direct observational inference of the population size challenging. Instead, birth rate modelling is used to predict the size of the population today, but as we have explored in appendix \hyperref[appendix:diffusion]{A.2}, stellar dynamics can influence the number of stars which remain in the GC at any one time, owing to the flow of stars away from the GC due to their large kick velocities and subsequent trajectories within the galactic gravitational potential. This suggests a closer study of stellar dynamics is warranted in future work. Given these uncertainties, in this work, we parametrise this ignorance by treating the number of stars at the Galactic Centre $N_{\rm GC}$ as a free parameter, whose range varies between $N_{\rm GC} \sim 1000$ (commensurate with birth rates alone) and $N_{\rm GC} \sim 20$ (which includes putative effects from stellar dynamics diluting the number of stars at the GC).

Using this simplified parametrisation we have re-compared constraints from observations of single objects, namely the GCM, and populations. We find that unless the population size at the GC is at or below $\sim 1000$, the GCM can be competitive, due to the narrower bandwidth and the large magnetic field associated to it. However, the uncertainty arising from the dependence of the signal on $(\alpha, \theta)$ is significantly larger than that associated to the largest population of stars, i.e., $N_{\rm GC} = 1000$. Although the statistical error on the constraint is low for large $N_{\rm GC}$, the true size of the population may be lower due to the velocities of neutron stars at birth, in which case the GCM could give stronger limits. For very for small values of $N_{\rm GC}$, one is no longer in the stochastic ``large-N" limit and the statistical method breaks down such that it becomes necessary to model the properties of individual stars in the population, defeating the purpose of a population analysis. Therefore, we conclude that without further observational evidence to support a larger value of $N_{\rm GC} \sim 1000$, the approach of modelling the population at the GC should be treated with some caution.   

These results suggest that it will be difficult to be totally confident in population constraints on $g_{a \gamma \gamma}$ from observations of the GC with a narrow beam (a few arcmin) which is only sensitive to details of the poorly understood population within $1\,{\rm pc}^3$. The original idea of using pulsar populations~\cite{Safdi2019} argued that using large numbers of NSs could reduce the sensitivity to the modelling of the magnetosphere. We have suggested that using observations of the GC with a wide beam $1-10\,{\rm deg}$, we will be sensitive to the signal predicted by {\tt PsrPopPy}. Note that unlike the GC, where the population size must be inferred indirectly, population modelling in {\tt PsrPopPy} is normalized against the observed population of pulsars, making it possible to predict a minimum signal as a function of $m_{\rm a}$ and $\gayy$. We have argued that such observations might be possible either using a bespoke instrument, or by adapting observations made using current and proposed instruments. Simple estimates suggest that signals for $\gayy\sim 10^{-11}\,{\rm GeV}$ could be detected or ruled to in the mass range $\sim 0.4-4\,\mu{\rm eV}$ corresponding to  $f_{\rm obs}\sim 100\,{\rm MHz}-1\,{\rm GHz}$. 

At these lower frequencies, such limits from e.g. MeerKAT and SKA-low are not competitive with constraints from  pulsar caps  \cite{Noordhuis:2022ljw} (which use current telescopes) or existing laboratory constraints from ADMX. It therefore seems that the most promising direction for placing constraints remains at higher frequencies using the magnetar or NS populations at the GC, where there are at present fewer lab constraints. However, in light of the astrophysical uncertainties on population sizes and the details of the GCM, it seems fairly evenly balanced as to whether the magnetar or the hidden GC population gives the strongest sensitivity, meaning that any future observing strategies with next-generation facilities should search for both types of signals. 

\section*{Acknowledgments}

 We thank Jonas Tjemsland and Sam Witte for help with code developed in ref.~\cite{Tjemsland:2023vvc} and the authors of {\tt PsrPopPy}~\cite{Bates:2013uma} for making their code freely available. We would like to thank Lina Levin-Preston and Ben Stappers for advice on the usage of {\tt PsrPopPy}, and also Mike Keith, Patrick Weltewrede and Rene Breton for helpful comments concerning pulsar populations. We have also benefited from helpful comments concerning ref.~\cite{Foster:2022fxn} from Sam Witte. SS was supported formerly by a George Rigg Scholarship and more recently by the UK Science and Technology Facilities Council (STFC) and an Alexander von Humboldt Fellowship at LMU.  JIM acknowledges support from STRC grants ST/T001038/1 and ST/X00077X/1.  For the purpose of open access, the authors have applied a Creative Commons Attribution (CC BY) licence to any Author Accepted Manuscript version arising.

\appendix

\section{Summary of NS birth rate calculations}
\label{ref:Appendix A}

In this section we review some aspects of the modelling of the NS population from ~\cite{Foster:2022fxn} that are relevant to the discussion in the rest of the paper. We will concentrate on  what they term population I which is a model for young NSs in the GC. In addition they consider a population of old NSs. Such a population may well exist, but the constituent NSs are likely to have relatively low magnetic fields and hence, unless they are very numerous, will not contribute significantly to the axion signal. In keeping with the philosophy of assessing the minimum signal advocated in the rest of the paper, we ignore this population.

\subsection{Number and distribution of neutron stars in the Galactic Centre} \label{ref:Appendix A.1}

The basis of the estimate of the number of NSs in the GC is the radius dependent birth function~\cite{Foster:2022fxn}
\begin{equation}
\Psi(r)=\Psi_0\left(\frac{r}{r_0}\right)^{-\beta}\exp\left[-\frac{r}{r_0}\right]\,,
\label{birth}
\end{equation}
where $\Psi_0=3.6\times 10^{-5}\,{\rm pc}^{-3}\,{\rm yr}^{-1}$, $r_0=0.5\,{\rm pc}$ and $\beta=1.93$. The overall birth rate can be calculated by integrating this over the volume element 
\begin{equation}
\psi=4\pi\int_0^{\infty}r^2dr\,\Psi(r)=4\pi\Psi_0r_0^3\Gamma(3-\beta)\,,
\label{birthrate}
\end{equation}
which  corresponds to a rate of $5.4\times 10^{-3}$ per century - note that $\Gamma(1.07)=0.964$. This number is very small compared to the overall pulsar birth rate in the Galaxy which is $\sim 2-3\,{\rm century}^{-1}$ (for example, see ref.~\cite{Keane:2008jj}) and, therefore, the population we are talking about is a small fraction of the NSs in the Galaxy, as one would expect since the GC is a very small volume. This implies that it could be difficult to constrain using observations.
In addition to this it is assumed that velocity distribution of the NSs is drawn from an isotropic Maxwellian distribution with a velocity dispersion, $v_0$, which depends on the $r$ according to 
\begin{equation}
v_0(r)\approx 188\,{\rm km}\,{\rm s}^{-1}\left(\frac{r}{r_0}\right)^{\sfrac{1}{2}}\,.
\label{vel}
\end{equation}

\begin{figure}
    \centering
    \includegraphics[width = 0.75\textwidth]{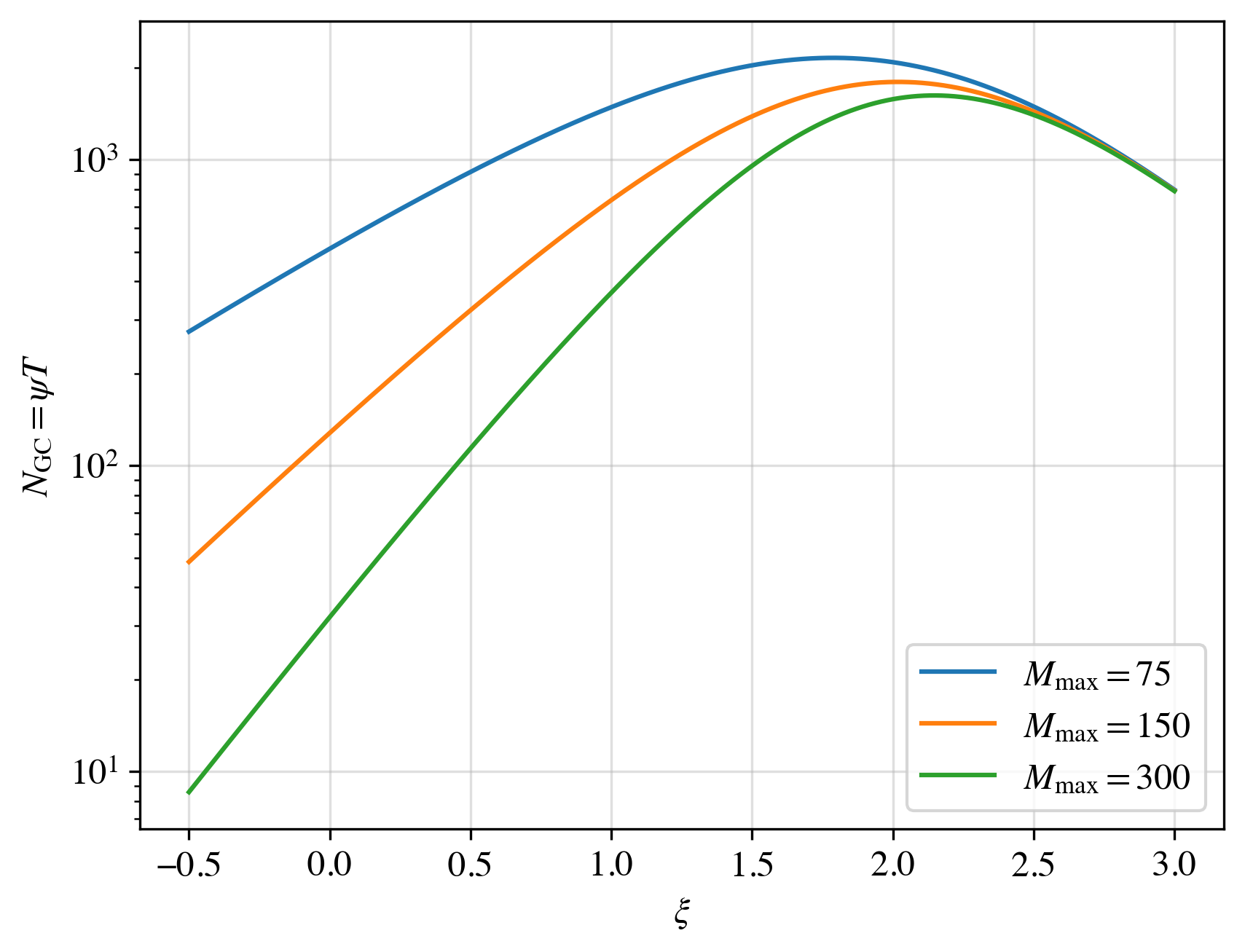}
    \caption{The number of NSs in the GC predicted by $N_{\rm GC}=\psi T$ where $T=30\,{\rm Myr}$ as a function of the IMF power law parameter $\xi$. We have also varied the upper limit on the total mass $M_{\rm max}$. It is clear that it is possible to reduce the number of NSs predicted by at least an order of magnitude. Note that the number of {\em observed} pulsars would need to be compared with this number divided by $\sim 10$ to take into account the beaming fraction which is $\sim 10\%$.}
    \label{fig:IMF}
\end{figure}

The birth rate (\ref{birth}) is calculated assuming that all stars between masses $M^{\rm NS}_{\rm min}=8\,M_\odot$ and $M^{\rm NS}_{\rm max}=20\,M_{\odot}$ form NSs. These stars with masses between $M_{\rm min}=1\,M_\odot$ and $M_{\rm max}=150\,M_{\odot}$ are produced with an Initial Mass Function (IMF) $dN/dM\propto M^{-\xi}$  and $\xi=1.7$~\cite{Lu:2013sn} (although see ref.~\cite{Bartko:2009qn} and note that the standard Salpeter value is $\xi=2.35$) which is normalised by the overall star formation star ${\dot M}_{\rm tot}=4\times 10^{-3}\,M_{\odot}\,{\rm yr}^{-1}$.  Under these assumptions the birth rate can be calculated 
\begin{align}
    \psi={\dot{M}_{\mathrm{tot}}} \frac{f_{\mathrm{NS}}}{\langle M\rangle}=\frac{\dot{M}_{\mathrm{tot}}}{M_{\odot}} 
    \frac{\int_{M_{\min }^{\mathrm{NS}} / M_{\odot}}^{M^{\mathrm{NS}}_{\mathrm{max}} / M_{\odot}} x^{-\xi} d x}{\int_{M_{\min } / M_{\odot}}^{M_{\text {max}} / M_{\odot}} x^{1-\xi} \; d x}=\frac{\dot{M}_{\text {tot }}}{M_{\odot}} \frac{2-\xi}{\xi-1}\left[\frac{ (x_{\min }^{\mathrm{NS}})^{1-\xi} - (x_{\max }^{\mathrm{NS}})^{1-\xi}}{x_{\mathrm{max}}^{2-\xi}-x_{\mathrm{min} }^{2-\xi}}\right] \, ,
    \label{IMF}
\end{align}
where $x=M/M_\odot$, $f_{\rm NS}$ is the fraction of NSs formed and $\langle M\rangle$ is the average mass of a star. This function has a strong dependence on the power law of the IMF, $\xi$, and depending on this value it can also be sensitive to the range assumed for the distribution of stars and also that of the stars assumed to become NSs. Moreover if $\xi\approx 0$ then $\psi\approx 2({\dot M}_{\rm tot }/M_\odot)(x^{\rm NS}_{\rm max}-x^{\rm NS}_{\rm min})/x_{\rm max}^2$  when $x_{\rm max}\gg x_{\rm min}$ illustrating that it can be very sensitive to the upper limit $M_{\rm max}$ for some values of $\xi$. We present the $\psi T$ as a function of $\xi$ for different values of $M_{\rm max}$ in fig.~\ref{fig:IMF}. 

The specific choices of $\xi$, $M_{\rm max}$ and ${\dot M}_{\rm tot}$, which are the key parameters, used in ref.~\cite{Foster:2022fxn} are not unreasonable, but it is difficult to believe that these parameters are known with any degree of certainty. In particular, ref.~\cite{Lu:2013sn} suggests $\xi=1.7\pm 0.2$, while ref.~\cite{Bartko:2009qn} find $\xi=0.45\pm 0.3$. Fig.~\ref{fig:IMF} illustrates that an order of magnitude or more reductions in the overall number of NSs are quite reasonable for the range of values allowed by observations.  Moreover, it seems that an efficiency factor, $\epsilon_{\rm NS}$,  should be attached to the formation of NSs, that is the number of NSs formed, is 
\begin{equation}
N_{\rm NS}=\epsilon_{\rm NS}\int_{M^{\rm NS}_{\rm min}}^{M^{\rm NS}_{\rm max}} \frac{dN}{dM}dM\,.
\end{equation} 
The assumption that $\epsilon_{\rm NS}=1$ is an extreme assumption, but it could quite reasonably be $\epsilon_{\rm NS}\ll 1$. 

Based on the arguments above, we conclude that there is at least a factor of ten uncertainty in the range of values of $N_{\rm GC}$ which are compatible with present observations. Nonetheless, it is difficult to reconcile these calculations with the {\tt PsrPopPy} prediction of $<1$ object in the central $1\,{\rm pc}^3$. 

\subsection{Neutron star diffusion}\label{appendix:diffusion}

\begin{figure}
    \centering
    \includegraphics[scale=0.6]{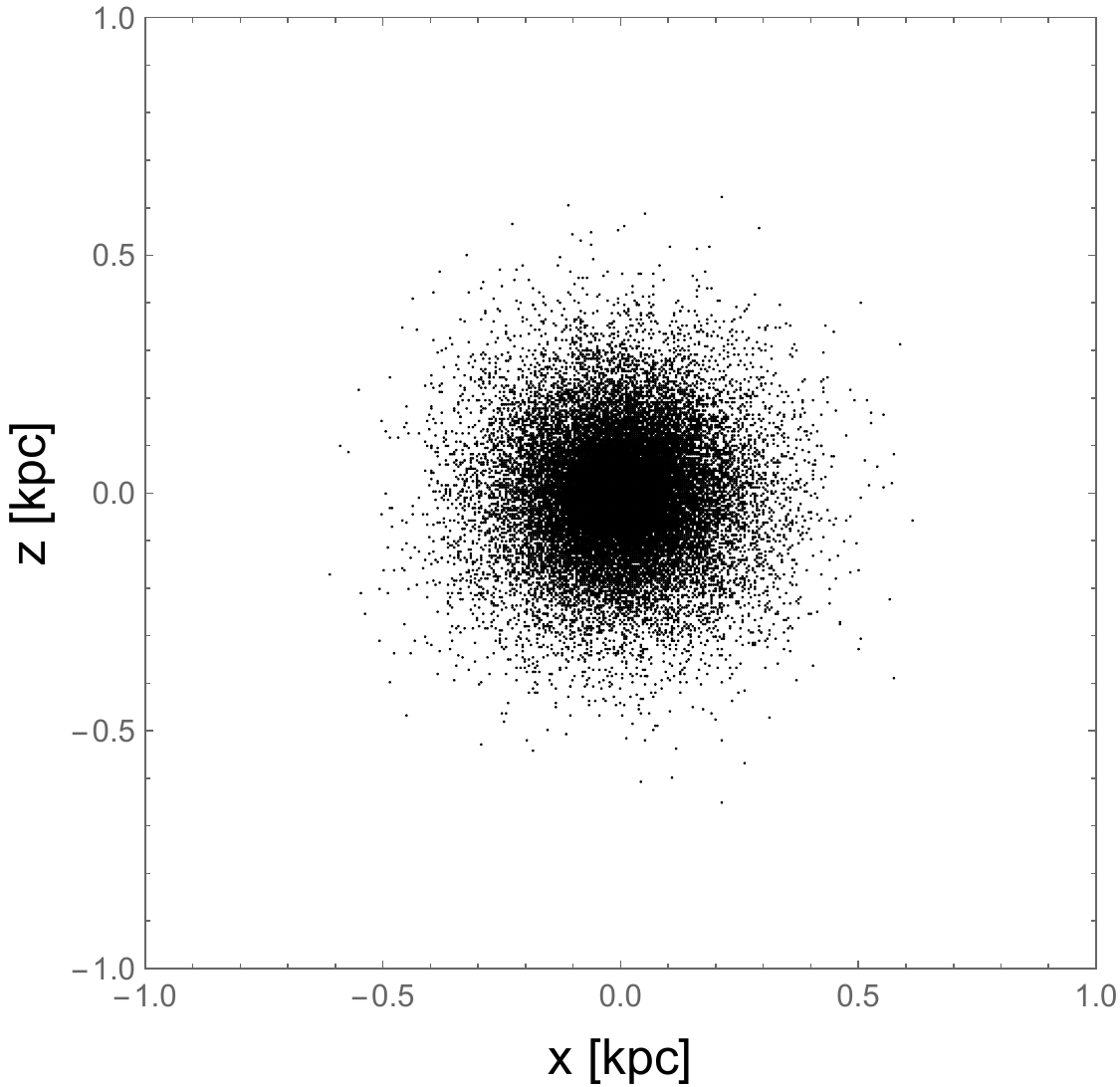}
    \caption{Distribution of NSs birthed within $1 {\rm pc}$ of the GC after $10 {\rm Myr}$ (projection onto $x$-$y$ plane). We took an initial Maxwell-Boltzmann distribution with velocity dispersion with $v_0 = 188\,{\rm km}\,{\rm s}^{-1}$. See text for details.}
    \label{fig:NSDistribution}
\end{figure}

Stars born with velocity $v_0$ with initial position relative the GC, $r_0$, can propagate outwards to a maximum possible distance of $r_{\rm max}$, which, by kinematics, satisfies
\begin{equation}\label{eq:Conservation}
   \Phi(r_{\rm max}) =  \Phi(r_0) + \frac{1}{2}v_0^2 \, ,
\end{equation}
where $\Phi$ is the gravitational potential in the Galaxy at $r=r_0$, which we shall model via \eqref{GGP-model} for the purpose of making estimates. In accordance with previous discussions, if we were to assume $r_0 \sim 1\,{\rm pc}$ and $v_0 \sim 200 {\rm km}\,{\rm s}^{-1}$ we find $r_{\rm max} \sim 100\,{\rm pc}$, by solving the equation above that for motion in the Galactic Plane ($z= 0$). Not all stars will reach this radius, which depends on the nature of their trajectories, but this scale nonetheless gives the characteristic size of the potential within which the stars are confined. To estimate the true distribution of stars, we released stars from $r_0\lesssim  1{\rm pc}$ at constant rate into the Galactic potential modelled by Eq.~\eqref{GGP-model}. We chose initial velocities given by a Maxwell-Boltzmann distribution with width set by $v_0$ and evolved them over the timescale of $\sim 10\,{\rm Myr}$. We find that after this time, the fraction of stars which remain in the inner $\sim 1\,{\rm pc}$ of the GC is $\sim 0.4 \%$ and that the average coordinate of stars is $\braket{|\textbf{x}|} \approx 150 {\rm pc}$ in accordance with the simple estimate above. We display these results in fig.~\ref{fig:NSDistribution}.

Of course, one may question whether other NSs are birthed beyond $r \sim 1\,{\rm pc}$, which could have diffused \textit{back into} the GC, but the presence of stars outside the GC is tightly constrained by the modelling of the ``normal" pulsar population within the Galaxy and indeed those simulations predict the number of pulsars to be low near the GC. In any case, all we attempt to do here is to evaluate the validity of the population modelling assumed in ref.~\cite{Foster:2022fxn} at face value and compute its consequences. 

Taking into account the diffusion of stars away into the Galactic potential, we can now calculate the number of stars within the GBT beam and the corresponding power, in view of the fact the dark matter density in which these stars reside will be lower, owing to it being farther from the GC. Assuming a beam width of $\theta_{\text{FWHM}} \sim \lambda/d$ with $\lambda = 5 \, \text{cm}$ in the middle of the C-band ($6 \, \text{GHz}$) and a diameter of $d=100 \, \text{m}$ for GBT, we find that the telescope beam covers distances of only $\approx 4 \, \text{pc}$ either side of the GC in which only $\sim 1\%$ of birthed stars remain, and hence this means that $N_{\rm GC}\approx 20$.

\begin{figure}
    \centering
    \includegraphics[scale=0.4]{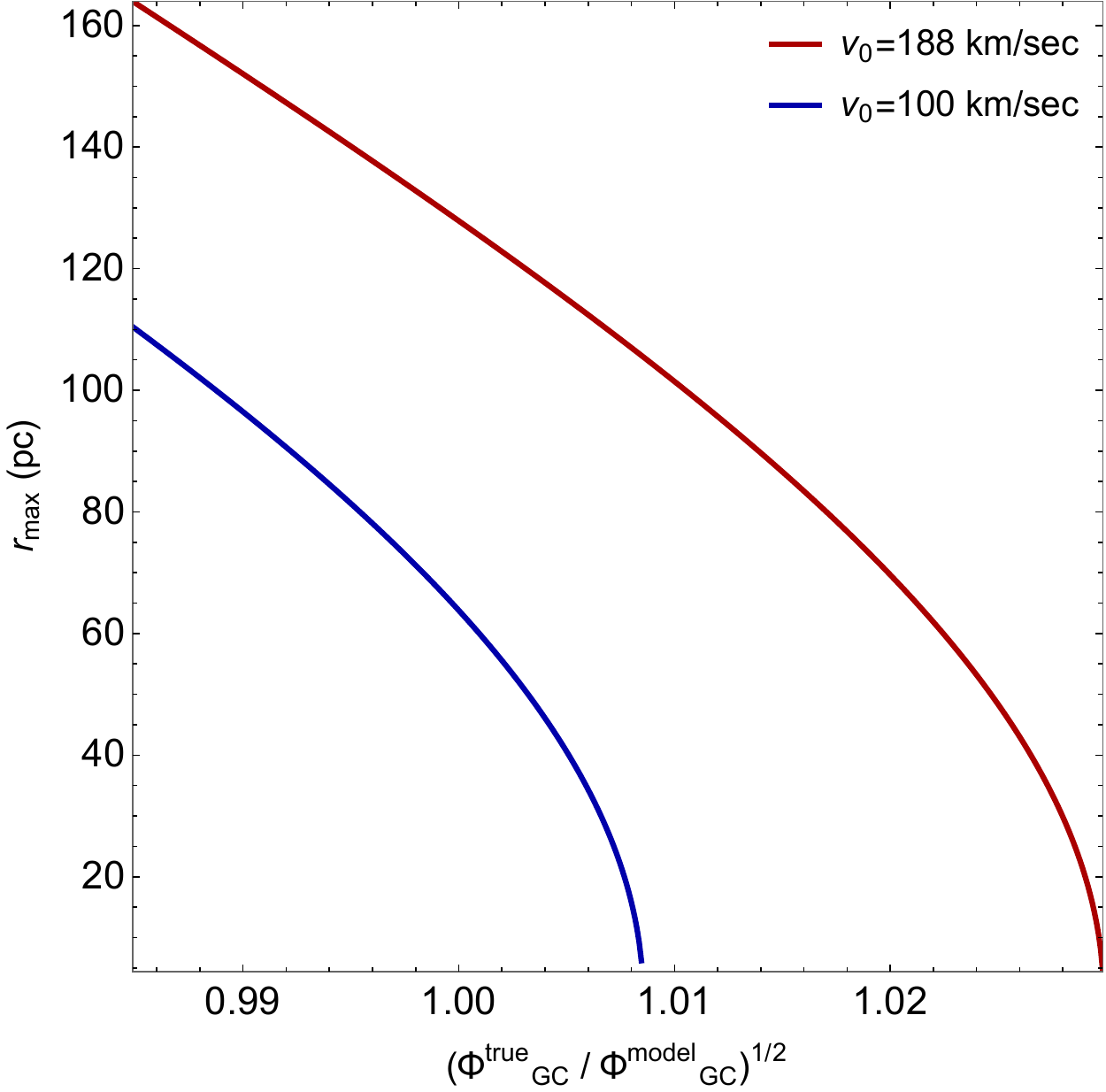}
    \caption{The confining radius $r_{\rm max}$ (in pc) of stars according to \eqref{eq:Conservation} by varying the value of $\Phi^{\rm true}$ the GC relative to the model value $\Phi^{\rm model}$ set by ~\eqref{GGP-model} and table \ref{tab:GP-model_constants}.  We chose two initial NS velocities of $v_0 = 188\,{\rm km}\,{\rm s}^{-1}$ and $v_0 = 100\,{\rm km}\,{\rm s}^{-1}$. }
    \label{fig:confinement_radius}
\end{figure}

In light of the fact that both the velocity distribution of neutron stars and the precise details of the gravitational potential near the GC are not precisely known, it is vital to understand the sensitivity of the above conclusions to these modelling details. To develop some sense of this, we plot the maximum confinement radius $r_{\max}$ by solving Eq.~\eqref{eq:Conservation} for $r_{\rm max}$, by inverting $\Phi(r_{\rm max})$ but treating $\Phi(r_0)$ as a variable quantity which characterises the depth of the potential at the GC. Results are shown in fig.~\ref{fig:confinement_radius}. From this we see that the confinement radius of neutron stars is highly sensitive to the details of the potential at the GC and, in turn, the initial kick velocity of neutron stars. Hence, any discussion of the number of neutron stars near the GC that is based on birth rates is very sensitive to the precise distributions of kick velocities and the shape of the Galactic potential, with the confinement radius varying by 100s of pc for changes of order a few percent in the GC potential. This sensitivity  arises from the kick velocities of NSs and the escape velocity of the galaxy being comparable in order of magnitude, with both characteristically being $\mathcal{O}(100s)\,{\rm km}\,{\rm sec}^{-1}$ . This makes the degree to which stars are confined near the GC very finely balanced.  Clearly this necessitates a closer study of stellar dynamics in future work. 

\subsection{Modelling the values of \texorpdfstring{$B_0$}{B0}, \texorpdfstring{$P$}{P} and \texorpdfstring{$\alpha$}{α}}

The model for the magnetic field, $B_0$, and the period, $P$, used in ref.~\cite{Foster:2022fxn} is based on the observed pulsar population and follows ref.~\cite{Aguilera:2007xk}. The initial distribution of the magnetic field of a NS is assumed to be a log normal in $B = B_0$, where $B_0$ is the magnetic field at birth, and is expressed as,   
\begin{align}
    \mathcal{P}\left(\log B_0 \mid \mu_{B}, \sigma_{B}\right)=\frac{1}{\sqrt{2 \pi} \sigma_{B}} \exp \left[-\frac{\left(\log B_0-\mu_{B}\right)^2}{2 \sigma_{B}^2}\right] \, ,
\end{align} 
where $\log B_0=\log_{10}(B_0/{\rm G})$, $\mu_{B}=13.2$, and $\sigma_{B}=0.62$. These initial values of $B$ are evolved in time using a first order differential equation given as
\begin{equation}
\label{B-decay}
\frac{dB}{dt} = - B\left[\frac{1}{t_{\rm Ohm}}+\frac{1}{t_{\rm  Hall}}\left( \frac{B}{B_0} \right)\right]\,,
\end{equation}
where $t_{\rm Ohm}$ is the Ohmic decay timescale and $t_{\rm Hall}$ is due to the Hall effect, both of which are assumed to be constant. (\ref{B-decay}) can be solved to give, 
\begin{equation}
B(t)=\frac{B_0 \, e^{\sfrac{-t}{t_{\rm Ohm}}}}{1+{\frac{t_{\rm Ohm}}{t_{\rm Hall}}} \left(1-e^{\sfrac{-t}{t_{\rm Ohm}}} \right)} \, .
\end{equation}
We see that for $t\ll t_{\rm Ohm}$ the magnetic field decays linearly
\begin{equation}
B\approx B_0\left(1- \frac{t}{t_{\rm comb}}\right)\,,
\end{equation}
where, $1/t_{\rm comb}=(1/t_{\rm Ohm})+(1/t_{\rm Hall})$ is the combined decay timescale. If $t\gg t_{\rm Ohm}$ then one finds exponential decay 
\begin{equation}
B\approx \frac{B_0t_{\rm Hall}}{t_{\rm Hall}+t_{\rm Ohm}} \; \exp \left[-\frac{t}{t_{\rm Ohm}} \right]\,.
\end{equation}
This line of argument suggests that the ``effective lifetime" in the context of it having a large magnetic field is dictated by $t_{\rm Ohm}$.

The amplitude of the decay timescales can be estimated for a region of characteristic size $L$ from the MHD equations  derived from $\nabla\times{\bf E}=-{\dot{\bf B}}$, $\nabla\times{\bf B}=\mu_0{\bf J}$ and Ohm's law including the effects of ambipolar diffusion to due the motion of positive and negative ions
\begin{equation}
{\bf E}=\frac{{\bf J}}{\sigma} + \frac{{\bf J}\times{\bf B}} {en_{\rm e}}\,,   
\end{equation}
where $\sigma$ is the conductivity and $n_{\rm e}$ is the electron density in the plasma. One finds that $\tau_{\rm Ohm}=\mu_0\sigma L^2\approx 10\,{\rm Myr}$ and $\tau_{\rm Hall}=\mu_0en_{\rm e}L^2/B_0\ll \tau_{\rm Ohm}$. Note that the Hall effect leads to slippage between the ions and electrons in the plasma allowing for a dissipative effect.

The initial periods, $P_0$, are drawn from a Gaussian distribution for $p_0=P_0/{\rm s}$ with mean $\mu_{p_0}=0.22$ and standard deviation $\sigma_{\rm p_0}=0.44$. These are then evolved in conjunction with the misalignment angle $\chi$ - assumed to have uniform probability distribution in $\cos\chi$ using~\cite{Philippov:2013aha}
\begin{equation}
{\dot P}=\gamma \frac{B^2}{P}(\kappa_0+\kappa_1\sin^2\chi)\,,\quad {\dot\chi}=-\gamma \frac{B^2}{P^2}\kappa_2\sin\chi\cos\chi\,,
\end{equation}
where $\gamma \sim [4.5 - 9] \times 10^{-40} \ \mathrm{G}^{-2} \; \mathrm{s}$ and $(\kappa_0,\kappa_1,\kappa_2)=(1,1.2,1)$. The overall process of evolution for $B$, $P$ and $\chi$ is very similar to that implemented in {\tt PsrPopPy} and hence we would expect that it reproduces similar distributions. Hence, their assumption is that, although the number of NSs in the GC is two orders of magnitude larger than is found by {\tt PsrPopPy}, the distribution $B$ and $P$ are similar to that of the active pulsar population - this is a very strong assumption.

\section{Aspects of the signal from the Galactic Centre}\label{appendix:BPCorellaiton}

\subsection{Calculating the luminosity}

One can estimate the total luminosity of $N_{\rm GC}$ NSs to be 
\begin{equation}
   L_{\rm GC}=N_{\rm GC} \langle L\rangle = \int L\,dN = \int L \; n(B,P,\alpha)d(\log_{10} B)d(\log_{10} P)d(\cos\alpha)\,,
   \label{L_integral}
\end{equation}
where $n(B,P,\alpha)$ is the number NSs in per $\log_{10} B$, $\log_{10} P$ and $\cos\alpha$. Using the assumptions above, but initially assuming no correlation between $B_0$ and $P$, that is taking $r_{\mathrm{corr}}=0$, we can rewrite this as
\begin{equation}
L_{\rm GC}={\frac{N_{\rm GC}}{2\pi}}\int_{-\infty}^{\infty} dX\int_{-\infty}^{\infty}dY e^{-\sfrac{(X^2+Y^2)}{2}}\int_0^{1}d(\cos\alpha)L(e^{X\sigma_B+\mu_B}\,{\rm G},e^{Y\sigma_P+\mu_P}\,{\rm G},\alpha)\,.
\end{equation}
Motivated by the analysis of section~\ref{sec:luminosity}, if we assume that $L(B,P,\alpha)\propto B^nP^m$ independent of $\alpha$ where $n\sim 1$ and $m\sim 1$, one finds that 
\begin{equation}
L_{\rm GC}=N_{\rm GC}L(10^{\mu_B}{\rm G},10^{\mu_P}{\rm s})\exp\left[\frac{1}{2}(\log_e10)^2(n^2\sigma_B^2+m^2\sigma_P^2)\right]\,,
\label{multilum}
\end{equation}
that is, $N_{\rm GC}$ multiplied by the luminosity of the average star, plus a correction for the distribution values of $B$ and $P$.  This suggests that in the more general case where $r_{\mathrm{corr}}\ne 0$ that 
\begin{equation}
    L_{\rm GC}\approx N_{\rm GC}L(10^{\mu_B}\,{\rm G},10^{\mu_P}\,{\rm s})F(\sigma_B,\sigma_P,r_{\mathrm{corr}})\,,
    \label{multilum2}
\end{equation}
where $F(\sigma_B,\sigma_P,r_{\mathrm{corr}})\sim {\cal O}(1)$ corrects for the distribution of stars, although including the effects of $r_{\mathrm{corr}}$ analytically is difficult.  

\begin{figure}
     \centering
     \includegraphics[width = 1.0\textwidth]{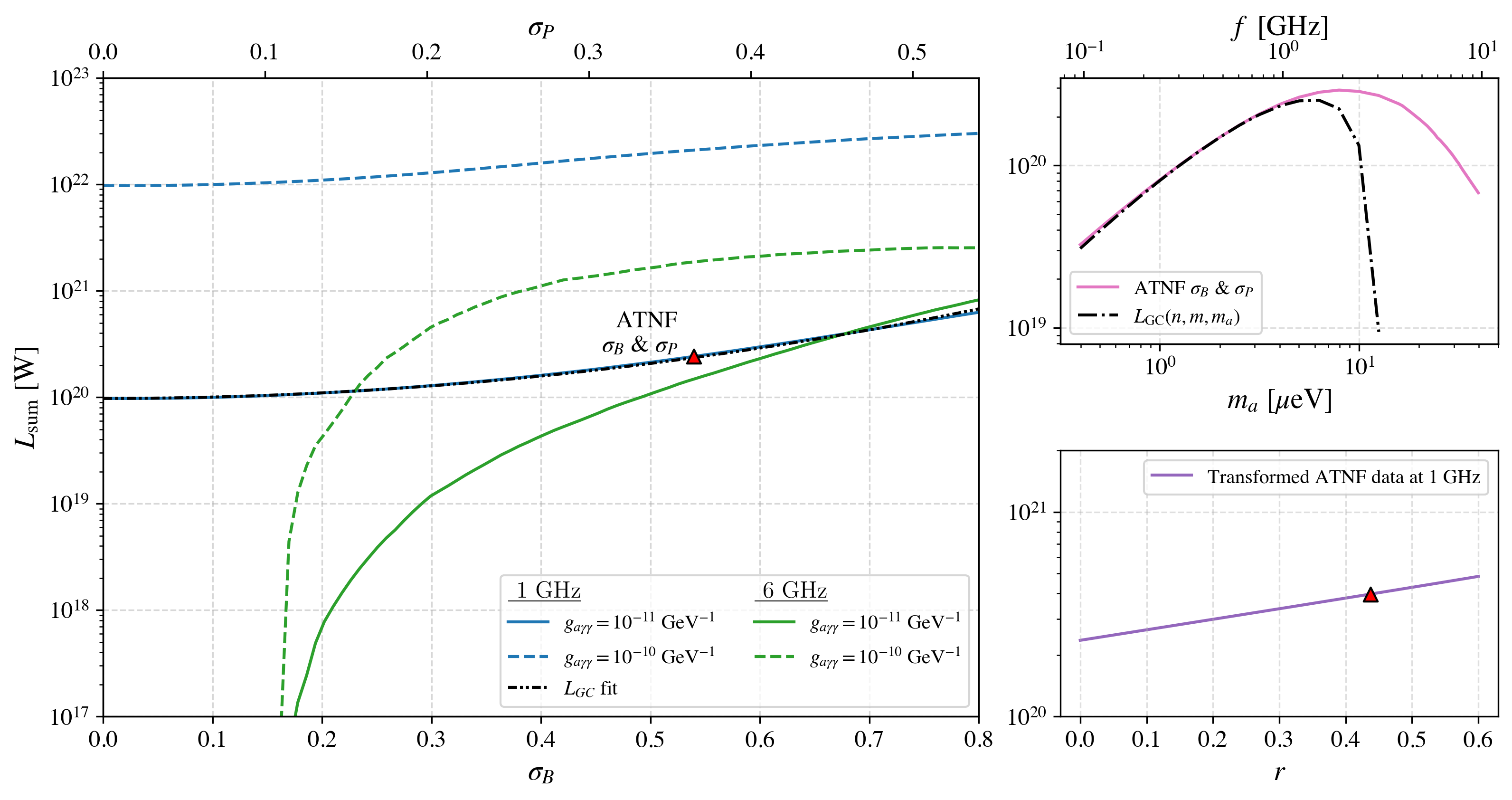}
     \caption{On the left is the total luminosity of the axion signal from the GC population calculated from realizations, in one case compared to \eqref{multilum}, as a function of the spread in the $B$ and $P$ distribution for $N_{\rm GC}=1000$. Taking $r_{\mathrm{corr}}=0$, we consider two frequencies $f_{\rm obs}=1\,{\rm GHz}$, which is the fiducial value we have used throughout the paper, and $6\,{\rm GHz}$ that is relevant to the analysis of the BL data~\cite{Foster:2022fxn}. We also consider two values $\gayy = 10^{-11}\,{\rm  GeV}^{-1}$ where the effects of multiple crossings are negligible and $\gayy=10^{-10}\,{\rm GeV}^{-1}$ where corrections for this effect need to be taken into account. The signal at $f_{\rm obs}=1\,{\rm GHz}$ and $\gayy=10^{-11}\,{\rm GeV}^{-1}$ is seen to follow the power law  expressed in (\ref{multilum}) with best-fit values $m=0.9$ and $n=0.88$. Meanwhile, for $f_{\rm obs}=6\,{\rm GHz}$ the effects of stars with low values of $B_0$ having no signal due to the critical surface dipping inside the star has an impact when $\sigma_B$ and $\sigma_P$ approach zero. When $\gayy=10^{-10}\,{\rm GeV}^{-1}$ there is less variation as function of $\sigma_B$ and $\sigma_P$ due to stars with large $B_0$ and $P$ being suppressed. The effect of varying the frequency is similar to that seen for $\gayy=10^{-11}\,{\rm GeV}^{-1}$. On the right, at the top, we present the axion signal for $N_{\rm GC}=1000$ and using the fiducial distributions for $B_0$ and $P$ as a function $m_{\rm a}$. We see that (\ref{multilum}) can be used to produce the expected signal for $m_{\rm a}\lesssim 5\,\mu{\rm eV}$. On the right at the bottom we demonstrate the effect of varying $r_{\mathrm{corr}}$, which leads to less than a factor of two variation over $0\le r_{\mathrm{corr}}<0.6$.}
     \label{fig:vary_sigbp}
 \end{figure}

We have tested this model by creating realizations for $N_{\rm GC}=1000$ using the fiducial parameters for the distributions of $B_0$ and $P$ but varying $\sigma_B$, $\sigma_P$ and $r_{\mathrm{corr}}$. The results are presented in fig.~\ref{fig:vary_sigbp}. 

First, let us consider the case where $\gayy=10^{-11}\,{\rm GeV}^{-1}$ and $f_{\rm obs}=1\,{\rm GHz}$. We see that with there is good agreement with (\ref{multilum}) when $n=1$ and $m=0.75$ which are close to the power law slopes in the luminosity that are seen fig.~\ref{fig:lvbpam}.  For the fiducial values of $\sigma_B$ and $\sigma_P$ the corrections due to the extended tails of the distribution in the log-normal distribution are $\sim 20\%$ and hence remarkably we see that, for values of $m_{\rm a}$ such that the power-law assumption is valid, the total luminosity is well approximated $N_{\rm GC}$ multiplied by that of the average luminosity for the distribution with an ${\cal O}(1)$ correction as suggested in (\ref{multilum}) and (\ref{multilum2}).

The picture is modified at higher frequencies and larger values of $\gayy$ for two reasons. At higher frequencies, for example, the case of $f_{\rm obs}=6\,{\rm GHz}$ shown in fig.~\ref{fig:vary_sigbp}, we see that as $\sigma_B$ and $\sigma_P$ go down toward zero the predicted luminosity rapidly falls off. The reason for this is that stars with the mean values of $B_0$ and $P$ at these frequencies do not emit due to axions because the critical surface is below the radius of the star. As a consequence of this, the only stars that emit are in the tails of the distribution which reduce significantly as $\sigma_B$ and $\sigma_P$ tend to zero, ultimately becoming zero very quickly for values $\sigma_B\sim 0.2$ and $\sigma_P\sim 0.1$. For larger values of $\gayy=10^{-10}\,{\rm GeV}^{-1}$, the effects of the $\braket{P_{a\gamma}}$ being large become important, and this leads to objects with large values of $B_0$, for example, which are more numerous when $\sigma_B$ is large, being excluded. Ultimately, this leads to reduced level of variation as a function $\sigma_B$ and $\sigma_P$ when compared to (\ref{multilum}). 

The variation of the signal with $m_{\rm a}$ is presented in the top right panel of fig.~\ref{fig:vary_sigbp} for the fiducial model and setting $r_{\mathrm{corr}}=0$. We see that the signal reflects the same shape seen in fig.~\ref{fig:lvbpam} and that this can be reproduced using (\ref{multilum}) using the values of $m$ and $n$ fixed at one value of $m_{\rm a}$ up to $m_{\rm a}\lesssim 5\,\mu{\rm eV}$ where additional effects kick in - see below.

We have also varied $r_{\mathrm{corr}}$ while keeping $\sigma_B$ and $\sigma_P$ fixed for $f_{\rm obs}=1\,{\rm GHz}$ and $N_{\rm GC}=1000$, and the results are presented in the bottom right panel of fig.~\ref{fig:vary_sigbp}. We see that in the range $0 \le r_{\mathrm{corr}} < 0.6$ there is a variation in the total luminosity of at most a factor of two.

Overall, we conclude that the basic picture of (\ref{multilum}) and $(\ref{multilum2})$ is correct for a wide range of parameters. However, there are important caveats at higher frequencies and larger values of $\gayy$. Ultimately, we will use full distribution for calculating the limits on $\gayy$, but it is useful to have an estimate of the size of signal we might expect from the GC population. Just taking the results of fig.~\ref{fig:vary_sigbp}, we estimate that for $\gayy<10^{-11}\,{\rm GeV}^{-1}$ and for $f_{\rm obs}<8\,{\rm GHz}$
\begin{equation}
    L_{\rm GC}\approx 3\times 10^{20}\,{\rm W}\left(\frac{\gayy}{10^{-11}\,{\rm GeV}^{-1}}\right)^2\left(\frac{N_{\rm GC}}{1000}\right)\,,
    \label{lum_gc} 
    \end{equation}
where there is an order of magnitude, $\approx 10^{\pm 0.5}$ uncertainty in the amplitude due to the imperfect knowledge of the parameters describing the distribution of $B_0$ and $P$. 

\subsection{Statistical errors from stellar orientations and ray-tracing}\label{sec:rayTracing}

\begin{figure}
    \centering
     \includegraphics[width=0.7\textwidth]{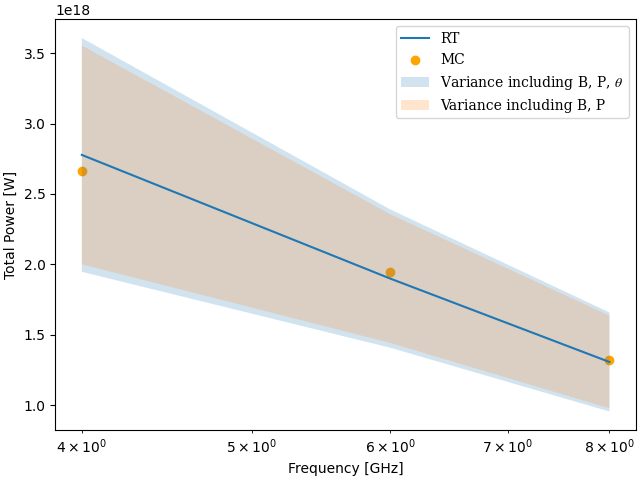}
    \caption{Statistical error in signal from variability in $B$,$P$ and $\theta$. We display the average luminosity and variance with and without $\theta$ variation. The contribution to the variance from $\theta$ is small in comparison to that from $(B, P)$ assuming that they have log-normal distributions as shown in fig.~\ref{fig:PsrPopPy_pulsar_properties}. It was noted in ref.~\cite{Battye:2023oac} that the luminosity signal may be modelled with $\mathcal{O}(1)$ uncertainty as a power law in $B$ and $P$. The above result confirms this hypothesis.  }
    \label{fig:SigmaComparison}
\end{figure}

 Stars whose spin axis is orientated in different directions relative to observers on earth will produce signals of different strengths. This orientation is characterised by the angle $\theta$ between the line of sight and the rotational axis, which a priori is assumed to be uniformly distributed with respect to $\cos \theta$. In deriving our constraints below, we use the power marginalised over $\theta$, $P$ and $B$, for which it is sufficient to work in terms of the integrated luminosity $L(B,P)$, where one has already marginalised over $\theta$. Here we discuss briefly the expected variance of the power from a randomly drawn star arising from variability in $\theta$. To do this, we must know the differential power $dP/d\Omega$, whose computation requires numerical ray-tracing \cite{leroy2020,Witte:2021arp,Battye:2021xvt,McDonald:2023shx,Tjemsland:2023vvc}. For a single star, we can define the luminosity by
\begin{align}
    L(B,P) &\equiv  \braket{d\mathcal{P}/d\Omega}_{\theta}
\end{align}
where $\braket{\cdots}_{\theta} =  \int d\Omega (\cdots)$. Using this, 
One can then define the \textit{full} variance $\sigma_{B,P,\theta}$, which includes variation of $B,P$ \textit{and} $\theta$ (which requires ray-tracing)
\begin{equation}
    \sigma_{B,P,\theta}^2 \equiv  \braket{ \, ( d\mathcal{P}/d\Omega - \braket{d\mathcal{P}/d\Omega} )^2\, }_{B,P,\theta}.
\end{equation}
Here, we define $\braket{\cdots}_{B,P,\theta} = \int dB dP \, n(B,P) \int d\Omega (\cdots)$. One can also define a partial variance $\sigma_{B,P}$ in which effectively marginalises over $\theta$ before taking the variance
\begin{equation}
    \sigma_{B,P}^2 = \braket{\,( L - \braket{L}_{B,P} )^2 \,}_{B,P} 
\end{equation}
In our constraint plot (fig.~\ref{fig:constraint}), we show error bars corresponding to $\sigma_{B,P}/\sqrt{N}$, where $N$ is the number of stars. However, it is important to compare this to the full error $\sigma_{B,P,\theta}$, which we do in fig.~\ref{fig:SigmaComparison}. We see therefore, that the variance is dominated by variability in $B,P$ rather than $\theta$, which has only a marginal effect on the sampling error. We therefore conclude that ray-tracing is not required to accurately characterise the sampling error in a population study, justifying our marginalisation over $\theta$ from the outset.

\section{Specifics of our usage of \tt{PsrPopPy}} 
\label{ref:Appendix C}

The package {\tt PsrPopPy} can be used in a number of different modes. The rate at which pulsars are added to the synthetic catalogue is such that the number observed by a particular survey is correct. For these purposes we use the PMBS survey which found 1206 pulsars.  In this section we will describe briefly our use of the {\tt evolve} option which we do largely with the default options. Since these may change in future releases we will outline them here so as to avoid possible confusion. The {\tt evolve} option creates a population of the observable pulsars in the Galaxy which are evolved from an initial state for the age of the Galaxy which we take to be $t_{\rm gal}=10^{9}\,{\rm years}$. We note that the properties of the population do not evolve much beyond this point.

\subsection{Evolution of the values of \texorpdfstring{$B_0$}{Bo}, \texorpdfstring{$P$}{P} and \texorpdfstring{$\alpha$}{α}}
\label{BPalpha_distribution}

\begin{figure}
    \centering
    \includegraphics[width = \textwidth]{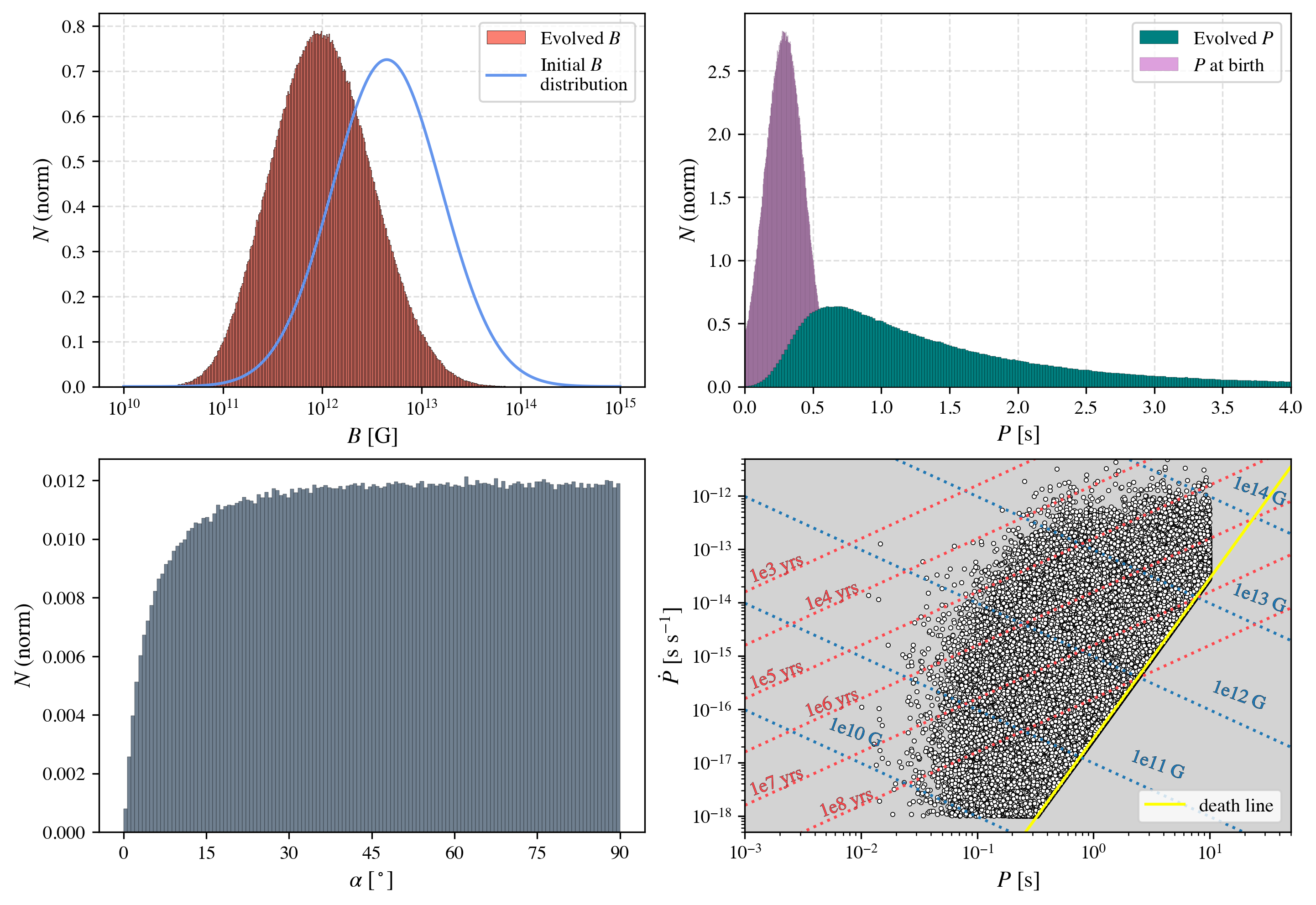}
    \caption{Output from a single realization of {\tt PsrPopPy} using the modelling and parameters discussed in the text and $t_{\rm gal}=10^{9}\,{\rm yr}$. The top-left plot is a histogram of the values for $B_0$ - included also is a curve for the initial log-normal distribution. The key feature to notice is the decline of the magnetic field with time. In the top-right we show the equivalent for $P$. Here, we see evolution of the initially normal distribution to what is more akin to a log-normal distribution. In the bottom left, is the final distribution of $\alpha$ which originally started as a uniform distribution in $\cos\alpha$. In the bottom right is the $P-{\dot P}$ diagram for this realization. } \label{fig:PsrPopPy_pulsar_properties}
\end{figure}

The initial distributions were chosen as follows~\cite{Bates:2013uma,Faucher-Giguere:2005dxp}.
\begin{itemize}
\item Magnetic field, $B_0$: a log normal distribution with mean $\mu_{\log B}=12.95$ and standard deviation $\sigma_{\log B}=0.55$.
\item Period, $P$: a normal distribution with mean $\mu_P=0.3$ and $\sigma_B=0.15$. Such a distribution allows for negative values which are removed and re-sampled, so  the distribution is forced to have $P>0$ and is not exactly Gaussian.
\item Misalignment angle, $\alpha$: uniform distribution in $\cos\alpha$.
\end{itemize}

\begin{table}
    \centering
    \begin{tabular}{|c||c|c|c|c|}
    \hline
      $k$   & 4.25 & 4.25 & 4.89 & 4.89 \\ \hline
       $r_0$  & 3.48 & 3.48 & 4.9 & 4.9 \\ \hline
       $\theta_0$  & 1.57 & 4.71 & 4.09 & 0.95 \\ \hline
    \end{tabular}
    \caption{Data sourced from ref.~\cite{spirals} defining the spiral arms of the Milky Way in conjunction with (\ref{spirals_eq}).}
    \label{tab:spiral_parameter_values}
\end{table}

These properties of the stars are evolved using the ``Magnetic Dipole Model" explained in ref.~\cite{ridley_isolated_2010}. According to the model, the pulsar pulse period is given by
\begin{equation}
    \label{P_dipole_model}
    P(t) = \left[P_0^{n-1} + \left(\frac{n-1}{2}\right) \, t_{\rm d} \,k \, B^2 \sin^2 \alpha_0 (1 - e^{-2t/t_{\rm d}})  \right]^{1/(n-1)}\,,
\end{equation}
where $P_0$ and $\alpha_0$ are the period and misalignment angle assigned at birth, respectively. $t_{\rm d} \sim 7\times 10^{7}\,{\rm yr}$ is the decay time-scale, and $n$ is the breaking index. The model considers the decay in the misalignment angle with time, governed by $\alpha = \alpha_0 \, e^{-t / t_{\rm d}}$. The constant $k$ and the breaking index $n$ originate from the pulsar "spin-down" equation, given by $\dot \nu = - K \nu^n$ which implies $P^{n-2} \dot P = K$
where $\nu$ is the rotation frequency of a pulsar. In an ideal scenario, it can be assumed that the energy loss in a rotating pulsar is due to pulse emission, resulting in a decrease in the rotational kinetic energy and an increase in the pulse period. Solving this yields the ideal "spin-down" equation as,
\begin{align}
    \label{ideal_spin_down}
    \dot \nu = - \left( \frac{8 \pi^2 B^2 R^6 \sin^2 \alpha}{3 I c^3} \right) \nu^3 \,. 
\end{align}
Hence, we can deduce that $n=3$ and $K = {8 \pi^2 B^2 R^6 \sin^2 \alpha / }{3 I c^3}$. Now, as $B = B(t)$ and $\alpha = \alpha (t)$, $K$ is modified to $K = k B^2 \sin^2 \alpha$. Assuming $R=10\,{\rm km}$ and the moment of inertia $I=10^{38}\,{\rm kg}\,{\rm m}^2$, $k$ comes out to be $9.8 \times 10^{-40}$. This case of pulsar "spin-down" is explained in detail in ref.~\cite{lorimer2012handbook}.

In reality it has been observed that only a tiny fraction of radio emission results in the energy loss of a pulsar so that  $n < 3$, but the limit of this equation occurs at the ``death line"~\cite{death_line} of a pulsar, after which the radio emission of the star ceases. This happens when
\begin{equation}\label{death_line}
    \frac{B_0}{P^2} = 1.7 \times 10^{11} \ \rm G \ \rm s^{-2}\,.
\end{equation}

\subsection{Modelling the pulsar positions in the Galaxy via the Galactic potential model}

{\tt PsrPopPy} adopts a realistic approach to simulate the spatial distribution of NSs. It dynamically evolves the positions of pulsars from initial random locations within the spirals, guided by a detailed Galactic potential model. This process yields the final positions and velocities for each pulsar.

This understanding is grounded in the fact that the birth regions of young and detectable NSs are closely tied to the spiral arm structure of the Milky Way. This association arises from the prevalence of supernova explosions in these regions that give birth to young NSs~\cite{Faucher-Giguere:2005dxp}. The representation of these spiral arm structures is depicted using
\begin{equation} \label{spirals_eq}
    \theta (r) = k \ln \left(\frac{r}{r_0} \right) + \theta_0,
\end{equation}
where, $r$ and $\theta$ respectively are the radial distance and the azimuth of a star from the GC in cylindrical coordinates. The values of $k, r_0$ and $\theta_0$ have been taken from ref.~\cite{spirals} and are presented in table~\ref{tab:spiral_parameter_values}.

The Galactic gravitational potential model consists three components: the disk-halo (dh), the bulge (b) and the nucleus (n)
\begin{equation}
\label{GGP-model}
    \Phi_{\rm G} (r,z) = \Phi_{\rm dh} (r,z)+\Phi_{\rm b}(r)+\Phi_{\rm n}(r),
\end{equation}
where 
\begin{eqnarray}
    \Phi_{\rm dh} (r,z) &=& \frac{-GM_{\rm dh}}{\sqrt{r^2 + b_{\rm dh}^2 + \left(a_{\rm G} + \sum^3_{i=1} \beta_i \sqrt{z^2 + h_i^2}\right)^2}}\,,\cr
    \Phi_{\rm b} (r) &=& \frac{-GM_{\rm b}}{\sqrt{r^2 + b_{\rm b}^2}}\,,\cr
    \Phi_{\rm n} (r) &=& \frac{-GM_{\rm n}}{\sqrt{r^2 + b_{\rm n}^2}}\,.
\end{eqnarray}
The constants take the values mentioned in Table \ref{tab:GP-model_constants}. 

\begin{table}
    \centering
    \begin{tabular}{|c|c|c|c|c|c|c|c|c|c|}
    \hline
     & $M_j/M_{\odot}$ & $\beta_1$ & $\beta_2$ & $\beta_3$ & $h_1$ & $h_2$ & $h_3$ & $a_{\rm G}$ & $b_j$\\ \hline 
    disk halo $(j = \mathrm{dh})$ & $1.45 \times 10^{11}$ & 0.4 & 0.5 & 0.1 & 0.325 & 0.09 & 0.125 & 2.4 & 5.5\\ \hline 
    bulge $(j = \mathrm{b})$ & $9.3 \times 10^{9}$ & & & & & & & & 0.25 \\ \hline
    nucleus $(j = \mathrm{n})$ & $1 \times 10^{10}$ & & & & & & & & 1.5 \\ \hline
    \end{tabular}
    \caption{Value of the constants in the equations of Galactic potential model~\cite{Faucher-Giguere:2005dxp}. The values for $h_1$, $h_2$, $h_3$, $a_{\rm G}$ and $b$ are given in ${\rm kpc}$.}
    \label{tab:GP-model_constants}
\end{table}

\begin{figure}
    \centering
    \includegraphics[width = \textwidth]{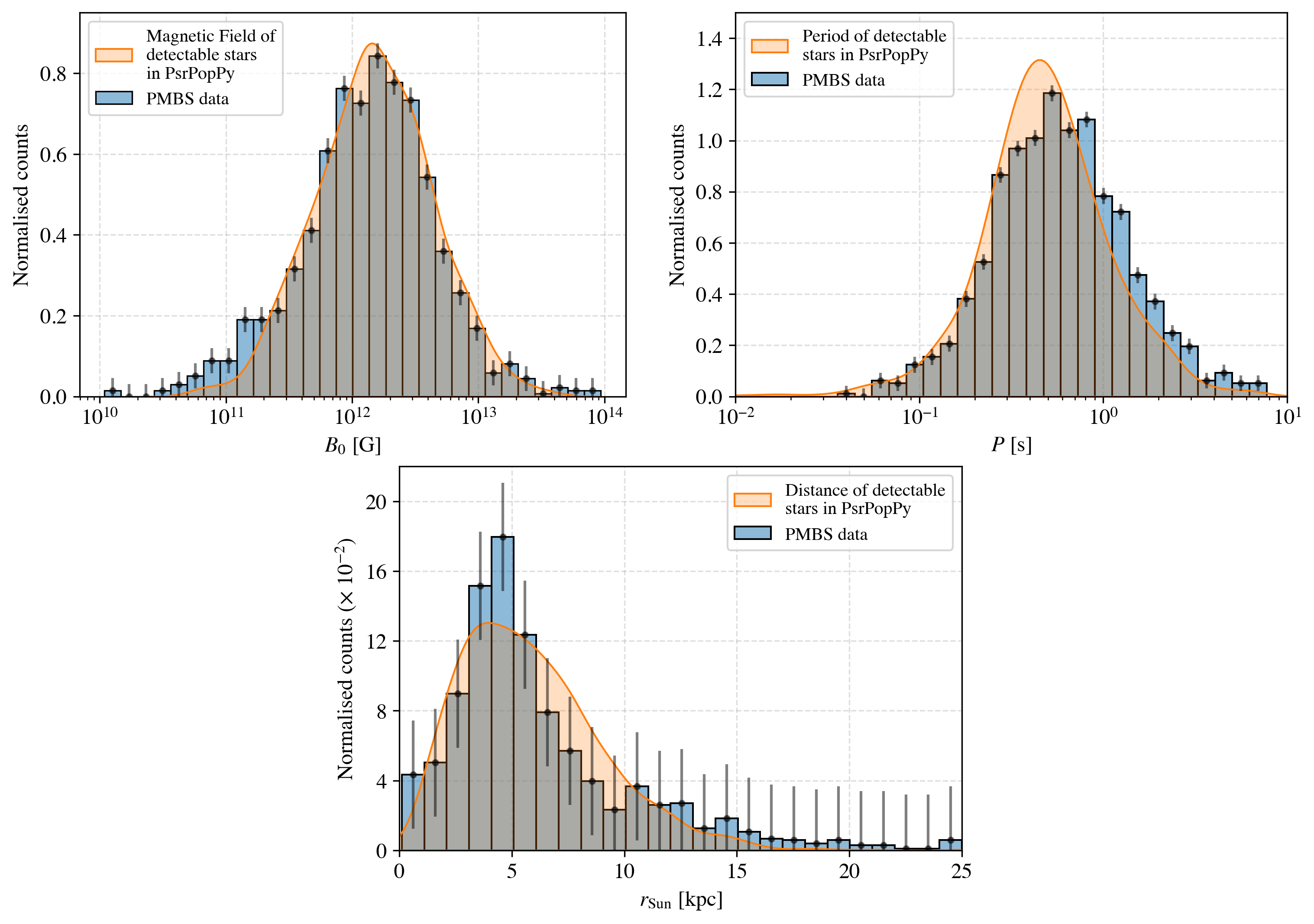}
    \caption{A comparison between the magnetic field, $B_0$, pulse period, $P$, and the distance of the ``detectable'' pulsars, $r_{\mathrm{Sun}}$ (see fig.~\ref{fig:cpdiag}) generated by \texttt{PsrPopPy} with the pulsars detected in the PMBS survey. There is good agreement between the two in each case.}
    \label{fig:compbpdist}
\end{figure}

\subsection{Pulsar luminosity}
\texttt{PsrPopPy} employs the luminosity model which is a power law in $P$ and $\dot P$ explained in ~\cite{Faucher-Giguere:2005dxp}, 
\begin{align}
    \label{pulse_luminosity}
    \log_{10}\left(\frac{L}{1 \  \mathrm{mJy\, kpc}^2}\right) = \log_{10}\left[\frac{L_0}{1 \  \mathrm{mJy \, kpc}^2} \ \left(\frac{P}{1 \ \mathrm{s}}\right)^a \ \left(\frac{\dot P}{10^{-15}} \right) ^b\right] + L_{\rm corr}.
\end{align}
Here, $L_{\mathrm{corr}}$ accounts for the distance measurement error of the pulsar, and its value is drawn from a normal distribution centered on zero. The default settings for these variables in \texttt{PsrPopPy} are configured with $a = -1.5$, $b = 0.5$, and $L_0 = 0.18$ mJy kpc$^2$, specifically for radio pulsars at 1.4 GHz. 

The power law model presented in (\ref{pulse_luminosity}) is exclusively applied within the \texttt{evolve} function in \texttt{PsrPopPy}. This is due to the fact that the computation of $\dot P$ is restricted to the \texttt{evolve} function. For other functions, such as \texttt{populate}, where $\dot P$ is not computed, \texttt{PsrPopPy} utilizes a log-normal distribution centered at -1.1 with a spread of 0.9 to determine the pulse luminosity, as explained in ref.~\cite{Bates:2013uma}. 

\subsection{Comparison with PMBS}
A crucial check to confirm the reliability of data generated by \texttt{PsrPopPy} is to compare it with actual observations. The PMBS, conducted by the ATNF~\cite{Manchester2001_PMBS}, serves as an ideal reference dataset for this purpose. In order to compare these datasets, we plot and compare the PDFs of the magnetic fields, periods and distances of the pulsars detected in the PMBS with those categorized as ``detectable'' in \texttt{PsrPopPy}. To ensure the validity of the comparison, we apply cuts to the ATNF data. In particular, pulsars that belong to binary systems or have unknown values for $B$ and $P$ are excluded from the analysis. Additionally, pulsars with magnetic fields less than $10^{10}\,{\rm G}$ (which most likely tend to be millisecond pulsars) are removed to maintain consistency.

\bibliography{References.bib}

\providecommand{\noopsort}[1]{}\providecommand{\singleletter}[1]{#1}%
\begin{thebibliography}{104}%
\makeatletter
\providecommand \@ifxundefined [1]{%
 \@ifx{#1\undefined}
}%
\providecommand \@ifnum [1]{%
 \ifnum #1\expandafter \@firstoftwo
 \else \expandafter \@secondoftwo
 \fi
}%
\providecommand \@ifx [1]{%
 \ifx #1\expandafter \@firstoftwo
 \else \expandafter \@secondoftwo
 \fi
}%
\providecommand \natexlab [1]{#1}%
\providecommand \enquote  [1]{``#1''}%
\providecommand \bibnamefont  [1]{#1}%
\providecommand \bibfnamefont [1]{#1}%
\providecommand \citenamefont [1]{#1}%
\providecommand \href@noop [0]{\@secondoftwo}%
\providecommand \href [0]{\begingroup \@sanitize@url \@href}%
\providecommand \@href[1]{\@@startlink{#1}\@@href}%
\providecommand \@@href[1]{\endgroup#1\@@endlink}%
\providecommand \@sanitize@url [0]{\catcode `\\12\catcode `\$12\catcode `\&12\catcode `\#12\catcode `\^12\catcode `\_12\catcode `\%12\relax}%
\providecommand \@@startlink[1]{}%
\providecommand \@@endlink[0]{}%
\providecommand \url  [0]{\begingroup\@sanitize@url \@url }%
\providecommand \@url [1]{\endgroup\@href {#1}{\urlprefix }}%
\providecommand \urlprefix  [0]{URL }%
\providecommand \Eprint [0]{\href }%
\providecommand \doibase [0]{http://dx.doi.org/}%
\providecommand \selectlanguage [0]{\@gobble}%
\providecommand \bibinfo  [0]{\@secondoftwo}%
\providecommand \bibfield  [0]{\@secondoftwo}%
\providecommand \translation [1]{[#1]}%
\providecommand \BibitemOpen [0]{}%
\providecommand \bibitemStop [0]{}%
\providecommand \bibitemNoStop [0]{.\EOS\space}%
\providecommand \EOS [0]{\spacefactor3000\relax}%
\providecommand \BibitemShut  [1]{\csname bibitem#1\endcsname}%
\let\auto@bib@innerbib\@empty
\bibitem [{\citenamefont {{Peccei}}\ and\ \citenamefont {{Quinn}}(1977)}]{ref:PQ}%
  \BibitemOpen
  \bibfield  {author} {\bibinfo {author} {\bibfnamefont {R.~D.}\ \bibnamefont {{Peccei}}}\ and\ \bibinfo {author} {\bibfnamefont {H.~R.}\ \bibnamefont {{Quinn}}},\ }\href {\doibase 10.1103/PhysRevD.16.1791} {\bibfield  {journal} {\bibinfo  {journal} {\prd}\ }\textbf {\bibinfo {volume} {16}},\ \bibinfo {pages} {1791} (\bibinfo {year} {1977})}\BibitemShut {NoStop}%
\bibitem [{\citenamefont {{Kim}}(1979)}]{ref:K}%
  \BibitemOpen
  \bibfield  {author} {\bibinfo {author} {\bibfnamefont {J.~E.}\ \bibnamefont {{Kim}}},\ }\href {\doibase 10.1103/PhysRevLett.43.103} {\bibfield  {journal} {\bibinfo  {journal} {Physical Review Letters}\ }\textbf {\bibinfo {volume} {43}},\ \bibinfo {pages} {103} (\bibinfo {year} {1979})}\BibitemShut {NoStop}%
\bibitem [{\citenamefont {{Shifman}}\ \emph {et~al.}(1980)\citenamefont {{Shifman}}, \citenamefont {{Vainshtein}},\ and\ \citenamefont {{Zakharov}}}]{ref:SVZ}%
  \BibitemOpen
  \bibfield  {author} {\bibinfo {author} {\bibfnamefont {M.~A.}\ \bibnamefont {{Shifman}}}, \bibinfo {author} {\bibfnamefont {A.~I.}\ \bibnamefont {{Vainshtein}}}, \ and\ \bibinfo {author} {\bibfnamefont {V.~I.}\ \bibnamefont {{Zakharov}}},\ }\href {\doibase 10.1016/0550-3213(80)90209-6} {\bibfield  {journal} {\bibinfo  {journal} {Nuclear Physics B}\ }\textbf {\bibinfo {volume} {166}},\ \bibinfo {pages} {493} (\bibinfo {year} {1980})}\BibitemShut {NoStop}%
\bibitem [{\citenamefont {{Dine}}\ \emph {et~al.}(1981)\citenamefont {{Dine}}, \citenamefont {{Fischler}},\ and\ \citenamefont {{Srednicki}}}]{ref:DFSZ}%
  \BibitemOpen
  \bibfield  {author} {\bibinfo {author} {\bibfnamefont {M.}~\bibnamefont {{Dine}}}, \bibinfo {author} {\bibfnamefont {W.}~\bibnamefont {{Fischler}}}, \ and\ \bibinfo {author} {\bibfnamefont {M.}~\bibnamefont {{Srednicki}}},\ }\href {\doibase 10.1016/0370-2693(81)90590-6} {\bibfield  {journal} {\bibinfo  {journal} {Physics Letters B}\ }\textbf {\bibinfo {volume} {104}},\ \bibinfo {pages} {199} (\bibinfo {year} {1981})}\BibitemShut {NoStop}%
\bibitem [{\citenamefont {Zhitnitsky}(1980)}]{ref:Zhit}%
  \BibitemOpen
  \bibfield  {author} {\bibinfo {author} {\bibfnamefont {A.~R.}\ \bibnamefont {Zhitnitsky}},\ }\href@noop {} {\bibfield  {journal} {\bibinfo  {journal} {Sov. J. Nucl. Phys.}\ }\textbf {\bibinfo {volume} {31}},\ \bibinfo {pages} {260} (\bibinfo {year} {1980})},\ \bibinfo {note} {[Yad. Fiz.31,497(1980)]}\BibitemShut {NoStop}%
\bibitem [{\citenamefont {Dine}\ and\ \citenamefont {Fischler}(1983)}]{ref:misalign1}%
  \BibitemOpen
  \bibfield  {author} {\bibinfo {author} {\bibfnamefont {M.}~\bibnamefont {Dine}}\ and\ \bibinfo {author} {\bibfnamefont {W.}~\bibnamefont {Fischler}},\ }\href {\doibase https://doi.org/10.1016/0370-2693(83)90639-1} {\bibfield  {journal} {\bibinfo  {journal} {Physics Letters B}\ }\textbf {\bibinfo {volume} {120}},\ \bibinfo {pages} {137 } (\bibinfo {year} {1983})}\BibitemShut {NoStop}%
\bibitem [{\citenamefont {Abbott}\ and\ \citenamefont {Sikivie}(1983)}]{ref:misalign2}%
  \BibitemOpen
  \bibfield  {author} {\bibinfo {author} {\bibfnamefont {L.}~\bibnamefont {Abbott}}\ and\ \bibinfo {author} {\bibfnamefont {P.}~\bibnamefont {Sikivie}},\ }\href {\doibase https://doi.org/10.1016/0370-2693(83)90638-X} {\bibfield  {journal} {\bibinfo  {journal} {Physics Letters B}\ }\textbf {\bibinfo {volume} {120}},\ \bibinfo {pages} {133 } (\bibinfo {year} {1983})}\BibitemShut {NoStop}%
\bibitem [{\citenamefont {Preskill}\ \emph {et~al.}(1983)\citenamefont {Preskill}, \citenamefont {Wise},\ and\ \citenamefont {Wilczek}}]{ref:misalign3}%
  \BibitemOpen
  \bibfield  {author} {\bibinfo {author} {\bibfnamefont {J.}~\bibnamefont {Preskill}}, \bibinfo {author} {\bibfnamefont {M.~B.}\ \bibnamefont {Wise}}, \ and\ \bibinfo {author} {\bibfnamefont {F.}~\bibnamefont {Wilczek}},\ }\href {\doibase https://doi.org/10.1016/0370-2693(83)90637-8} {\bibfield  {journal} {\bibinfo  {journal} {Physics Letters B}\ }\textbf {\bibinfo {volume} {120}},\ \bibinfo {pages} {127 } (\bibinfo {year} {1983})}\BibitemShut {NoStop}%
\bibitem [{\citenamefont {Arvanitaki}\ \emph {et~al.}(2010)\citenamefont {Arvanitaki}, \citenamefont {Dimopoulos}, \citenamefont {Dubovsky}, \citenamefont {Kaloper},\ and\ \citenamefont {March-Russell}}]{Arvanitaki:2009fg}%
  \BibitemOpen
  \bibfield  {author} {\bibinfo {author} {\bibfnamefont {A.}~\bibnamefont {Arvanitaki}}, \bibinfo {author} {\bibfnamefont {S.}~\bibnamefont {Dimopoulos}}, \bibinfo {author} {\bibfnamefont {S.}~\bibnamefont {Dubovsky}}, \bibinfo {author} {\bibfnamefont {N.}~\bibnamefont {Kaloper}}, \ and\ \bibinfo {author} {\bibfnamefont {J.}~\bibnamefont {March-Russell}},\ }\href {\doibase 10.1103/PhysRevD.81.123530} {\bibfield  {journal} {\bibinfo  {journal} {Phys. Rev. D}\ }\textbf {\bibinfo {volume} {81}},\ \bibinfo {pages} {123530} (\bibinfo {year} {2010})},\ \Eprint {http://arxiv.org/abs/0905.4720} {arXiv:0905.4720 [hep-th]} \BibitemShut {NoStop}%
\bibitem [{\citenamefont {Svrcek}\ and\ \citenamefont {Witten}(2006)}]{Svrcek:2006yi}%
  \BibitemOpen
  \bibfield  {author} {\bibinfo {author} {\bibfnamefont {P.}~\bibnamefont {Svrcek}}\ and\ \bibinfo {author} {\bibfnamefont {E.}~\bibnamefont {Witten}},\ }\href {\doibase 10.1088/1126-6708/2006/06/051} {\bibfield  {journal} {\bibinfo  {journal} {JHEP}\ }\textbf {\bibinfo {volume} {06}},\ \bibinfo {pages} {051} (\bibinfo {year} {2006})},\ \Eprint {http://arxiv.org/abs/hep-th/0605206} {arXiv:hep-th/0605206} \BibitemShut {NoStop}%
\bibitem [{\citenamefont {DePanfilis}\ \emph {et~al.}(1987)\citenamefont {DePanfilis}, \citenamefont {Melissinos}, \citenamefont {Moskowitz}, \citenamefont {Rogers}, \citenamefont {Semertzidis}, \citenamefont {Wuensch}, \citenamefont {Halama}, \citenamefont {Prodell}, \citenamefont {Fowler},\ and\ \citenamefont {Nezrick}}]{RBF}%
  \BibitemOpen
  \bibfield  {author} {\bibinfo {author} {\bibfnamefont {S.}~\bibnamefont {DePanfilis}}, \bibinfo {author} {\bibfnamefont {A.~C.}\ \bibnamefont {Melissinos}}, \bibinfo {author} {\bibfnamefont {B.~E.}\ \bibnamefont {Moskowitz}}, \bibinfo {author} {\bibfnamefont {J.~T.}\ \bibnamefont {Rogers}}, \bibinfo {author} {\bibfnamefont {Y.~K.}\ \bibnamefont {Semertzidis}}, \bibinfo {author} {\bibfnamefont {W.~U.}\ \bibnamefont {Wuensch}}, \bibinfo {author} {\bibfnamefont {H.~J.}\ \bibnamefont {Halama}}, \bibinfo {author} {\bibfnamefont {A.~G.}\ \bibnamefont {Prodell}}, \bibinfo {author} {\bibfnamefont {W.~B.}\ \bibnamefont {Fowler}}, \ and\ \bibinfo {author} {\bibfnamefont {F.~A.}\ \bibnamefont {Nezrick}},\ }\href {\doibase 10.1103/PhysRevLett.59.839} {\bibfield  {journal} {\bibinfo  {journal} {Phys. Rev. Lett.}\ }\textbf {\bibinfo {volume} {59}},\ \bibinfo {pages} {839} (\bibinfo {year} {1987})}\BibitemShut {NoStop}%
\bibitem [{\citenamefont {Wuensch}\ \emph {et~al.}(1989)\citenamefont {Wuensch}, \citenamefont {De~Panfilis-Wuensch}, \citenamefont {Semertzidis}, \citenamefont {Rogers}, \citenamefont {Melissinos}, \citenamefont {Halama}, \citenamefont {Moskowitz}, \citenamefont {Prodell}, \citenamefont {Fowler},\ and\ \citenamefont {Nezrick}}]{Wuensch:1989sa}%
  \BibitemOpen
  \bibfield  {author} {\bibinfo {author} {\bibfnamefont {W.}~\bibnamefont {Wuensch}}, \bibinfo {author} {\bibfnamefont {S.}~\bibnamefont {De~Panfilis-Wuensch}}, \bibinfo {author} {\bibfnamefont {Y.~K.}\ \bibnamefont {Semertzidis}}, \bibinfo {author} {\bibfnamefont {J.~T.}\ \bibnamefont {Rogers}}, \bibinfo {author} {\bibfnamefont {A.~C.}\ \bibnamefont {Melissinos}}, \bibinfo {author} {\bibfnamefont {H.~J.}\ \bibnamefont {Halama}}, \bibinfo {author} {\bibfnamefont {B.~E.}\ \bibnamefont {Moskowitz}}, \bibinfo {author} {\bibfnamefont {A.~G.}\ \bibnamefont {Prodell}}, \bibinfo {author} {\bibfnamefont {W.~B.}\ \bibnamefont {Fowler}}, \ and\ \bibinfo {author} {\bibfnamefont {F.~A.}\ \bibnamefont {Nezrick}},\ }\href {\doibase 10.1103/PhysRevD.40.3153} {\bibfield  {journal} {\bibinfo  {journal} {Phys. Rev. D}\ }\textbf {\bibinfo {volume} {40}},\ \bibinfo {pages} {3153} (\bibinfo {year} {1989})}\BibitemShut {NoStop}%
\bibitem [{\citenamefont {Hagmann}\ \emph {et~al.}(1990)\citenamefont {Hagmann}, \citenamefont {Sikivie}, \citenamefont {Sullivan},\ and\ \citenamefont {Tanner}}]{UF}%
  \BibitemOpen
  \bibfield  {author} {\bibinfo {author} {\bibfnamefont {C.}~\bibnamefont {Hagmann}}, \bibinfo {author} {\bibfnamefont {P.}~\bibnamefont {Sikivie}}, \bibinfo {author} {\bibfnamefont {N.~S.}\ \bibnamefont {Sullivan}}, \ and\ \bibinfo {author} {\bibfnamefont {D.~B.}\ \bibnamefont {Tanner}},\ }\href {\doibase 10.1103/PhysRevD.42.1297} {\bibfield  {journal} {\bibinfo  {journal} {Phys. Rev. D}\ }\textbf {\bibinfo {volume} {42}},\ \bibinfo {pages} {1297} (\bibinfo {year} {1990})}\BibitemShut {NoStop}%
\bibitem [{\citenamefont {Asztalos}\ \emph {et~al.}(2010)\citenamefont {Asztalos} \emph {et~al.}}]{ADMX:2009iij}%
  \BibitemOpen
  \bibfield  {author} {\bibinfo {author} {\bibfnamefont {S.~J.}\ \bibnamefont {Asztalos}} \emph {et~al.} (\bibinfo {collaboration} {ADMX}),\ }\href {\doibase 10.1103/PhysRevLett.104.041301} {\bibfield  {journal} {\bibinfo  {journal} {Phys. Rev. Lett.}\ }\textbf {\bibinfo {volume} {104}},\ \bibinfo {pages} {041301} (\bibinfo {year} {2010})},\ \Eprint {http://arxiv.org/abs/0910.5914} {arXiv:0910.5914 [astro-ph.CO]} \BibitemShut {NoStop}%
\bibitem [{\citenamefont {Du}\ \emph {et~al.}(2018)\citenamefont {Du} \emph {et~al.}}]{ref:ADMX2018}%
  \BibitemOpen
  \bibfield  {author} {\bibinfo {author} {\bibfnamefont {N.}~\bibnamefont {Du}} \emph {et~al.} (\bibinfo {collaboration} {ADMX Collaboration}),\ }\href {\doibase 10.1103/PhysRevLett.120.151301} {\bibfield  {journal} {\bibinfo  {journal} {Phys. Rev. Lett.}\ }\textbf {\bibinfo {volume} {120}},\ \bibinfo {pages} {151301} (\bibinfo {year} {2018})}\BibitemShut {NoStop}%
\bibitem [{\citenamefont {Boutan}\ \emph {et~al.}(2018)\citenamefont {Boutan} \emph {et~al.}}]{ADMX:2018ogs}%
  \BibitemOpen
  \bibfield  {author} {\bibinfo {author} {\bibfnamefont {C.}~\bibnamefont {Boutan}} \emph {et~al.} (\bibinfo {collaboration} {ADMX}),\ }\href {\doibase 10.1103/PhysRevLett.121.261302} {\bibfield  {journal} {\bibinfo  {journal} {Phys. Rev. Lett.}\ }\textbf {\bibinfo {volume} {121}},\ \bibinfo {pages} {261302} (\bibinfo {year} {2018})},\ \Eprint {http://arxiv.org/abs/1901.00920} {arXiv:1901.00920 [hep-ex]} \BibitemShut {NoStop}%
\bibitem [{\citenamefont {Braine}\ \emph {et~al.}(2020)\citenamefont {Braine} \emph {et~al.}}]{ADMX:2019uok}%
  \BibitemOpen
  \bibfield  {author} {\bibinfo {author} {\bibfnamefont {T.}~\bibnamefont {Braine}} \emph {et~al.} (\bibinfo {collaboration} {ADMX}),\ }\href {\doibase 10.1103/PhysRevLett.124.101303} {\bibfield  {journal} {\bibinfo  {journal} {Phys. Rev. Lett.}\ }\textbf {\bibinfo {volume} {124}},\ \bibinfo {pages} {101303} (\bibinfo {year} {2020})},\ \Eprint {http://arxiv.org/abs/1910.08638} {arXiv:1910.08638 [hep-ex]} \BibitemShut {NoStop}%
\bibitem [{\citenamefont {Bartram}\ \emph {et~al.}(2021)\citenamefont {Bartram} \emph {et~al.}}]{ADMX:2021nhd}%
  \BibitemOpen
  \bibfield  {author} {\bibinfo {author} {\bibfnamefont {C.}~\bibnamefont {Bartram}} \emph {et~al.} (\bibinfo {collaboration} {ADMX}),\ }\href {\doibase 10.1103/PhysRevLett.127.261803} {\bibfield  {journal} {\bibinfo  {journal} {Phys. Rev. Lett.}\ }\textbf {\bibinfo {volume} {127}},\ \bibinfo {pages} {261803} (\bibinfo {year} {2021})},\ \Eprint {http://arxiv.org/abs/2110.06096} {arXiv:2110.06096 [hep-ex]} \BibitemShut {NoStop}%
\bibitem [{\citenamefont {Bartram}\ \emph {et~al.}(2023)\citenamefont {Bartram}, \citenamefont {Braine}, \citenamefont {Cervantes} \emph {et~al.}}]{ADMX:2021mio}%
  \BibitemOpen
  \bibfield  {author} {\bibinfo {author} {\bibfnamefont {C.}~\bibnamefont {Bartram}}, \bibinfo {author} {\bibfnamefont {T.}~\bibnamefont {Braine}}, \bibinfo {author} {\bibnamefont {Cervantes}},  \emph {et~al.},\ }\href {\doibase 10.1063/5.0122907} {\bibfield  {journal} {\bibinfo  {journal} {Review of Scientific Instruments}\ }\textbf {\bibinfo {volume} {94}} (\bibinfo {year} {2023}),\ 10.1063/5.0122907}\BibitemShut {NoStop}%
\bibitem [{\citenamefont {Crisosto}\ \emph {et~al.}(2020)\citenamefont {Crisosto}, \citenamefont {Sikivie}, \citenamefont {Sullivan}, \citenamefont {Tanner}, \citenamefont {Yang},\ and\ \citenamefont {Rybka}}]{Crisosto:2019fcj}%
  \BibitemOpen
  \bibfield  {author} {\bibinfo {author} {\bibfnamefont {N.}~\bibnamefont {Crisosto}}, \bibinfo {author} {\bibfnamefont {P.}~\bibnamefont {Sikivie}}, \bibinfo {author} {\bibfnamefont {N.~S.}\ \bibnamefont {Sullivan}}, \bibinfo {author} {\bibfnamefont {D.~B.}\ \bibnamefont {Tanner}}, \bibinfo {author} {\bibfnamefont {J.}~\bibnamefont {Yang}}, \ and\ \bibinfo {author} {\bibfnamefont {G.}~\bibnamefont {Rybka}},\ }\href {\doibase 10.1103/PhysRevLett.124.241101} {\bibfield  {journal} {\bibinfo  {journal} {Phys. Rev. Lett.}\ }\textbf {\bibinfo {volume} {124}},\ \bibinfo {pages} {241101} (\bibinfo {year} {2020})},\ \Eprint {http://arxiv.org/abs/1911.05772} {arXiv:1911.05772 [astro-ph.CO]} \BibitemShut {NoStop}%
\bibitem [{\citenamefont {Lee}\ \emph {et~al.}(2020)\citenamefont {Lee}, \citenamefont {Ahn}, \citenamefont {Choi}, \citenamefont {Ko},\ and\ \citenamefont {Semertzidis}}]{CAPP:Lee_2020cfj}%
  \BibitemOpen
  \bibfield  {author} {\bibinfo {author} {\bibfnamefont {S.}~\bibnamefont {Lee}}, \bibinfo {author} {\bibfnamefont {S.}~\bibnamefont {Ahn}}, \bibinfo {author} {\bibfnamefont {J.}~\bibnamefont {Choi}}, \bibinfo {author} {\bibfnamefont {B.~R.}\ \bibnamefont {Ko}}, \ and\ \bibinfo {author} {\bibfnamefont {Y.~K.}\ \bibnamefont {Semertzidis}},\ }\href {\doibase 10.1103/PhysRevLett.124.101802} {\bibfield  {journal} {\bibinfo  {journal} {Phys. Rev. Lett.}\ }\textbf {\bibinfo {volume} {124}},\ \bibinfo {pages} {101802} (\bibinfo {year} {2020})},\ \Eprint {http://arxiv.org/abs/2001.05102} {arXiv:2001.05102 [hep-ex]} \BibitemShut {NoStop}%
\bibitem [{\citenamefont {Jeong}\ \emph {et~al.}(2020)\citenamefont {Jeong}, \citenamefont {Youn}, \citenamefont {Bae}, \citenamefont {Kim}, \citenamefont {Seong}, \citenamefont {Kim},\ and\ \citenamefont {Semertzidis}}]{CAPP:Jeong_2020cwz}%
  \BibitemOpen
  \bibfield  {author} {\bibinfo {author} {\bibfnamefont {J.}~\bibnamefont {Jeong}}, \bibinfo {author} {\bibfnamefont {S.}~\bibnamefont {Youn}}, \bibinfo {author} {\bibfnamefont {S.}~\bibnamefont {Bae}}, \bibinfo {author} {\bibfnamefont {J.}~\bibnamefont {Kim}}, \bibinfo {author} {\bibfnamefont {T.}~\bibnamefont {Seong}}, \bibinfo {author} {\bibfnamefont {J.~E.}\ \bibnamefont {Kim}}, \ and\ \bibinfo {author} {\bibfnamefont {Y.~K.}\ \bibnamefont {Semertzidis}},\ }\href {\doibase 10.1103/physrevlett.125.221302} {\bibfield  {journal} {\bibinfo  {journal} {Physical Review Letters}\ }\textbf {\bibinfo {volume} {125}} (\bibinfo {year} {2020}),\ 10.1103/physrevlett.125.221302}\BibitemShut {NoStop}%
\bibitem [{\citenamefont {Lee}\ \emph {et~al.}(2022)\citenamefont {Lee}, \citenamefont {Yang}, \citenamefont {Yoon}, \citenamefont {Ahn}, \citenamefont {Park}, \citenamefont {Min}, \citenamefont {Kim},\ and\ \citenamefont {Yoo}}]{CAPP:Lee_2022mnc}%
  \BibitemOpen
  \bibfield  {author} {\bibinfo {author} {\bibfnamefont {Y.}~\bibnamefont {Lee}}, \bibinfo {author} {\bibfnamefont {B.}~\bibnamefont {Yang}}, \bibinfo {author} {\bibfnamefont {H.}~\bibnamefont {Yoon}}, \bibinfo {author} {\bibfnamefont {M.}~\bibnamefont {Ahn}}, \bibinfo {author} {\bibfnamefont {H.}~\bibnamefont {Park}}, \bibinfo {author} {\bibfnamefont {B.}~\bibnamefont {Min}}, \bibinfo {author} {\bibfnamefont {D.}~\bibnamefont {Kim}}, \ and\ \bibinfo {author} {\bibfnamefont {J.}~\bibnamefont {Yoo}},\ }\href {\doibase 10.1103/physrevlett.128.241805} {\bibfield  {journal} {\bibinfo  {journal} {Physical Review Letters}\ }\textbf {\bibinfo {volume} {128}} (\bibinfo {year} {2022}),\ 10.1103/physrevlett.128.241805}\BibitemShut {NoStop}%
\bibitem [{\citenamefont {Kim}\ \emph {et~al.}(2022)\citenamefont {Kim}, \citenamefont {Kwon}, \citenamefont {Kutlu}, \citenamefont {Chung}, \citenamefont {Matlashov}, \citenamefont {Uchaikin}, \citenamefont {van Loo}, \citenamefont {Nakamura}, \citenamefont {Oh}, \citenamefont {Byun}, \citenamefont {Ahn},\ and\ \citenamefont {Semertzidis}}]{CAPP:Kim2022}%
  \BibitemOpen
  \bibfield  {author} {\bibinfo {author} {\bibfnamefont {J.}~\bibnamefont {Kim}}, \bibinfo {author} {\bibfnamefont {O.}~\bibnamefont {Kwon}}, \bibinfo {author} {\bibfnamefont {C.}~\bibnamefont {Kutlu}}, \bibinfo {author} {\bibfnamefont {W.}~\bibnamefont {Chung}}, \bibinfo {author} {\bibfnamefont {A.}~\bibnamefont {Matlashov}}, \bibinfo {author} {\bibfnamefont {S.}~\bibnamefont {Uchaikin}}, \bibinfo {author} {\bibfnamefont {A.~F.}\ \bibnamefont {van Loo}}, \bibinfo {author} {\bibfnamefont {Y.}~\bibnamefont {Nakamura}}, \bibinfo {author} {\bibfnamefont {S.}~\bibnamefont {Oh}}, \bibinfo {author} {\bibfnamefont {H.}~\bibnamefont {Byun}}, \bibinfo {author} {\bibfnamefont {D.}~\bibnamefont {Ahn}}, \ and\ \bibinfo {author} {\bibfnamefont {Y.~K.}\ \bibnamefont {Semertzidis}},\ }\href {\doibase 10.48550/ARXIV.2207.13597} {\enquote {\bibinfo {title} {Near-quantum-noise axion dark matter search at capp},}\ } (\bibinfo {year} {2022})\BibitemShut {NoStop}%
\bibitem [{\citenamefont {{Brubaker}}\ \emph {et~al.}(2017)\citenamefont {{Brubaker}}, \citenamefont {{Zhong}}, \citenamefont {{Lamoreaux}}, \citenamefont {{Lehnert}},\ and\ \citenamefont {van Bibber}}]{HAYSTAC:Brubaker2017}%
  \BibitemOpen
  \bibfield  {author} {\bibinfo {author} {\bibfnamefont {B.}~\bibnamefont {{Brubaker}}}, \bibinfo {author} {\bibfnamefont {L.}~\bibnamefont {{Zhong}}}, \bibinfo {author} {\bibfnamefont {S.}~\bibnamefont {{Lamoreaux}}}, \bibinfo {author} {\bibfnamefont {K.}~\bibnamefont {{Lehnert}}}, \ and\ \bibinfo {author} {\bibfnamefont {K.}~\bibnamefont {van Bibber}},\ }\href {\doibase 10.1103/physrevd.96.123008} {\bibfield  {journal} {\bibinfo  {journal} {Physical Review D}\ }\textbf {\bibinfo {volume} {96}} (\bibinfo {year} {2017}),\ 10.1103/physrevd.96.123008}\BibitemShut {NoStop}%
\bibitem [{\citenamefont {Zhong}\ \emph {et~al.}(2018)\citenamefont {Zhong} \emph {et~al.}}]{HAYSTAC:2018rwy}%
  \BibitemOpen
  \bibfield  {author} {\bibinfo {author} {\bibfnamefont {L.}~\bibnamefont {Zhong}} \emph {et~al.} (\bibinfo {collaboration} {HAYSTAC}),\ }\href {\doibase 10.1103/PhysRevD.97.092001} {\bibfield  {journal} {\bibinfo  {journal} {Phys. Rev. D}\ }\textbf {\bibinfo {volume} {97}},\ \bibinfo {pages} {092001} (\bibinfo {year} {2018})},\ \Eprint {http://arxiv.org/abs/1803.03690} {arXiv:1803.03690 [hep-ex]} \BibitemShut {NoStop}%
\bibitem [{\citenamefont {Backes}\ \emph {et~al.}(2021)\citenamefont {Backes} \emph {et~al.}}]{HAYSTAC:2020kwv}%
  \BibitemOpen
  \bibfield  {author} {\bibinfo {author} {\bibfnamefont {K.~M.}\ \bibnamefont {Backes}} \emph {et~al.} (\bibinfo {collaboration} {HAYSTAC}),\ }\href {\doibase 10.1038/s41586-021-03226-7} {\bibfield  {journal} {\bibinfo  {journal} {Nature}\ }\textbf {\bibinfo {volume} {590}},\ \bibinfo {pages} {238} (\bibinfo {year} {2021})},\ \Eprint {http://arxiv.org/abs/2008.01853} {arXiv:2008.01853 [quant-ph]} \BibitemShut {NoStop}%
\bibitem [{\citenamefont {Alesini}\ \emph {et~al.}(2019)\citenamefont {Alesini} \emph {et~al.}}]{Alesini:2019ajt}%
  \BibitemOpen
  \bibfield  {author} {\bibinfo {author} {\bibfnamefont {D.}~\bibnamefont {Alesini}} \emph {et~al.},\ }\href {\doibase 10.1103/PhysRevD.99.101101} {\bibfield  {journal} {\bibinfo  {journal} {Phys. Rev. D}\ }\textbf {\bibinfo {volume} {99}},\ \bibinfo {pages} {101101} (\bibinfo {year} {2019})},\ \Eprint {http://arxiv.org/abs/1903.06547} {arXiv:1903.06547 [physics.ins-det]} \BibitemShut {NoStop}%
\bibitem [{\citenamefont {Alesini}\ \emph {et~al.}(2021)\citenamefont {Alesini} \emph {et~al.}}]{Alesini:2020vny}%
  \BibitemOpen
  \bibfield  {author} {\bibinfo {author} {\bibfnamefont {D.}~\bibnamefont {Alesini}} \emph {et~al.},\ }\href {\doibase 10.1103/PhysRevD.103.102004} {\bibfield  {journal} {\bibinfo  {journal} {Phys. Rev. D}\ }\textbf {\bibinfo {volume} {103}},\ \bibinfo {pages} {102004} (\bibinfo {year} {2021})},\ \Eprint {http://arxiv.org/abs/2012.09498} {arXiv:2012.09498 [hep-ex]} \BibitemShut {NoStop}%
\bibitem [{\citenamefont {McAllister}\ \emph {et~al.}(2017)\citenamefont {McAllister}, \citenamefont {Flower}, \citenamefont {Ivanov}, \citenamefont {Goryachev}, \citenamefont {Bourhill},\ and\ \citenamefont {Tobar}}]{ORGAN1}%
  \BibitemOpen
  \bibfield  {author} {\bibinfo {author} {\bibfnamefont {B.~T.}\ \bibnamefont {McAllister}}, \bibinfo {author} {\bibfnamefont {G.}~\bibnamefont {Flower}}, \bibinfo {author} {\bibfnamefont {E.~N.}\ \bibnamefont {Ivanov}}, \bibinfo {author} {\bibfnamefont {M.}~\bibnamefont {Goryachev}}, \bibinfo {author} {\bibfnamefont {J.}~\bibnamefont {Bourhill}}, \ and\ \bibinfo {author} {\bibfnamefont {M.~E.}\ \bibnamefont {Tobar}},\ }\href {\doibase 10.1016/j.dark.2017.09.010} {\bibfield  {journal} {\bibinfo  {journal} {Physics of the Dark Universe}\ }\textbf {\bibinfo {volume} {18}},\ \bibinfo {pages} {67} (\bibinfo {year} {2017})}\BibitemShut {NoStop}%
\bibitem [{\citenamefont {Quiskamp}\ \emph {et~al.}(2022)\citenamefont {Quiskamp}, \citenamefont {McAllister}, \citenamefont {Altin}, \citenamefont {Ivanov}, \citenamefont {Goryachev},\ and\ \citenamefont {Tobar}}]{ORGAN2}%
  \BibitemOpen
  \bibfield  {author} {\bibinfo {author} {\bibfnamefont {A.}~\bibnamefont {Quiskamp}}, \bibinfo {author} {\bibfnamefont {B.~T.}\ \bibnamefont {McAllister}}, \bibinfo {author} {\bibfnamefont {P.}~\bibnamefont {Altin}}, \bibinfo {author} {\bibfnamefont {E.~N.}\ \bibnamefont {Ivanov}}, \bibinfo {author} {\bibfnamefont {M.}~\bibnamefont {Goryachev}}, \ and\ \bibinfo {author} {\bibfnamefont {M.~E.}\ \bibnamefont {Tobar}},\ }\href {\doibase 10.1126/sciadv.abq3765} {\bibfield  {journal} {\bibinfo  {journal} {Science Advances}\ }\textbf {\bibinfo {volume} {8}} (\bibinfo {year} {2022}),\ 10.1126/sciadv.abq3765}\BibitemShut {NoStop}%
\bibitem [{\citenamefont {Melc\'on}\ \emph {et~al.}(2020)\citenamefont {Melc\'on} \emph {et~al.}}]{CAST:2020rlf}%
  \BibitemOpen
  \bibfield  {author} {\bibinfo {author} {\bibfnamefont {A.~A.}\ \bibnamefont {Melc\'on}} \emph {et~al.} (\bibinfo {collaboration} {CAST}),\ }\href {\doibase 10.1007/JHEP10(2021)075} {\bibfield  {journal} {\bibinfo  {journal} {JHEP}\ }\textbf {\bibinfo {volume} {21}},\ \bibinfo {pages} {075} (\bibinfo {year} {2020})},\ \Eprint {http://arxiv.org/abs/2104.13798} {arXiv:2104.13798 [hep-ex]} \BibitemShut {NoStop}%
\bibitem [{\citenamefont {Chang}\ \emph {et~al.}(2022)\citenamefont {Chang} \emph {et~al.}}]{TASEH:2022vvu}%
  \BibitemOpen
  \bibfield  {author} {\bibinfo {author} {\bibfnamefont {H.}~\bibnamefont {Chang}} \emph {et~al.} (\bibinfo {collaboration} {TASEH}),\ }\href {\doibase 10.1103/PhysRevLett.129.111802} {\bibfield  {journal} {\bibinfo  {journal} {Phys. Rev. Lett.}\ }\textbf {\bibinfo {volume} {129}},\ \bibinfo {pages} {111802} (\bibinfo {year} {2022})},\ \Eprint {http://arxiv.org/abs/2205.05574} {arXiv:2205.05574 [hep-ex]} \BibitemShut {NoStop}%
\bibitem [{\citenamefont {Grenet}\ \emph {et~al.}(2021)\citenamefont {Grenet}, \citenamefont {Ballou}, \citenamefont {Basto}, \citenamefont {Martineau}, \citenamefont {Perrier}, \citenamefont {Pugnat}, \citenamefont {Quevillon}, \citenamefont {Roch},\ and\ \citenamefont {Smith}}]{Grenet:2021vbb}%
  \BibitemOpen
  \bibfield  {author} {\bibinfo {author} {\bibfnamefont {T.}~\bibnamefont {Grenet}}, \bibinfo {author} {\bibfnamefont {R.}~\bibnamefont {Ballou}}, \bibinfo {author} {\bibfnamefont {Q.}~\bibnamefont {Basto}}, \bibinfo {author} {\bibfnamefont {K.}~\bibnamefont {Martineau}}, \bibinfo {author} {\bibfnamefont {P.}~\bibnamefont {Perrier}}, \bibinfo {author} {\bibfnamefont {P.}~\bibnamefont {Pugnat}}, \bibinfo {author} {\bibfnamefont {J.}~\bibnamefont {Quevillon}}, \bibinfo {author} {\bibfnamefont {N.}~\bibnamefont {Roch}}, \ and\ \bibinfo {author} {\bibfnamefont {C.}~\bibnamefont {Smith}},\ }\href@noop {} {\enquote {\bibinfo {title} {The grenoble axion haloscope platform (grahal): development plan and first results},}\ } (\bibinfo {year} {2021}),\ \Eprint {http://arxiv.org/abs/2110.14406} {arXiv:2110.14406 [hep-ex]} \BibitemShut {NoStop}%
\bibitem [{\citenamefont {Pshirkov}\ and\ \citenamefont {Popov}(2009)}]{Pshirkov:2007st}%
  \BibitemOpen
  \bibfield  {author} {\bibinfo {author} {\bibfnamefont {M.~S.}\ \bibnamefont {Pshirkov}}\ and\ \bibinfo {author} {\bibfnamefont {S.~B.}\ \bibnamefont {Popov}},\ }\href {\doibase 10.1134/S1063776109030030} {\bibfield  {journal} {\bibinfo  {journal} {J. Exp. Theor. Phys.}\ }\textbf {\bibinfo {volume} {108}},\ \bibinfo {pages} {384} (\bibinfo {year} {2009})},\ \Eprint {http://arxiv.org/abs/0711.1264} {arXiv:0711.1264 [astro-ph]} \BibitemShut {NoStop}%
\bibitem [{\citenamefont {Huang}\ \emph {et~al.}(2018)\citenamefont {Huang}, \citenamefont {Kadota}, \citenamefont {Sekiguchi},\ and\ \citenamefont {Tashiro}}]{Huang:2018lxq}%
  \BibitemOpen
  \bibfield  {author} {\bibinfo {author} {\bibfnamefont {F.~P.}\ \bibnamefont {Huang}}, \bibinfo {author} {\bibfnamefont {K.}~\bibnamefont {Kadota}}, \bibinfo {author} {\bibfnamefont {T.}~\bibnamefont {Sekiguchi}}, \ and\ \bibinfo {author} {\bibfnamefont {H.}~\bibnamefont {Tashiro}},\ }\href {\doibase 10.1103/PhysRevD.97.123001} {\bibfield  {journal} {\bibinfo  {journal} {Phys. Rev. D}\ }\textbf {\bibinfo {volume} {97}},\ \bibinfo {pages} {123001} (\bibinfo {year} {2018})},\ \Eprint {http://arxiv.org/abs/1803.08230} {arXiv:1803.08230 [hep-ph]} \BibitemShut {NoStop}%
\bibitem [{\citenamefont {Hook}\ \emph {et~al.}(2018)\citenamefont {Hook}, \citenamefont {Kahn}, \citenamefont {Safdi},\ and\ \citenamefont {Sun}}]{hook2018}%
  \BibitemOpen
  \bibfield  {author} {\bibinfo {author} {\bibfnamefont {A.}~\bibnamefont {Hook}}, \bibinfo {author} {\bibfnamefont {Y.}~\bibnamefont {Kahn}}, \bibinfo {author} {\bibfnamefont {B.}~\bibnamefont {Safdi}}, \ and\ \bibinfo {author} {\bibfnamefont {Z.}~\bibnamefont {Sun}},\ }\href@noop {} {\bibfield  {journal} {\bibinfo  {journal} {Phys.\ Rev.\ Lett.}\ }\textbf {\bibinfo {volume} {121}},\ \bibinfo {pages} {241102} (\bibinfo {year} {2018})}\BibitemShut {NoStop}%
\bibitem [{\citenamefont {Battye}\ \emph {et~al.}(2020)\citenamefont {Battye}, \citenamefont {Garbrecht}, \citenamefont {McDonald}, \citenamefont {Pace},\ and\ \citenamefont {Srinivasan}}]{Battye_2020}%
  \BibitemOpen
  \bibfield  {author} {\bibinfo {author} {\bibfnamefont {R.}~\bibnamefont {Battye}}, \bibinfo {author} {\bibfnamefont {B.}~\bibnamefont {Garbrecht}}, \bibinfo {author} {\bibfnamefont {J.}~\bibnamefont {McDonald}}, \bibinfo {author} {\bibfnamefont {F.}~\bibnamefont {Pace}}, \ and\ \bibinfo {author} {\bibfnamefont {S.}~\bibnamefont {Srinivasan}},\ }\href {\doibase 10.1103/physrevd.102.023504} {\bibfield  {journal} {\bibinfo  {journal} {Physical Review D}\ }\textbf {\bibinfo {volume} {102}} (\bibinfo {year} {2020}),\ 10.1103/physrevd.102.023504}\BibitemShut {NoStop}%
\bibitem [{\citenamefont {Walters}\ \emph {et~al.}(2024)\citenamefont {Walters}, \citenamefont {Shroyer}, \citenamefont {Edenton}, \citenamefont {Agrawal}, \citenamefont {Johnson}, \citenamefont {Kavanagh}, \citenamefont {Marsh},\ and\ \citenamefont {Visinelli}}]{walters2024axions}%
  \BibitemOpen
  \bibfield  {author} {\bibinfo {author} {\bibfnamefont {L.}~\bibnamefont {Walters}}, \bibinfo {author} {\bibfnamefont {J.}~\bibnamefont {Shroyer}}, \bibinfo {author} {\bibfnamefont {M.}~\bibnamefont {Edenton}}, \bibinfo {author} {\bibfnamefont {P.}~\bibnamefont {Agrawal}}, \bibinfo {author} {\bibfnamefont {B.}~\bibnamefont {Johnson}}, \bibinfo {author} {\bibfnamefont {B.~J.}\ \bibnamefont {Kavanagh}}, \bibinfo {author} {\bibfnamefont {D.~J.}\ \bibnamefont {Marsh}}, \ and\ \bibinfo {author} {\bibfnamefont {L.}~\bibnamefont {Visinelli}},\ }\href@noop {} {\bibfield  {journal} {\bibinfo  {journal} {arXiv preprint arXiv:2407.13060}\ } (\bibinfo {year} {2024})}\BibitemShut {NoStop}%
\bibitem [{\citenamefont {{Leroy}}\ \emph {et~al.}(2020)\citenamefont {{Leroy}}, \citenamefont {{Chianese}}, \citenamefont {{Edwards}},\ and\ \citenamefont {{Weniger}}}]{leroy2020}%
  \BibitemOpen
  \bibfield  {author} {\bibinfo {author} {\bibfnamefont {M.}~\bibnamefont {{Leroy}}}, \bibinfo {author} {\bibfnamefont {M.}~\bibnamefont {{Chianese}}}, \bibinfo {author} {\bibfnamefont {T.~D.~P.}\ \bibnamefont {{Edwards}}}, \ and\ \bibinfo {author} {\bibfnamefont {C.}~\bibnamefont {{Weniger}}},\ }\href {\doibase 10.1103/PhysRevD.101.123003} {\bibfield  {journal} {\bibinfo  {journal} {\prd}\ }\textbf {\bibinfo {volume} {101}},\ \bibinfo {eid} {123003} (\bibinfo {year} {2020})},\ \Eprint {http://arxiv.org/abs/1912.08815} {arXiv:1912.08815 [hep-ph]} \BibitemShut {NoStop}%
\bibitem [{\citenamefont {Battye}\ \emph {et~al.}(2021)\citenamefont {Battye}, \citenamefont {Garbrecht}, \citenamefont {McDonald},\ and\ \citenamefont {Srinivasan}}]{Battye:2021xvt}%
  \BibitemOpen
  \bibfield  {author} {\bibinfo {author} {\bibfnamefont {R.~A.}\ \bibnamefont {Battye}}, \bibinfo {author} {\bibfnamefont {B.}~\bibnamefont {Garbrecht}}, \bibinfo {author} {\bibfnamefont {J.~I.}\ \bibnamefont {McDonald}}, \ and\ \bibinfo {author} {\bibfnamefont {S.}~\bibnamefont {Srinivasan}},\ }\href {\doibase 10.1007/JHEP09(2021)105} {\bibfield  {journal} {\bibinfo  {journal} {JHEP}\ }\textbf {\bibinfo {volume} {09}},\ \bibinfo {pages} {105} (\bibinfo {year} {2021})},\ \Eprint {http://arxiv.org/abs/2104.08290} {arXiv:2104.08290 [hep-ph]} \BibitemShut {NoStop}%
\bibitem [{\citenamefont {{Witte}}\ \emph {et~al.}(2021)\citenamefont {{Witte}}, \citenamefont {{Noordhuis}}, \citenamefont {{Edwards}},\ and\ \citenamefont {{Weniger}}}]{Witte:2021arp}%
  \BibitemOpen
  \bibfield  {author} {\bibinfo {author} {\bibfnamefont {S.~J.}\ \bibnamefont {{Witte}}}, \bibinfo {author} {\bibfnamefont {D.}~\bibnamefont {{Noordhuis}}}, \bibinfo {author} {\bibfnamefont {T.~D.~P.}\ \bibnamefont {{Edwards}}}, \ and\ \bibinfo {author} {\bibfnamefont {C.}~\bibnamefont {{Weniger}}},\ }\href {\doibase 10.1103/PhysRevD.104.103030} {\bibfield  {journal} {\bibinfo  {journal} {\prd}\ }\textbf {\bibinfo {volume} {104}},\ \bibinfo {eid} {103030} (\bibinfo {year} {2021})},\ \Eprint {http://arxiv.org/abs/2104.07670} {arXiv:2104.07670 [hep-ph]} \BibitemShut {NoStop}%
\bibitem [{\citenamefont {McDonald}\ and\ \citenamefont {Witte}(2023)}]{McDonald:2023shx}%
  \BibitemOpen
  \bibfield  {author} {\bibinfo {author} {\bibfnamefont {J.~I.}\ \bibnamefont {McDonald}}\ and\ \bibinfo {author} {\bibfnamefont {S.~J.}\ \bibnamefont {Witte}},\ }\href {\doibase 10.1103/PhysRevD.108.103021} {\bibfield  {journal} {\bibinfo  {journal} {Phys. Rev. D}\ }\textbf {\bibinfo {volume} {108}},\ \bibinfo {pages} {103021} (\bibinfo {year} {2023})}\BibitemShut {NoStop}%
\bibitem [{\citenamefont {Goldreich}\ and\ \citenamefont {Julian}(1969)}]{goldreich1969}%
  \BibitemOpen
  \bibfield  {author} {\bibinfo {author} {\bibfnamefont {P.}~\bibnamefont {Goldreich}}\ and\ \bibinfo {author} {\bibfnamefont {W.~H.}\ \bibnamefont {Julian}},\ }\href@noop {} {\bibfield  {journal} {\bibinfo  {journal} {ApJ}\ }\textbf {\bibinfo {volume} {157}},\ \bibinfo {pages} {869} (\bibinfo {year} {1969})}\BibitemShut {NoStop}%
\bibitem [{\citenamefont {{Foster}}\ \emph {et~al.}(2020)\citenamefont {{Foster}}, \citenamefont {{Kahn}}, \citenamefont {{Macias}}, \citenamefont {{Sun}}, \citenamefont {{Eatough}}, \citenamefont {{Kondratiev}}, \citenamefont {{Peters}}, \citenamefont {{Weniger}},\ and\ \citenamefont {{Safdi}}}]{foster2020}%
  \BibitemOpen
  \bibfield  {author} {\bibinfo {author} {\bibfnamefont {J.~W.}\ \bibnamefont {{Foster}}}, \bibinfo {author} {\bibfnamefont {Y.}~\bibnamefont {{Kahn}}}, \bibinfo {author} {\bibfnamefont {O.}~\bibnamefont {{Macias}}}, \bibinfo {author} {\bibfnamefont {Z.}~\bibnamefont {{Sun}}}, \bibinfo {author} {\bibfnamefont {R.~P.}\ \bibnamefont {{Eatough}}}, \bibinfo {author} {\bibfnamefont {V.~I.}\ \bibnamefont {{Kondratiev}}}, \bibinfo {author} {\bibfnamefont {W.~M.}\ \bibnamefont {{Peters}}}, \bibinfo {author} {\bibfnamefont {C.}~\bibnamefont {{Weniger}}}, \ and\ \bibinfo {author} {\bibfnamefont {B.~R.}\ \bibnamefont {{Safdi}}},\ }\href {\doibase 10.1103/PhysRevLett.125.171301} {\bibfield  {journal} {\bibinfo  {journal} {\prl}\ }\textbf {\bibinfo {volume} {125}},\ \bibinfo {eid} {171301} (\bibinfo {year} {2020})},\ \Eprint {http://arxiv.org/abs/2004.00011} {arXiv:2004.00011 [astro-ph.CO]} \BibitemShut {NoStop}%
\bibitem [{\citenamefont {Darling}(2020)}]{Darling:2020plz}%
  \BibitemOpen
  \bibfield  {author} {\bibinfo {author} {\bibfnamefont {J.}~\bibnamefont {Darling}},\ }\href {\doibase 10.1103/PhysRevLett.125.121103} {\bibfield  {journal} {\bibinfo  {journal} {Phys. Rev. Lett.}\ }\textbf {\bibinfo {volume} {125}},\ \bibinfo {pages} {121103} (\bibinfo {year} {2020})},\ \Eprint {http://arxiv.org/abs/2008.01877} {arXiv:2008.01877 [astro-ph.CO]} \BibitemShut {NoStop}%
\bibitem [{\citenamefont {{Darling}}(2020)}]{darling2020apj}%
  \BibitemOpen
  \bibfield  {author} {\bibinfo {author} {\bibfnamefont {J.}~\bibnamefont {{Darling}}},\ }\href {\doibase 10.3847/2041-8213/abb23f} {\bibfield  {journal} {\bibinfo  {journal} {\apjl}\ }\textbf {\bibinfo {volume} {900}},\ \bibinfo {eid} {L28} (\bibinfo {year} {2020})},\ \Eprint {http://arxiv.org/abs/2008.11188} {arXiv:2008.11188 [astro-ph.CO]} \BibitemShut {NoStop}%
\bibitem [{\citenamefont {{Battye}}\ \emph {et~al.}(2022)\citenamefont {{Battye}}, \citenamefont {{Darling}}, \citenamefont {{McDonald}},\ and\ \citenamefont {{Srinivasan}}}]{Battye2022}%
  \BibitemOpen
  \bibfield  {author} {\bibinfo {author} {\bibfnamefont {R.~A.}\ \bibnamefont {{Battye}}}, \bibinfo {author} {\bibfnamefont {J.}~\bibnamefont {{Darling}}}, \bibinfo {author} {\bibfnamefont {J.~I.}\ \bibnamefont {{McDonald}}}, \ and\ \bibinfo {author} {\bibfnamefont {S.}~\bibnamefont {{Srinivasan}}},\ }\href {\doibase 10.1103/PhysRevD.105.L021305} {\bibfield  {journal} {\bibinfo  {journal} {\prd}\ }\textbf {\bibinfo {volume} {105}},\ \bibinfo {eid} {L021305} (\bibinfo {year} {2022})},\ \Eprint {http://arxiv.org/abs/2107.01225} {arXiv:2107.01225 [astro-ph.CO]} \BibitemShut {NoStop}%
\bibitem [{\citenamefont {Battye}\ \emph {et~al.}(2023)\citenamefont {Battye}, \citenamefont {Keith}, \citenamefont {McDonald}, \citenamefont {Srinivasan}, \citenamefont {Stappers},\ and\ \citenamefont {Weltevrede}}]{Battye:2023oac}%
  \BibitemOpen
  \bibfield  {author} {\bibinfo {author} {\bibfnamefont {R.~A.}\ \bibnamefont {Battye}}, \bibinfo {author} {\bibfnamefont {M.~J.}\ \bibnamefont {Keith}}, \bibinfo {author} {\bibfnamefont {J.~I.}\ \bibnamefont {McDonald}}, \bibinfo {author} {\bibfnamefont {S.}~\bibnamefont {Srinivasan}}, \bibinfo {author} {\bibfnamefont {B.~W.}\ \bibnamefont {Stappers}}, \ and\ \bibinfo {author} {\bibfnamefont {P.}~\bibnamefont {Weltevrede}},\ }\href {\doibase 10.1103/PhysRevD.108.063001} {\bibfield  {journal} {\bibinfo  {journal} {Phys. Rev. D}\ }\textbf {\bibinfo {volume} {108}},\ \bibinfo {pages} {063001} (\bibinfo {year} {2023})},\ \Eprint {http://arxiv.org/abs/2303.11792} {arXiv:2303.11792 [astro-ph.CO]} \BibitemShut {NoStop}%
\bibitem [{\citenamefont {Safdi}\ \emph {et~al.}(2019)\citenamefont {Safdi}, \citenamefont {Sun},\ and\ \citenamefont {Chen}}]{Safdi2019}%
  \BibitemOpen
  \bibfield  {author} {\bibinfo {author} {\bibfnamefont {B.~R.}\ \bibnamefont {Safdi}}, \bibinfo {author} {\bibfnamefont {Z.}~\bibnamefont {Sun}}, \ and\ \bibinfo {author} {\bibfnamefont {A.~Y.}\ \bibnamefont {Chen}},\ }\href {\doibase 10.1103/PhysRevD.99.123021} {\bibfield  {journal} {\bibinfo  {journal} {Physical Review D}\ }\textbf {\bibinfo {volume} {99}} (\bibinfo {year} {2019}),\ 10.1103/PhysRevD.99.123021},\ \Eprint {http://arxiv.org/abs/1811.01020} {arXiv:1811.01020} \BibitemShut {NoStop}%
\bibitem [{\citenamefont {Foster}\ \emph {et~al.}(2022)\citenamefont {Foster}, \citenamefont {Witte}, \citenamefont {Lawson}, \citenamefont {Linden}, \citenamefont {Gajjar}, \citenamefont {Weniger},\ and\ \citenamefont {Safdi}}]{Foster:2022fxn}%
  \BibitemOpen
  \bibfield  {author} {\bibinfo {author} {\bibfnamefont {J.~W.}\ \bibnamefont {Foster}}, \bibinfo {author} {\bibfnamefont {S.~J.}\ \bibnamefont {Witte}}, \bibinfo {author} {\bibfnamefont {M.}~\bibnamefont {Lawson}}, \bibinfo {author} {\bibfnamefont {T.}~\bibnamefont {Linden}}, \bibinfo {author} {\bibfnamefont {V.}~\bibnamefont {Gajjar}}, \bibinfo {author} {\bibfnamefont {C.}~\bibnamefont {Weniger}}, \ and\ \bibinfo {author} {\bibfnamefont {B.~R.}\ \bibnamefont {Safdi}},\ }\href {\doibase 10.1103/PhysRevLett.129.251102} {\bibfield  {journal} {\bibinfo  {journal} {Phys. Rev. Lett.}\ }\textbf {\bibinfo {volume} {129}},\ \bibinfo {pages} {251102} (\bibinfo {year} {2022})},\ \Eprint {http://arxiv.org/abs/2202.08274} {arXiv:2202.08274 [astro-ph.CO]} \BibitemShut {NoStop}%
\bibitem [{\citenamefont {{Jurado}}\ \emph {et~al.}(2023)\citenamefont {{Jurado}}, \citenamefont {{Naoz}}, \citenamefont {{Lam}},\ and\ \citenamefont {{Hoang}}}]{Jurado:2023lkg}%
  \BibitemOpen
  \bibfield  {author} {\bibinfo {author} {\bibfnamefont {C.}~\bibnamefont {{Jurado}}}, \bibinfo {author} {\bibfnamefont {S.}~\bibnamefont {{Naoz}}}, \bibinfo {author} {\bibfnamefont {C.~Y.}\ \bibnamefont {{Lam}}}, \ and\ \bibinfo {author} {\bibfnamefont {B.-M.}\ \bibnamefont {{Hoang}}},\ }\href {\doibase 10.48550/arXiv.2310.17707} {\bibfield  {journal} {\bibinfo  {journal} {arXiv e-prints}\ ,\ \bibinfo {eid} {arXiv:2310.17707}} (\bibinfo {year} {2023})},\ \Eprint {http://arxiv.org/abs/2310.17707} {arXiv:2310.17707 [astro-ph.GA]} \BibitemShut {NoStop}%
\bibitem [{\citenamefont {{Manchester}}(2001)}]{Manchester2001_PMBS}%
  \BibitemOpen
  \bibfield  {author} {\bibinfo {author} {\bibfnamefont {R.~N.}\ \bibnamefont {{Manchester}}},\ }\href {\doibase 10.1071/AS01002} {\bibfield  {journal} {\bibinfo  {journal} {\pasa}\ }\textbf {\bibinfo {volume} {18}},\ \bibinfo {pages} {1} (\bibinfo {year} {2001})},\ \Eprint {http://arxiv.org/abs/astro-ph/0009405} {arXiv:astro-ph/0009405 [astro-ph]} \BibitemShut {NoStop}%
\bibitem [{\citenamefont {Noordhuis}\ \emph {et~al.}(2023)\citenamefont {Noordhuis}, \citenamefont {Prabhu}, \citenamefont {Witte}, \citenamefont {Chen}, \citenamefont {Cruz},\ and\ \citenamefont {Weniger}}]{Noordhuis:2022ljw}%
  \BibitemOpen
  \bibfield  {author} {\bibinfo {author} {\bibfnamefont {D.}~\bibnamefont {Noordhuis}}, \bibinfo {author} {\bibfnamefont {A.}~\bibnamefont {Prabhu}}, \bibinfo {author} {\bibfnamefont {S.~J.}\ \bibnamefont {Witte}}, \bibinfo {author} {\bibfnamefont {A.~Y.}\ \bibnamefont {Chen}}, \bibinfo {author} {\bibfnamefont {F.}~\bibnamefont {Cruz}}, \ and\ \bibinfo {author} {\bibfnamefont {C.}~\bibnamefont {Weniger}},\ }\href {\doibase 10.1103/PhysRevLett.131.111004} {\bibfield  {journal} {\bibinfo  {journal} {Phys. Rev. Lett.}\ }\textbf {\bibinfo {volume} {131}},\ \bibinfo {pages} {111004} (\bibinfo {year} {2023})}\BibitemShut {NoStop}%
\bibitem [{\citenamefont {{Tjemsland}}\ \emph {et~al.}(2023)\citenamefont {{Tjemsland}}, \citenamefont {{McDonald}},\ and\ \citenamefont {{Witte}}}]{Tjemsland:2023vvc}%
  \BibitemOpen
  \bibfield  {author} {\bibinfo {author} {\bibfnamefont {J.}~\bibnamefont {{Tjemsland}}}, \bibinfo {author} {\bibfnamefont {J.}~\bibnamefont {{McDonald}}}, \ and\ \bibinfo {author} {\bibfnamefont {S.~J.}\ \bibnamefont {{Witte}}},\ }\href {\doibase 10.48550/arXiv.2310.18403} {\bibfield  {journal} {\bibinfo  {journal} {arXiv e-prints}\ ,\ \bibinfo {eid} {arXiv:2310.18403}} (\bibinfo {year} {2023})},\ \Eprint {http://arxiv.org/abs/2310.18403} {arXiv:2310.18403 [hep-ph]} \BibitemShut {NoStop}%
\bibitem [{\citenamefont {Carenza}\ and\ \citenamefont {Marsh}(2023)}]{Carenza:2023nck}%
  \BibitemOpen
  \bibfield  {author} {\bibinfo {author} {\bibfnamefont {P.}~\bibnamefont {Carenza}}\ and\ \bibinfo {author} {\bibfnamefont {M.~D.}\ \bibnamefont {Marsh}},\ }\href {\doibase 10.1088/1475-7516/2023/04/021} {\bibfield  {journal} {\bibinfo  {journal} {Journal of Cosmology and Astroparticle Physics}\ }\textbf {\bibinfo {volume} {2023}},\ \bibinfo {pages} {021} (\bibinfo {year} {2023})}\BibitemShut {NoStop}%
\bibitem [{\citenamefont {McDonald}\ \emph {et~al.}(2023)\citenamefont {McDonald}, \citenamefont {Garbrecht},\ and\ \citenamefont {Millington}}]{McDonald:2023ohd}%
  \BibitemOpen
  \bibfield  {author} {\bibinfo {author} {\bibfnamefont {J.}~\bibnamefont {McDonald}}, \bibinfo {author} {\bibfnamefont {B.}~\bibnamefont {Garbrecht}}, \ and\ \bibinfo {author} {\bibfnamefont {P.}~\bibnamefont {Millington}},\ }\href {\doibase 10.1088/1475-7516/2023/12/031} {\bibfield  {journal} {\bibinfo  {journal} {Journal of Cosmology and Astroparticle Physics}\ }\textbf {\bibinfo {volume} {2023}},\ \bibinfo {pages} {031} (\bibinfo {year} {2023})}\BibitemShut {NoStop}%
\bibitem [{\citenamefont {{McDonald}}\ and\ \citenamefont {{Millington}}(2024)}]{McDonald:2024uuh}%
  \BibitemOpen
  \bibfield  {author} {\bibinfo {author} {\bibfnamefont {J.~I.}\ \bibnamefont {{McDonald}}}\ and\ \bibinfo {author} {\bibfnamefont {P.}~\bibnamefont {{Millington}}},\ }\href {\doibase 10.48550/arXiv.2407.11192} {\bibfield  {journal} {\bibinfo  {journal} {arXiv e-prints}\ ,\ \bibinfo {eid} {arXiv:2407.11192}} (\bibinfo {year} {2024})},\ \Eprint {http://arxiv.org/abs/2407.11192} {arXiv:2407.11192 [hep-ph]} \BibitemShut {NoStop}%
\bibitem [{\citenamefont {{Utrilla Gin{\'e}s}}\ \emph {et~al.}(2024)\citenamefont {{Utrilla Gin{\'e}s}}, \citenamefont {{Noordhuis}}, \citenamefont {{Weniger}},\ and\ \citenamefont {{Witte}}}]{Gines:2024ekm}%
  \BibitemOpen
  \bibfield  {author} {\bibinfo {author} {\bibfnamefont {E.}~\bibnamefont {{Utrilla Gin{\'e}s}}}, \bibinfo {author} {\bibfnamefont {D.}~\bibnamefont {{Noordhuis}}}, \bibinfo {author} {\bibfnamefont {C.}~\bibnamefont {{Weniger}}}, \ and\ \bibinfo {author} {\bibfnamefont {S.~J.}\ \bibnamefont {{Witte}}},\ }\href {\doibase 10.48550/arXiv.2405.08865} {\bibfield  {journal} {\bibinfo  {journal} {arXiv e-prints}\ ,\ \bibinfo {eid} {arXiv:2405.08865}} (\bibinfo {year} {2024})},\ \Eprint {http://arxiv.org/abs/2405.08865} {arXiv:2405.08865 [hep-ph]} \BibitemShut {NoStop}%
\bibitem [{\citenamefont {Philippov}\ \emph {et~al.}(2015)\citenamefont {Philippov}, \citenamefont {Spitkovsky},\ and\ \citenamefont {Cerutti}}]{Philippov:2014mqa}%
  \BibitemOpen
  \bibfield  {author} {\bibinfo {author} {\bibfnamefont {A.~A.}\ \bibnamefont {Philippov}}, \bibinfo {author} {\bibfnamefont {A.}~\bibnamefont {Spitkovsky}}, \ and\ \bibinfo {author} {\bibfnamefont {B.}~\bibnamefont {Cerutti}},\ }\href {\doibase 10.1088/2041-8205/801/1/L19} {\bibfield  {journal} {\bibinfo  {journal} {Astrophys. J. Lett.}\ }\textbf {\bibinfo {volume} {801}},\ \bibinfo {pages} {L19} (\bibinfo {year} {2015})},\ \Eprint {http://arxiv.org/abs/1412.0673} {arXiv:1412.0673 [astro-ph.HE]} \BibitemShut {NoStop}%
\bibitem [{\citenamefont {Hu}\ and\ \citenamefont {Beloborodov}(2022)}]{Hu:2021nxu}%
  \BibitemOpen
  \bibfield  {author} {\bibinfo {author} {\bibfnamefont {R.}~\bibnamefont {Hu}}\ and\ \bibinfo {author} {\bibfnamefont {A.~M.}\ \bibnamefont {Beloborodov}},\ }\href {\doibase 10.3847/1538-4357/ac961d} {\bibfield  {journal} {\bibinfo  {journal} {Astrophys. J.}\ }\textbf {\bibinfo {volume} {939}},\ \bibinfo {pages} {42} (\bibinfo {year} {2022})},\ \Eprint {http://arxiv.org/abs/2109.03935} {arXiv:2109.03935 [astro-ph.HE]} \BibitemShut {NoStop}%
\bibitem [{\citenamefont {Bates}\ \emph {et~al.}(2014)\citenamefont {Bates}, \citenamefont {Lorimer}, \citenamefont {Rane},\ and\ \citenamefont {Swiggum}}]{Bates:2013uma}%
  \BibitemOpen
  \bibfield  {author} {\bibinfo {author} {\bibfnamefont {S.}~\bibnamefont {Bates}}, \bibinfo {author} {\bibfnamefont {D.}~\bibnamefont {Lorimer}}, \bibinfo {author} {\bibfnamefont {A.}~\bibnamefont {Rane}}, \ and\ \bibinfo {author} {\bibfnamefont {J.}~\bibnamefont {Swiggum}},\ }\href {\doibase 10.1093/mnras/stu157} {\bibfield  {journal} {\bibinfo  {journal} {Mon. Not. Roy. Astron. Soc.}\ }\textbf {\bibinfo {volume} {439}},\ \bibinfo {pages} {2893} (\bibinfo {year} {2014})},\ \Eprint {http://arxiv.org/abs/1311.3427} {arXiv:1311.3427 [astro-ph.IM]} \BibitemShut {NoStop}%
\bibitem [{\citenamefont {{Lorimer}}\ \emph {et~al.}(2006)\citenamefont {{Lorimer}}, \citenamefont {{Faulkner}}, \citenamefont {{Lyne}}, \citenamefont {{Manchester}}, \citenamefont {{Kramer}}, \citenamefont {{McLaughlin}}, \citenamefont {{Hobbs}}, \citenamefont {{Possenti}}, \citenamefont {{Stairs}}, \citenamefont {{Camilo}}, \citenamefont {{Burgay}}, \citenamefont {{D’Amico}}, \citenamefont {{Corongiu}},\ and\ \citenamefont {{Crawford}}}]{lorimer_pmps}%
  \BibitemOpen
  \bibfield  {author} {\bibinfo {author} {\bibfnamefont {D.}~\bibnamefont {{Lorimer}}}, \bibinfo {author} {\bibfnamefont {A.}~\bibnamefont {{Faulkner}}}, \bibinfo {author} {\bibfnamefont {A.}~\bibnamefont {{Lyne}}}, \bibinfo {author} {\bibfnamefont {R.}~\bibnamefont {{Manchester}}}, \bibinfo {author} {\bibfnamefont {M.}~\bibnamefont {{Kramer}}}, \bibinfo {author} {\bibfnamefont {M.}~\bibnamefont {{McLaughlin}}}, \bibinfo {author} {\bibfnamefont {G.}~\bibnamefont {{Hobbs}}}, \bibinfo {author} {\bibfnamefont {A.}~\bibnamefont {{Possenti}}}, \bibinfo {author} {\bibfnamefont {I.}~\bibnamefont {{Stairs}}}, \bibinfo {author} {\bibfnamefont {F.}~\bibnamefont {{Camilo}}}, \bibinfo {author} {\bibfnamefont {M.}~\bibnamefont {{Burgay}}}, \bibinfo {author} {\bibfnamefont {N.}~\bibnamefont {{D’Amico}}}, \bibinfo {author} {\bibfnamefont {A.}~\bibnamefont {{Corongiu}}}, \ and\ \bibinfo {author} {\bibfnamefont {F.}~\bibnamefont {{Crawford}}},\ }\href@noop {} {\bibfield  {journal} {\bibinfo  {journal} {Monthly Notices of
  the Royal Astronomical Society 372, 777–800}\ } (\bibinfo {year} {2006})}\BibitemShut {NoStop}%
\bibitem [{\citenamefont {Faucher-Giguere}\ and\ \citenamefont {Kaspi}(2006)}]{Faucher-Giguere:2005dxp}%
  \BibitemOpen
  \bibfield  {author} {\bibinfo {author} {\bibfnamefont {C.-A.}\ \bibnamefont {Faucher-Giguere}}\ and\ \bibinfo {author} {\bibfnamefont {V.~M.}\ \bibnamefont {Kaspi}},\ }\href {\doibase 10.1086/501516} {\bibfield  {journal} {\bibinfo  {journal} {Astrophys. J.}\ }\textbf {\bibinfo {volume} {643}},\ \bibinfo {pages} {332} (\bibinfo {year} {2006})},\ \Eprint {http://arxiv.org/abs/astro-ph/0512585} {arXiv:astro-ph/0512585} \BibitemShut {NoStop}%
\bibitem [{\citenamefont {{Tauris}}\ and\ \citenamefont {{Manchester}}(1998)}]{1998MNRAS.298..625T}%
  \BibitemOpen
  \bibfield  {author} {\bibinfo {author} {\bibfnamefont {T.~M.}\ \bibnamefont {{Tauris}}}\ and\ \bibinfo {author} {\bibfnamefont {R.~N.}\ \bibnamefont {{Manchester}}},\ }\href {\doibase 10.1046/j.1365-8711.1998.01369.x} {\bibfield  {journal} {\bibinfo  {journal} {\mnras}\ }\textbf {\bibinfo {volume} {298}},\ \bibinfo {pages} {625} (\bibinfo {year} {1998})}\BibitemShut {NoStop}%
\bibitem [{\citenamefont {G\'orski}\ \emph {et~al.}(2005)\citenamefont {G\'orski}, \citenamefont {Hivon}, \citenamefont {Banday}, \citenamefont {Wandelt}, \citenamefont {Hansen}, \citenamefont {Reinecke},\ and\ \citenamefont {Bartelman}}]{Gorski:2004by}%
  \BibitemOpen
  \bibfield  {author} {\bibinfo {author} {\bibfnamefont {K.~M.}\ \bibnamefont {G\'orski}}, \bibinfo {author} {\bibfnamefont {E.}~\bibnamefont {Hivon}}, \bibinfo {author} {\bibfnamefont {A.~J.}\ \bibnamefont {Banday}}, \bibinfo {author} {\bibfnamefont {B.~D.}\ \bibnamefont {Wandelt}}, \bibinfo {author} {\bibfnamefont {F.~K.}\ \bibnamefont {Hansen}}, \bibinfo {author} {\bibfnamefont {M.}~\bibnamefont {Reinecke}}, \ and\ \bibinfo {author} {\bibfnamefont {M.}~\bibnamefont {Bartelman}},\ }\href {\doibase 10.1086/427976} {\bibfield  {journal} {\bibinfo  {journal} {Astrophys. J.}\ }\textbf {\bibinfo {volume} {622}},\ \bibinfo {pages} {759} (\bibinfo {year} {2005})},\ \Eprint {http://arxiv.org/abs/astro-ph/0409513} {arXiv:astro-ph/0409513} \BibitemShut {NoStop}%
\bibitem [{\citenamefont {{Johnston}}\ \emph {et~al.}(2006)\citenamefont {{Johnston}}, \citenamefont {{Kramer}}, \citenamefont {{Lorimer}}, \citenamefont {{Lyne}}, \citenamefont {{McLaughlin}}, \citenamefont {{Klein}},\ and\ \citenamefont {{Manchester}}}]{2006MNRAS.373L...6J}%
  \BibitemOpen
  \bibfield  {author} {\bibinfo {author} {\bibfnamefont {S.}~\bibnamefont {{Johnston}}}, \bibinfo {author} {\bibfnamefont {M.}~\bibnamefont {{Kramer}}}, \bibinfo {author} {\bibfnamefont {D.~R.}\ \bibnamefont {{Lorimer}}}, \bibinfo {author} {\bibfnamefont {A.~G.}\ \bibnamefont {{Lyne}}}, \bibinfo {author} {\bibfnamefont {M.}~\bibnamefont {{McLaughlin}}}, \bibinfo {author} {\bibfnamefont {B.}~\bibnamefont {{Klein}}}, \ and\ \bibinfo {author} {\bibfnamefont {R.~N.}\ \bibnamefont {{Manchester}}},\ }\href {\doibase 10.1111/j.1745-3933.2006.00232.x} {\bibfield  {journal} {\bibinfo  {journal} {\mnras}\ }\textbf {\bibinfo {volume} {373}},\ \bibinfo {pages} {L6} (\bibinfo {year} {2006})},\ \Eprint {http://arxiv.org/abs/astro-ph/0606465} {arXiv:astro-ph/0606465 [astro-ph]} \BibitemShut {NoStop}%
\bibitem [{\citenamefont {Deneva}\ \emph {et~al.}(2009)\citenamefont {Deneva}, \citenamefont {Cordes},\ and\ \citenamefont {Lazio}}]{Deneva:2009mx}%
  \BibitemOpen
  \bibfield  {author} {\bibinfo {author} {\bibfnamefont {J.~S.}\ \bibnamefont {Deneva}}, \bibinfo {author} {\bibfnamefont {J.~M.}\ \bibnamefont {Cordes}}, \ and\ \bibinfo {author} {\bibfnamefont {T.~J.~W.}\ \bibnamefont {Lazio}},\ }\href {\doibase 10.1088/0004-637X/702/2/L177} {\bibfield  {journal} {\bibinfo  {journal} {Astrophys. J. Lett.}\ }\textbf {\bibinfo {volume} {702}},\ \bibinfo {pages} {L177} (\bibinfo {year} {2009})},\ \Eprint {http://arxiv.org/abs/0908.1331} {arXiv:0908.1331 [astro-ph.SR]} \BibitemShut {NoStop}%
\bibitem [{\citenamefont {Bates}\ \emph {et~al.}(2011)\citenamefont {Bates} \emph {et~al.}}]{Bates:2010wb}%
  \BibitemOpen
  \bibfield  {author} {\bibinfo {author} {\bibfnamefont {S.~D.}\ \bibnamefont {Bates}} \emph {et~al.},\ }\href {\doibase 10.1111/j.1365-2966.2010.17790.x} {\bibfield  {journal} {\bibinfo  {journal} {Mon. Not. Roy. Astron. Soc.}\ }\textbf {\bibinfo {volume} {411}},\ \bibinfo {pages} {1575} (\bibinfo {year} {2011})},\ \Eprint {http://arxiv.org/abs/1009.5873} {arXiv:1009.5873 [astro-ph.SR]} \BibitemShut {NoStop}%
\bibitem [{\citenamefont {{Abuter, R.}}\ \emph {et~al.}(2022)\citenamefont {{Abuter, R.}}, \citenamefont {{Aimar, N.}}, \citenamefont {{Amorim, A.}} \emph {et~al.}}]{GRAVITY_coll_2022}%
  \BibitemOpen
  \bibfield  {author} {\bibinfo {author} {\bibnamefont {{Abuter, R.}}}, \bibinfo {author} {\bibnamefont {{Aimar, N.}}}, \bibinfo {author} {\bibnamefont {{Amorim, A.}}},  \emph {et~al.} (\bibinfo {collaboration} {GRAVITY collaboration}),\ }\href {\doibase 10.1051/0004-6361/202142465} {\bibfield  {journal} {\bibinfo  {journal} {\aap}\ }\textbf {\bibinfo {volume} {657}},\ \bibinfo {eid} {L12} (\bibinfo {year} {2022})},\ \Eprint {http://arxiv.org/abs/2112.07478} {arXiv:2112.07478 [astro-ph.GA]} \BibitemShut {NoStop}%
\bibitem [{\citenamefont {Akiyama}\ \emph {et~al.}(2022)\citenamefont {Akiyama}, \citenamefont {Alberdi}, \citenamefont {Alef} \emph {et~al.}}]{EHT_SagA}%
  \BibitemOpen
  \bibfield  {author} {\bibinfo {author} {\bibfnamefont {K.}~\bibnamefont {Akiyama}}, \bibinfo {author} {\bibfnamefont {A.}~\bibnamefont {Alberdi}}, \bibinfo {author} {\bibfnamefont {W.}~\bibnamefont {Alef}},  \emph {et~al.} (\bibinfo {collaboration} {EHT Collaboration}),\ }\href {\doibase 10.3847/2041-8213/ac6674} {\bibfield  {journal} {\bibinfo  {journal} {The Astrophysical Journal Letters}\ }\textbf {\bibinfo {volume} {930}},\ \bibinfo {pages} {L12} (\bibinfo {year} {2022})}\BibitemShut {NoStop}%
\bibitem [{\citenamefont {Pfahl}\ and\ \citenamefont {Loeb}(2004)}]{Pfahl:2003tf}%
  \BibitemOpen
  \bibfield  {author} {\bibinfo {author} {\bibfnamefont {E.}~\bibnamefont {Pfahl}}\ and\ \bibinfo {author} {\bibfnamefont {A.}~\bibnamefont {Loeb}},\ }\href {\doibase 10.1086/423975} {\bibfield  {journal} {\bibinfo  {journal} {Astrophys. J.}\ }\textbf {\bibinfo {volume} {615}},\ \bibinfo {pages} {253} (\bibinfo {year} {2004})},\ \Eprint {http://arxiv.org/abs/astro-ph/0309744} {arXiv:astro-ph/0309744} \BibitemShut {NoStop}%
\bibitem [{\citenamefont {Dexter}\ and\ \citenamefont {O'Leary}(2014)}]{Dexter:2013xga}%
  \BibitemOpen
  \bibfield  {author} {\bibinfo {author} {\bibfnamefont {J.}~\bibnamefont {Dexter}}\ and\ \bibinfo {author} {\bibfnamefont {R.~M.}\ \bibnamefont {O'Leary}},\ }\href {\doibase 10.1088/2041-8205/783/1/L7} {\bibfield  {journal} {\bibinfo  {journal} {Astrophys. J. Lett.}\ }\textbf {\bibinfo {volume} {783}},\ \bibinfo {pages} {L7} (\bibinfo {year} {2014})},\ \Eprint {http://arxiv.org/abs/1310.7022} {arXiv:1310.7022 [astro-ph.GA]} \BibitemShut {NoStop}%
\bibitem [{\citenamefont {Zhang}\ \emph {et~al.}(2014)\citenamefont {Zhang}, \citenamefont {Lu},\ and\ \citenamefont {Yu}}]{Zhang:2014kva}%
  \BibitemOpen
  \bibfield  {author} {\bibinfo {author} {\bibfnamefont {F.}~\bibnamefont {Zhang}}, \bibinfo {author} {\bibfnamefont {Y.}~\bibnamefont {Lu}}, \ and\ \bibinfo {author} {\bibfnamefont {Q.}~\bibnamefont {Yu}},\ }\href {\doibase 10.1088/0004-637X/784/2/106} {\bibfield  {journal} {\bibinfo  {journal} {Astrophys. J.}\ }\textbf {\bibinfo {volume} {784}},\ \bibinfo {pages} {106} (\bibinfo {year} {2014})},\ \Eprint {http://arxiv.org/abs/1402.2505} {arXiv:1402.2505 [astro-ph.GA]} \BibitemShut {NoStop}%
\bibitem [{\citenamefont {Wharton}\ \emph {et~al.}(2012)\citenamefont {Wharton}, \citenamefont {Chatterjee}, \citenamefont {Cordes}, \citenamefont {Deneva},\ and\ \citenamefont {Lazio}}]{Wharton:2011dv}%
  \BibitemOpen
  \bibfield  {author} {\bibinfo {author} {\bibfnamefont {R.~S.}\ \bibnamefont {Wharton}}, \bibinfo {author} {\bibfnamefont {S.}~\bibnamefont {Chatterjee}}, \bibinfo {author} {\bibfnamefont {J.~M.}\ \bibnamefont {Cordes}}, \bibinfo {author} {\bibfnamefont {J.~S.}\ \bibnamefont {Deneva}}, \ and\ \bibinfo {author} {\bibfnamefont {T.~J.~W.}\ \bibnamefont {Lazio}},\ }\href {\doibase 10.1088/0004-637X/753/2/108} {\bibfield  {journal} {\bibinfo  {journal} {Astrophys. J.}\ }\textbf {\bibinfo {volume} {753}},\ \bibinfo {pages} {108} (\bibinfo {year} {2012})},\ \Eprint {http://arxiv.org/abs/1111.4216} {arXiv:1111.4216 [astro-ph.HE]} \BibitemShut {NoStop}%
\bibitem [{\citenamefont {Chennamangalam}\ and\ \citenamefont {Lorimer}(2014)}]{Chennamangalam:2013zja}%
  \BibitemOpen
  \bibfield  {author} {\bibinfo {author} {\bibfnamefont {J.}~\bibnamefont {Chennamangalam}}\ and\ \bibinfo {author} {\bibfnamefont {D.~R.}\ \bibnamefont {Lorimer}},\ }\href {\doibase 10.1093/mnrasl/slu025} {\bibfield  {journal} {\bibinfo  {journal} {Mon. Not. Roy. Astron. Soc.}\ }\textbf {\bibinfo {volume} {440}},\ \bibinfo {pages} {86} (\bibinfo {year} {2014})},\ \Eprint {http://arxiv.org/abs/1311.4846} {arXiv:1311.4846 [astro-ph.HE]} \BibitemShut {NoStop}%
\bibitem [{\citenamefont {{Mori}}\ \emph {et~al.}(2013)\citenamefont {{Mori}}, \citenamefont {{Gotthelf}}, \citenamefont {{Zhang}} \emph {et~al.}}]{2013ApJ...770L..23M}%
  \BibitemOpen
  \bibfield  {author} {\bibinfo {author} {\bibfnamefont {K.}~\bibnamefont {{Mori}}}, \bibinfo {author} {\bibfnamefont {E.~V.}\ \bibnamefont {{Gotthelf}}}, \bibinfo {author} {\bibnamefont {{Zhang}}},  \emph {et~al.},\ }\href {\doibase 10.1088/2041-8205/770/2/L23} {\bibfield  {journal} {\bibinfo  {journal} {\apjl}\ }\textbf {\bibinfo {volume} {770}},\ \bibinfo {eid} {L23} (\bibinfo {year} {2013})},\ \Eprint {http://arxiv.org/abs/1305.1945} {arXiv:1305.1945 [astro-ph.HE]} \BibitemShut {NoStop}%
\bibitem [{\citenamefont {{Kennea}}\ \emph {et~al.}(2013)\citenamefont {{Kennea}}, \citenamefont {{Burrows}}, \citenamefont {{Kouveliotou}} \emph {et~al.}}]{2013ApJ...770L..24K}%
  \BibitemOpen
  \bibfield  {author} {\bibinfo {author} {\bibfnamefont {J.~A.}\ \bibnamefont {{Kennea}}}, \bibinfo {author} {\bibfnamefont {D.~N.}\ \bibnamefont {{Burrows}}}, \bibinfo {author} {\bibfnamefont {C.}~\bibnamefont {{Kouveliotou}}},  \emph {et~al.},\ }\href {\doibase 10.1088/2041-8205/770/2/L24} {\bibfield  {journal} {\bibinfo  {journal} {\apjl}\ }\textbf {\bibinfo {volume} {770}},\ \bibinfo {eid} {L24} (\bibinfo {year} {2013})},\ \Eprint {http://arxiv.org/abs/1305.2128} {arXiv:1305.2128 [astro-ph.HE]} \BibitemShut {NoStop}%
\bibitem [{\citenamefont {Cordes}\ and\ \citenamefont {Lazio}(1997)}]{Cordes:1996bt}%
  \BibitemOpen
  \bibfield  {author} {\bibinfo {author} {\bibfnamefont {J.~M.}\ \bibnamefont {Cordes}}\ and\ \bibinfo {author} {\bibfnamefont {T.~J.~W.}\ \bibnamefont {Lazio}},\ }\href {\doibase 10.1086/303569} {\bibfield  {journal} {\bibinfo  {journal} {Astrophys. J.}\ }\textbf {\bibinfo {volume} {475}},\ \bibinfo {pages} {557} (\bibinfo {year} {1997})},\ \Eprint {http://arxiv.org/abs/astro-ph/9608028} {arXiv:astro-ph/9608028} \BibitemShut {NoStop}%
\bibitem [{\citenamefont {Spitler}\ \emph {et~al.}(2014)\citenamefont {Spitler} \emph {et~al.}}]{Spitler:2013uva}%
  \BibitemOpen
  \bibfield  {author} {\bibinfo {author} {\bibfnamefont {L.~G.}\ \bibnamefont {Spitler}} \emph {et~al.},\ }\href {\doibase 10.1088/2041-8205/780/1/L3} {\bibfield  {journal} {\bibinfo  {journal} {Astrophys. J. Lett.}\ }\textbf {\bibinfo {volume} {780}},\ \bibinfo {pages} {L3} (\bibinfo {year} {2014})},\ \Eprint {http://arxiv.org/abs/1309.4673} {arXiv:1309.4673 [astro-ph.HE]} \BibitemShut {NoStop}%
\bibitem [{\citenamefont {Macquart}\ \emph {et~al.}(2010)\citenamefont {Macquart}, \citenamefont {Kanekar}, \citenamefont {Frail},\ and\ \citenamefont {Ransom}}]{Macquart:2010vf}%
  \BibitemOpen
  \bibfield  {author} {\bibinfo {author} {\bibfnamefont {J.-P.}\ \bibnamefont {Macquart}}, \bibinfo {author} {\bibfnamefont {N.}~\bibnamefont {Kanekar}}, \bibinfo {author} {\bibfnamefont {D.}~\bibnamefont {Frail}}, \ and\ \bibinfo {author} {\bibfnamefont {S.}~\bibnamefont {Ransom}},\ }\href {\doibase 10.1088/0004-637X/715/2/939} {\bibfield  {journal} {\bibinfo  {journal} {Astrophys. J.}\ }\textbf {\bibinfo {volume} {715}},\ \bibinfo {pages} {939} (\bibinfo {year} {2010})},\ \Eprint {http://arxiv.org/abs/1004.1643} {arXiv:1004.1643 [astro-ph.GA]} \BibitemShut {NoStop}%
\bibitem [{\citenamefont {{Siemion}}\ \emph {et~al.}(2013)\citenamefont {{Siemion}}, \citenamefont {{Bailes}}, \citenamefont {{Bower}} \emph {et~al.}}]{2013IAUS..291...57S}%
  \BibitemOpen
  \bibfield  {author} {\bibinfo {author} {\bibfnamefont {A.}~\bibnamefont {{Siemion}}}, \bibinfo {author} {\bibfnamefont {M.}~\bibnamefont {{Bailes}}}, \bibinfo {author} {\bibnamefont {{Bower}}},  \emph {et~al.},\ }in\ \href {\doibase 10.1017/S1743921312023149} {\emph {\bibinfo {booktitle} {Neutron Stars and Pulsars: Challenges and Opportunities after 80 years}}},\ Vol.\ \bibinfo {volume} {291},\ \bibinfo {editor} {edited by\ \bibinfo {editor} {\bibfnamefont {J.}~\bibnamefont {{van Leeuwen}}}}\ (\bibinfo {year} {2013})\ pp.\ \bibinfo {pages} {57--57}\BibitemShut {NoStop}%
\bibitem [{\citenamefont {{Eatough}}\ \emph {et~al.}(2013)\citenamefont {{Eatough}}, \citenamefont {{Falcke}}, \citenamefont {{Karuppusamy}} \emph {et~al.}}]{2013Natur.501..391E}%
  \BibitemOpen
  \bibfield  {author} {\bibinfo {author} {\bibfnamefont {R.~P.}\ \bibnamefont {{Eatough}}}, \bibinfo {author} {\bibfnamefont {H.}~\bibnamefont {{Falcke}}}, \bibinfo {author} {\bibfnamefont {R.}~\bibnamefont {{Karuppusamy}}},  \emph {et~al.},\ }\href {\doibase 10.1038/nature12499} {\bibfield  {journal} {\bibinfo  {journal} {\nat}\ }\textbf {\bibinfo {volume} {501}},\ \bibinfo {pages} {391} (\bibinfo {year} {2013})},\ \Eprint {http://arxiv.org/abs/1308.3147} {arXiv:1308.3147 [astro-ph.GA]} \BibitemShut {NoStop}%
\bibitem [{\citenamefont {Torne}\ \emph {et~al.}(2023)\citenamefont {Torne} \emph {et~al.}}]{EHT:2023hcj}%
  \BibitemOpen
  \bibfield  {author} {\bibinfo {author} {\bibfnamefont {P.}~\bibnamefont {Torne}} \emph {et~al.} (\bibinfo {collaboration} {EHT}),\ }\href {\doibase 10.3847/1538-4357/acf4f2} {\bibfield  {journal} {\bibinfo  {journal} {Astrophys. J.}\ }\textbf {\bibinfo {volume} {959}},\ \bibinfo {pages} {14} (\bibinfo {year} {2023})},\ \Eprint {http://arxiv.org/abs/2308.15381} {arXiv:2308.15381 [astro-ph.HE]} \BibitemShut {NoStop}%
\bibitem [{\citenamefont {Macquart}\ and\ \citenamefont {Kanekar}(2015)}]{Macquart:2015jfa}%
  \BibitemOpen
  \bibfield  {author} {\bibinfo {author} {\bibfnamefont {J.-P.}\ \bibnamefont {Macquart}}\ and\ \bibinfo {author} {\bibfnamefont {N.}~\bibnamefont {Kanekar}},\ }\href {\doibase 10.1088/0004-637X/805/2/172} {\bibfield  {journal} {\bibinfo  {journal} {Astrophys. J.}\ }\textbf {\bibinfo {volume} {805}},\ \bibinfo {pages} {172} (\bibinfo {year} {2015})},\ \Eprint {http://arxiv.org/abs/1504.02492} {arXiv:1504.02492 [astro-ph.HE]} \BibitemShut {NoStop}%
\bibitem [{\citenamefont {Xie}\ \emph {et~al.}(2024)\citenamefont {Xie}, \citenamefont {Wang}, \citenamefont {Wang}, \citenamefont {Manchester},\ and\ \citenamefont {Hobbs}}]{Xie:2024rru}%
  \BibitemOpen
  \bibfield  {author} {\bibinfo {author} {\bibfnamefont {J.~T.}\ \bibnamefont {Xie}}, \bibinfo {author} {\bibfnamefont {J.~B.}\ \bibnamefont {Wang}}, \bibinfo {author} {\bibfnamefont {N.}~\bibnamefont {Wang}}, \bibinfo {author} {\bibfnamefont {R.}~\bibnamefont {Manchester}}, \ and\ \bibinfo {author} {\bibfnamefont {G.}~\bibnamefont {Hobbs}},\ }\href {\doibase 10.3847/2041-8213/ad2850} {\bibfield  {journal} {\bibinfo  {journal} {Astrophys. J. Lett.}\ }\textbf {\bibinfo {volume} {963}},\ \bibinfo {pages} {L39} (\bibinfo {year} {2024})},\ \Eprint {http://arxiv.org/abs/2402.11428} {arXiv:2402.11428 [astro-ph.HE]} \BibitemShut {NoStop}%
\bibitem [{\citenamefont {{Navarro}}\ \emph {et~al.}(1996)\citenamefont {{Navarro}}, \citenamefont {{Frenk}},\ and\ \citenamefont {{White}}}]{NFW}%
  \BibitemOpen
  \bibfield  {author} {\bibinfo {author} {\bibfnamefont {J.~F.}\ \bibnamefont {{Navarro}}}, \bibinfo {author} {\bibfnamefont {C.~S.}\ \bibnamefont {{Frenk}}}, \ and\ \bibinfo {author} {\bibfnamefont {S.~D.~M.}\ \bibnamefont {{White}}},\ }\href {\doibase 10.1086/177173} {\bibfield  {journal} {\bibinfo  {journal} {ApJ}\ }\textbf {\bibinfo {volume} {462}},\ \bibinfo {pages} {563} (\bibinfo {year} {1996})},\ \Eprint {http://arxiv.org/abs/astro-ph/9508025} {arXiv:astro-ph/9508025 [astro-ph]} \BibitemShut {NoStop}%
\bibitem [{\citenamefont {{Navarro}}\ \emph {et~al.}(1997)\citenamefont {{Navarro}}, \citenamefont {{Frenk}},\ and\ \citenamefont {{White}}}]{1997ApJ...490..493N}%
  \BibitemOpen
  \bibfield  {author} {\bibinfo {author} {\bibfnamefont {J.~F.}\ \bibnamefont {{Navarro}}}, \bibinfo {author} {\bibfnamefont {C.~S.}\ \bibnamefont {{Frenk}}}, \ and\ \bibinfo {author} {\bibfnamefont {S.~D.~M.}\ \bibnamefont {{White}}},\ }\href {\doibase 10.1086/304888} {\bibfield  {journal} {\bibinfo  {journal} {\apj}\ }\textbf {\bibinfo {volume} {490}},\ \bibinfo {pages} {493} (\bibinfo {year} {1997})},\ \Eprint {http://arxiv.org/abs/astro-ph/9611107} {arXiv:astro-ph/9611107 [astro-ph]} \BibitemShut {NoStop}%
\bibitem [{\citenamefont {Lin}\ and\ \citenamefont {Li}(2019)}]{lin2019dark}%
  \BibitemOpen
  \bibfield  {author} {\bibinfo {author} {\bibfnamefont {H.-N.}\ \bibnamefont {Lin}}\ and\ \bibinfo {author} {\bibfnamefont {X.}~\bibnamefont {Li}},\ }\href@noop {} {\bibfield  {journal} {\bibinfo  {journal} {Monthly Notices of the Royal Astronomical Society}\ }\textbf {\bibinfo {volume} {487}},\ \bibinfo {pages} {5679} (\bibinfo {year} {2019})}\BibitemShut {NoStop}%
\bibitem [{\citenamefont {Adair}\ \emph {et~al.}(2022)\citenamefont {Adair}, \citenamefont {Altenmüller}, \citenamefont {Anastassopoulos} \emph {et~al.}}]{Adair_2022}%
  \BibitemOpen
  \bibfield  {author} {\bibinfo {author} {\bibfnamefont {C.~M.}\ \bibnamefont {Adair}}, \bibinfo {author} {\bibfnamefont {K.}~\bibnamefont {Altenmüller}}, \bibinfo {author} {\bibfnamefont {V.}~\bibnamefont {Anastassopoulos}},  \emph {et~al.},\ }\href {\doibase 10.1038/s41467-022-33913-6} {\bibfield  {journal} {\bibinfo  {journal} {Nature Communications}\ }\textbf {\bibinfo {volume} {13}} (\bibinfo {year} {2022}),\ 10.1038/s41467-022-33913-6}\BibitemShut {NoStop}%
\bibitem [{\citenamefont {Remazeilles}\ \emph {et~al.}(2015)\citenamefont {Remazeilles}, \citenamefont {Dickinson}, \citenamefont {Banday}, \citenamefont {Bigot-Sazy},\ and\ \citenamefont {Ghosh}}]{Remazeilles:2014mba}%
  \BibitemOpen
  \bibfield  {author} {\bibinfo {author} {\bibfnamefont {M.}~\bibnamefont {Remazeilles}}, \bibinfo {author} {\bibfnamefont {C.}~\bibnamefont {Dickinson}}, \bibinfo {author} {\bibfnamefont {A.~J.}\ \bibnamefont {Banday}}, \bibinfo {author} {\bibfnamefont {M.~A.}\ \bibnamefont {Bigot-Sazy}}, \ and\ \bibinfo {author} {\bibfnamefont {T.}~\bibnamefont {Ghosh}},\ }\href {\doibase 10.1093/mnras/stv1274} {\bibfield  {journal} {\bibinfo  {journal} {Mon. Not. Roy. Astron. Soc.}\ }\textbf {\bibinfo {volume} {451}},\ \bibinfo {pages} {4311} (\bibinfo {year} {2015})},\ \Eprint {http://arxiv.org/abs/1411.3628} {arXiv:1411.3628 [astro-ph.IM]} \BibitemShut {NoStop}%
\bibitem [{\citenamefont {{Cleary}}\ \emph {et~al.}(2022)\citenamefont {{Cleary}}, \citenamefont {{Borowska}}, \citenamefont {{Breysse}} \emph {et~al.}}]{Cleary:2021dsp}%
  \BibitemOpen
  \bibfield  {author} {\bibinfo {author} {\bibfnamefont {K.~A.}\ \bibnamefont {{Cleary}}}, \bibinfo {author} {\bibfnamefont {J.}~\bibnamefont {{Borowska}}}, \bibinfo {author} {\bibnamefont {{Breysse}}},  \emph {et~al.},\ }\href {\doibase 10.3847/1538-4357/ac63cc} {\bibfield  {journal} {\bibinfo  {journal} {\apj}\ }\textbf {\bibinfo {volume} {933}},\ \bibinfo {eid} {182} (\bibinfo {year} {2022})},\ \Eprint {http://arxiv.org/abs/2111.05927} {arXiv:2111.05927 [astro-ph.CO]} \BibitemShut {NoStop}%
\bibitem [{\citenamefont {Bacon}\ \emph {et~al.}(2020)\citenamefont {Bacon} \emph {et~al.}}]{SKA:2018ckk}%
  \BibitemOpen
  \bibfield  {author} {\bibinfo {author} {\bibfnamefont {D.~J.}\ \bibnamefont {Bacon}} \emph {et~al.} (\bibinfo {collaboration} {SKA}),\ }\href {\doibase 10.1017/pasa.2019.51} {\bibfield  {journal} {\bibinfo  {journal} {Publ. Astron. Soc. Austral.}\ }\textbf {\bibinfo {volume} {37}},\ \bibinfo {pages} {e007} (\bibinfo {year} {2020})},\ \Eprint {http://arxiv.org/abs/1811.02743} {arXiv:1811.02743 [astro-ph.CO]} \BibitemShut {NoStop}%
\bibitem [{\citenamefont {Bull}\ \emph {et~al.}(2015)\citenamefont {Bull}, \citenamefont {Ferreira}, \citenamefont {Patel},\ and\ \citenamefont {Santos}}]{Bull:2014rha}%
  \BibitemOpen
  \bibfield  {author} {\bibinfo {author} {\bibfnamefont {P.}~\bibnamefont {Bull}}, \bibinfo {author} {\bibfnamefont {P.~G.}\ \bibnamefont {Ferreira}}, \bibinfo {author} {\bibfnamefont {P.}~\bibnamefont {Patel}}, \ and\ \bibinfo {author} {\bibfnamefont {M.~G.}\ \bibnamefont {Santos}},\ }\href {\doibase 10.1088/0004-637X/803/1/21} {\bibfield  {journal} {\bibinfo  {journal} {Astrophys. J.}\ }\textbf {\bibinfo {volume} {803}},\ \bibinfo {pages} {21} (\bibinfo {year} {2015})},\ \Eprint {http://arxiv.org/abs/1405.1452} {arXiv:1405.1452 [astro-ph.CO]} \BibitemShut {NoStop}%
\bibitem [{\citenamefont {Abdurashidova}\ \emph {et~al.}(2022)\citenamefont {Abdurashidova} \emph {et~al.}}]{HERA:2021bsv}%
  \BibitemOpen
  \bibfield  {author} {\bibinfo {author} {\bibfnamefont {Z.}~\bibnamefont {Abdurashidova}} \emph {et~al.} (\bibinfo {collaboration} {HERA}),\ }\href {\doibase 10.3847/1538-4357/ac1c78} {\bibfield  {journal} {\bibinfo  {journal} {Astrophys. J.}\ }\textbf {\bibinfo {volume} {925}},\ \bibinfo {pages} {221} (\bibinfo {year} {2022})},\ \Eprint {http://arxiv.org/abs/2108.02263} {arXiv:2108.02263 [astro-ph.CO]} \BibitemShut {NoStop}%
\bibitem [{\citenamefont {Keane}\ and\ \citenamefont {Kramer}(2008)}]{Keane:2008jj}%
  \BibitemOpen
  \bibfield  {author} {\bibinfo {author} {\bibfnamefont {E.~F.}\ \bibnamefont {Keane}}\ and\ \bibinfo {author} {\bibfnamefont {M.}~\bibnamefont {Kramer}},\ }\href {\doibase 10.1111/j.1365-2966.2008.14045.x} {\bibfield  {journal} {\bibinfo  {journal} {Mon. Not. Roy. Astron. Soc.}\ }\textbf {\bibinfo {volume} {391}},\ \bibinfo {pages} {2009} (\bibinfo {year} {2008})},\ \Eprint {http://arxiv.org/abs/0810.1512} {arXiv:0810.1512 [astro-ph]} \BibitemShut {NoStop}%
\bibitem [{\citenamefont {Lu}\ \emph {et~al.}(2013)\citenamefont {Lu}, \citenamefont {Do}, \citenamefont {Ghez}, \citenamefont {Morris}, \citenamefont {Yelda},\ and\ \citenamefont {Matthews}}]{Lu:2013sn}%
  \BibitemOpen
  \bibfield  {author} {\bibinfo {author} {\bibfnamefont {J.~R.}\ \bibnamefont {Lu}}, \bibinfo {author} {\bibfnamefont {T.}~\bibnamefont {Do}}, \bibinfo {author} {\bibfnamefont {A.~M.}\ \bibnamefont {Ghez}}, \bibinfo {author} {\bibfnamefont {M.~R.}\ \bibnamefont {Morris}}, \bibinfo {author} {\bibfnamefont {S.}~\bibnamefont {Yelda}}, \ and\ \bibinfo {author} {\bibfnamefont {K.}~\bibnamefont {Matthews}},\ }\href {\doibase 10.1088/0004-637X/764/2/155} {\bibfield  {journal} {\bibinfo  {journal} {Astrophys. J.}\ }\textbf {\bibinfo {volume} {764}},\ \bibinfo {pages} {155} (\bibinfo {year} {2013})},\ \Eprint {http://arxiv.org/abs/1301.0540} {arXiv:1301.0540 [astro-ph.SR]} \BibitemShut {NoStop}%
\bibitem [{\citenamefont {Bartko}\ \emph {et~al.}(2010)\citenamefont {Bartko} \emph {et~al.}}]{Bartko:2009qn}%
  \BibitemOpen
  \bibfield  {author} {\bibinfo {author} {\bibfnamefont {H.}~\bibnamefont {Bartko}} \emph {et~al.},\ }\href {\doibase 10.1088/0004-637X/708/1/834} {\bibfield  {journal} {\bibinfo  {journal} {Astrophys. J.}\ }\textbf {\bibinfo {volume} {708}},\ \bibinfo {pages} {834} (\bibinfo {year} {2010})},\ \Eprint {http://arxiv.org/abs/0908.2177} {arXiv:0908.2177 [astro-ph.GA]} \BibitemShut {NoStop}%
\bibitem [{\citenamefont {Aguilera}\ \emph {et~al.}(2008)\citenamefont {Aguilera}, \citenamefont {Pons},\ and\ \citenamefont {Miralles}}]{Aguilera:2007xk}%
  \BibitemOpen
  \bibfield  {author} {\bibinfo {author} {\bibfnamefont {D.~N.}\ \bibnamefont {Aguilera}}, \bibinfo {author} {\bibfnamefont {J.~A.}\ \bibnamefont {Pons}}, \ and\ \bibinfo {author} {\bibfnamefont {J.~A.}\ \bibnamefont {Miralles}},\ }\href {\doibase 10.1051/0004-6361:20078786} {\bibfield  {journal} {\bibinfo  {journal} {Astron. Astrophys.}\ }\textbf {\bibinfo {volume} {486}},\ \bibinfo {pages} {255} (\bibinfo {year} {2008})},\ \Eprint {http://arxiv.org/abs/0710.0854} {arXiv:0710.0854 [astro-ph]} \BibitemShut {NoStop}%
\bibitem [{\citenamefont {Philippov}\ \emph {et~al.}(2014)\citenamefont {Philippov}, \citenamefont {Tchekhovskoy},\ and\ \citenamefont {Li}}]{Philippov:2013aha}%
  \BibitemOpen
  \bibfield  {author} {\bibinfo {author} {\bibfnamefont {A.}~\bibnamefont {Philippov}}, \bibinfo {author} {\bibfnamefont {A.}~\bibnamefont {Tchekhovskoy}}, \ and\ \bibinfo {author} {\bibfnamefont {J.~G.}\ \bibnamefont {Li}},\ }\href {\doibase 10.1093/mnras/stu591} {\bibfield  {journal} {\bibinfo  {journal} {Mon. Not. Roy. Astron. Soc.}\ }\textbf {\bibinfo {volume} {441}},\ \bibinfo {pages} {1879} (\bibinfo {year} {2014})},\ \Eprint {http://arxiv.org/abs/1311.1513} {arXiv:1311.1513 [astro-ph.HE]} \BibitemShut {NoStop}%
\bibitem [{\citenamefont {Wainscoat}\ \emph {et~al.}(1992)\citenamefont {Wainscoat}, \citenamefont {Cohen}, \citenamefont {Volk}, \citenamefont {Walker},\ and\ \citenamefont {Schwartz}}]{spirals}%
  \BibitemOpen
  \bibfield  {author} {\bibinfo {author} {\bibfnamefont {R.}~\bibnamefont {Wainscoat}}, \bibinfo {author} {\bibfnamefont {M.}~\bibnamefont {Cohen}}, \bibinfo {author} {\bibfnamefont {K.}~\bibnamefont {Volk}}, \bibinfo {author} {\bibfnamefont {H.}~\bibnamefont {Walker}}, \ and\ \bibinfo {author} {\bibfnamefont {D.}~\bibnamefont {Schwartz}},\ }\href {\doibase 10.1086/191733} {\bibfield  {journal} {\bibinfo  {journal} {The Astrophysical Journal Supplement Series}\ }\textbf {\bibinfo {volume} {83}} (\bibinfo {year} {1992}),\ 10.1086/191733}\BibitemShut {NoStop}%
\bibitem [{\citenamefont {Ridley}\ and\ \citenamefont {Lorimer}(2010)}]{ridley_isolated_2010}%
  \BibitemOpen
  \bibfield  {author} {\bibinfo {author} {\bibfnamefont {J.~P.}\ \bibnamefont {Ridley}}\ and\ \bibinfo {author} {\bibfnamefont {D.~R.}\ \bibnamefont {Lorimer}},\ }\href {\doibase 10.1111/j.1365-2966.2010.16342.x} {\bibfield  {journal} {\bibinfo  {journal} {Monthly Notices of the Royal Astronomical Society}\ }\textbf {\bibinfo {volume} {404}},\ \bibinfo {pages} {1081} (\bibinfo {year} {2010})},\ \bibinfo {note} {arXiv:1001.2483 [astro-ph]}\BibitemShut {NoStop}%
\bibitem [{\citenamefont {Lorimer}\ and\ \citenamefont {Kramer}(2012)}]{lorimer2012handbook}%
  \BibitemOpen
  \bibfield  {author} {\bibinfo {author} {\bibfnamefont {D.~R.}\ \bibnamefont {Lorimer}}\ and\ \bibinfo {author} {\bibfnamefont {M.}~\bibnamefont {Kramer}},\ }\href@noop {} {\emph {\bibinfo {title} {Handbook of Pulsar Astronomy}}}\ (\bibinfo  {publisher} {Cambridge University Press},\ \bibinfo {year} {2012})\BibitemShut {NoStop}%
\bibitem [{\citenamefont {{Bhattacharya}}\ \emph {et~al.}(1992)\citenamefont {{Bhattacharya}}, \citenamefont {{Wijers}}, \citenamefont {{Hartman}},\ and\ \citenamefont {{Verbunt}}}]{death_line}%
  \BibitemOpen
  \bibfield  {author} {\bibinfo {author} {\bibfnamefont {D.}~\bibnamefont {{Bhattacharya}}}, \bibinfo {author} {\bibfnamefont {R.~A.~M.~J.}\ \bibnamefont {{Wijers}}}, \bibinfo {author} {\bibfnamefont {J.~W.}\ \bibnamefont {{Hartman}}}, \ and\ \bibinfo {author} {\bibfnamefont {F.}~\bibnamefont {{Verbunt}}},\ }\href@noop {} {\bibfield  {journal} {\bibinfo  {journal} {\aap}\ }\textbf {\bibinfo {volume} {254}},\ \bibinfo {pages} {198} (\bibinfo {year} {1992})}\BibitemShut {NoStop}%
\end{thebibliography}%
\bibliographystyle{apsrev4-1}

\end{document}